\documentclass[aps,prb,twocolumn,superscriptaddress,longbibliography]{revtex4-2}

\usepackage{mhchem,graphicx,longtable,xfrac}
\usepackage[colorlinks=true, citecolor=blue, linkcolor=blue, breaklinks=true]{hyperref}
\usepackage{amssymb}
\usepackage{graphicx}
\usepackage{amsmath}
\usepackage{multirow}
\usepackage{tabularx}
\usepackage{array}

\usepackage{times}
\usepackage[usenames, dvipsnames]{color}

\usepackage{bm}
\usepackage{ulem}

\usepackage{dcolumn}

\begin{document}

\newcommand {\beq} {\begin{equation}}
\newcommand {\eeq} {\end{equation}}
\newcommand {\bqa} {\begin{eqnarray}}
\newcommand {\eqa} {\end{eqnarray}}
\newcommand {\ba} {\ensuremath{b^\dagger}}
\newcommand {\Ma} {\ensuremath{M^\dagger}}
\newcommand {\psia} {\ensuremath{\psi^\dagger}}
\newcommand {\psita} {\ensuremath{\tilde{\psi}^\dagger}}
\newcommand{\lp} {\ensuremath{{\lambda '}}}
\newcommand{\A} {\ensuremath{{\bf A}}}
\newcommand{\Q} {\ensuremath{{\bf Q}}}
\newcommand{\kk} {\ensuremath{{\bf k}}}
\newcommand{\qq} {\ensuremath{{\bf q}}}
\newcommand{\kp} {\ensuremath{{\bf k'}}}
\newcommand{\rr} {\ensuremath{{\bf r}}}
\newcommand{\rp} {\ensuremath{{\bf r'}}}
\newcommand {\ep} {\ensuremath{\epsilon}}
\newcommand{\nbr} {\ensuremath{\langle ij \rangle}}
\newcommand {\no} {\nonumber}
\newcommand{\up} {\ensuremath{\uparrow}}
\newcommand{\dn} {\ensuremath{\downarrow}}

\title{Microscopic investigation of enhanced Pauli paramagnetism in metallic Pu$_2$C$_3$}

\author{R. Yamamoto}
\thanks{These authors contributed equally to this work.}
\affiliation{Materials Physics and Applications $-$ Quantum, Los Alamos National Laboratory, Los Alamos, New Mexico  87545, USA}

\author{M. S. Cook}
\thanks{These authors contributed equally to this work.}
\affiliation{MST-16 $-$ Nuclear Materials Science, Los Alamos National Laboratory, Los Alamos, New Mexico  87545, USA}

\author{A. R. Altenhof}
\thanks{These authors contributed equally to this work.}
\affiliation{Materials Physics and Applications $-$ Quantum, Los Alamos National Laboratory, Los Alamos, New Mexico  87545, USA}

\author{P. Sherpa}
\affiliation{Materials Physics and Applications $-$ Quantum, Los Alamos National Laboratory, Los Alamos, New Mexico 87545, USA}
\affiliation{Department of Physics and Astronomy, University of California, Davis, California 95616, USA}

\author{S. Park}
\affiliation{Materials Physics and Applications $-$ Quantum, Los Alamos National Laboratory, Los Alamos, New Mexico 87545, USA}

\author{J. D. Thompson}
\affiliation{Materials Physics and Applications $-$ Quantum, Los Alamos National Laboratory, Los Alamos, New Mexico 87545, USA}

\author{H. E. Mason}
\affiliation{Chemistry Division, Los Alamos National Laboratory, Los Alamos, New Mexico  87545, USA}

\author{D. C. Arellano}
\affiliation{MST-16 $-$ Nuclear Materials Science, Los Alamos National Laboratory, Los Alamos, New Mexico  87545, USA}

\author{D. V. Prada}
\affiliation{MST-16 $-$ Nuclear Materials Science, Los Alamos National Laboratory, Los Alamos, New Mexico  87545, USA}

\author{P. H. Tobash}
\affiliation{MST-16 $-$ Nuclear Materials Science, Los Alamos National Laboratory, Los Alamos, New Mexico  87545, USA}

\author{F. Ronning}
\affiliation{Materials Physics and Applications $-$ Quantum, Los Alamos National Laboratory, Los Alamos, New Mexico  87545, USA}

\author{E. D. Bauer}
\affiliation{Materials Physics and Applications $-$ Quantum, Los Alamos National Laboratory, Los Alamos, New Mexico  87545, USA}

\author{N. Harrison}
\affiliation{National High Magnetic Field Laboratory, Los Alamos National Laboratory, Los Alamos, New Mexico  87545, USA}

\author{W. A. Phelan}
\affiliation{MST-16 $-$ Nuclear Materials Science, Los Alamos National Laboratory, Los Alamos, New Mexico  87545, USA}

\author{A. P. Dioguardi}\thanks{apd@lanl.gov} 
\affiliation{Materials Physics and Applications $-$ Quantum, Los Alamos National Laboratory, Los Alamos, New Mexico  87545, USA}

\author{M. Hirata}\thanks{mhirata@lanl.gov}
\affiliation{Materials Physics and Applications $-$ Quantum, Los Alamos National Laboratory, Los Alamos, New Mexico  87545, USA}

\date{\today}

\begin{abstract}
A combined study of the structural and electronic properties of polycrystalline Pu$_2$C$_3$ is reported based on x-ray diffraction, specific heat, magnetic susceptibility, ${}^{13}$C nuclear magnetic resonance (NMR), and band structure calculations. X-ray diffraction reveals a global noncentrosymmetric cubic lattice, with a nearest-neighbor C--C bond length of $r=$~1.38\,\r{A}. ${}^{13}$C NMR measurements indicate that the global cubic symmetry is locally broken, revealing two unique carbon environments. Magnetic susceptibility suggests enhanced Pauli paramagnetism, and specific heat reveals a moderately large electronic Sommerfeld coefficient $\gamma$ = 45 mJ\,mol$_{\mathrm{Pu}}^{-1}$\,K$^{-2}$, with a Wilson ratio $R_W \approx 1.3$ further indicating moderate correlations. ${}^{13}$C nuclear spin-lattice relaxation rate ($1/T_1$) and Knight shift ($K$) measurements find metallic Korringa behavior (i.e., $T_1TK^2=$~const.) with modest ferromagnetic spin fluctuations at low temperature. Taken together, the data point to a delocalized nature of a narrow 5$f$-electron band with weak electronic correlations. Density functional theory band-structure calculations confirm the appearance of such narrow 5$f$ bands near the Fermi level. Our data provide prime evidence for a plutonium-based metallic system with weak electronic correlations, which sheds new light on the understanding of complex paramagnetism in actinide-based metallic compounds.
\end{abstract}


\maketitle

\section{Introduction}
Electron correlations in $d$- and $f$-electron materials provide a plethora of intriguing phenomena that often originate from a competition between localization and itinerancy. For the 5$f$ electrons of plutonium this competition is nearly balanced, resulting in a variety of ground state properties with rich and complex nature, ranging from enhanced electron mass and magnetism to unconventional superconductivity~\cite{Bauer_2015_PlutoniumBasedHeavy, Sarrao_2002_Plutoniumbasedsuperconductivity, Shim_2007_Fluctuatingvalencecorrelated}.

As compared with other lanthanide and actinide compounds, the---often anomalous---electronic properties of Pu-based metals are relatively unexplored. $\delta$-Pu offers one such example~\cite{Shim_2007_Fluctuatingvalencecorrelated, Harrison_2024_Indicationsflatbands}, which is a paramagnetic metal realized on the verge of magnetic order and exhibits the largest quasiparticle mass enhancement of any elemental metal, with a residual specific heat (Sommerfeld) coefficient $\gamma = 30-60$\,mJ\,mol$_{\mathrm{Pu}}^{-1}$\,K$^{-2}$~\cite{Lashley_2003_ExperimentalElectronicHeat, Javorsky_2006_SpecificHeat}. Strong fluctuations of Pu valence [between Pu$^{2+}(5f^6)$, Pu$^{3+}(5f^5)$, and Pu$^{4+}(5f^4)$] were theoretically predicted~\cite{Shim_2007_Fluctuatingvalencecorrelated} and later confirmed experimentally~\cite{Janoschek_2015_valencefluctuatingground} to cause intense spin fluctuations in a narrow 5$f$-electron band near the Fermi level $E_\mathrm{F}$. Valence fluctuations seem to play an equally important role in many other Pu-based metals including PuCoGa$_5$~\cite{Ramshaw_2015_Avoidedvalencetransition}, which shows a similar mass enhancement ($\gamma=$ 77 mJ mol$_{\mathrm{Pu}}^{-1}$ K$^{-2}$) and exhibits the highest superconducting transition temperature among heavy-fermion compounds ($T_c=$~18.5~K)~\cite{Sarrao_2002_Plutoniumbasedsuperconductivity}.

Plutonium carbides have been studied for nuclear fuel applications~\cite{Morss_2011_ChemistryActinideTransactinide} and are often assumed to be metallic. Four plutonium carbide binary compounds are known to exist: Pu$_3$C$_2$, PuC$_{1-\mathrm{x}}$, Pu$_2$C$_3$, and
PuC$_2$~\cite{Clark_2006_PlutoniumPlutoniumCompounds}. Two of these can be synthesized in pure form: the first is face-centered-cubic (fcc) plutonium monocarbide, PuC$_{1-\mathrm{x}}$, which has an NaCl-type structure, and is a Type-I antiferromagnet that orders at around 100\,K~\cite{Green_1970_Crystallographicmagneticordering, Ellinger_1968_ConstitutionPlutoniumAlloys}. The second binary compound is body-centered-cubic (bcc) plutonium sesquicarbide, Pu$_2$C$_3$, that remains paramagnetic down to 4\,K~\cite{Green_1970_Crystallographicmagneticordering, Raphael_1969_Susceptibilitiesmagnetiquesdes}. No transport and thermodynamic data exist at low temperatures, and the ground state properties of Pu$_2$C$_3$ are not well understood. The photoemission spectrum of Pu$_2$C$_3$ is similar to that of $\delta$-Pu~\cite{Gouder_2007_Variability5fstates, Joyce_2003_PhotoemissionElectronicStructure}, which is used to claim that the similar physical properties of compounds---together with the normal state of PuCoGa$_5$---can be explained by an admixture of $5f$-electron configurations~\cite{Joyce_2003_PhotoemissionElectronicStructure, Havela_2009_Magneticpropertiesplutonium}.

The crystal structure of the sesquicarbide family $R_2$C$_3$ (with $R$ being rare earth or actinide) offers in general unique electronic properties, linked to their bcc lattice without inversion symmetry~\cite{Atoji_1961_Neutron‐DiffractionStudiesLa_2C_3}. A recent band-structure calculation based on density functional theory (DFT) reveals that in rare-earth $R_2$C$_3$ this crystal structure leads to topological band crossings with unconventional multifold Weyl points (although locating far away from $E_\mathrm{F}$)~\cite{Jin_2021_Sixfoldfourfoldthreefold}. The absence of inversion symmetry also causes exotic non-centrosymmetric superconductivity in La$_2$C$_3$~\cite{Kim_2007_Strongelectronphonon, Sugawara_2007_Anomaloussuperconductinggap} and Y$_2$C$_3$~\cite{Akutagawa_2007_SuperconductivityY_2C_3Investigated, Chen_2011_Evidencenodalgap}, which combined with spin-orbit coupling offers an anomalous mixing of spin-singlet and triplet channels in the Cooper pairing~\cite{Smidman_2017_Superconductivityspin–orbitcoupling}. Uranium sesquicarbide U$_2$C$_3$, although first reported as an antiferromagnet~\cite{DeNovion_1965_Existencedunetransition,Raphael_1969_Susceptibilitiesmagnetiquesdes}, has been shown to be a good metal with enhanced Pauli paramagnetism~\cite{Koelling_1985_felectronhybridization}, with notable spin fluctuations developing at low temperature~\cite{Cornelius_1999_ElectronicpropertiesUX_3} reminiscent of the intermetallic actinide fluctuator family, such as cubic AuCu$_3$-type UAl$_3$ and USn$_3$ as well as cubic Laves phase UAl$_2$~\cite{Jullien_1976_Existencedunetransition, BealMonod_1968_TemperatureDependenceSpin, Trainor_1975_SpecificHeatSpin, Cornelius_1999_ElectronicpropertiesUX_3}.

Here, we report structural and electronic properties studies of polycrystalline Pu$_2$C$_3$ by combining x-ray diffraction (XRD), magnetic susceptibility, specific heat, nuclear magnetic resonance (NMR), and DFT band-structure calculations. XRD-based structural analysis reveals a global cubic lattice symmetry, with a nearest-neighbor C--C bond length of $r=$~1.38\,\r{A} and a nearest-neighbor Pu distance 3.364\,\r{A}, which is slightly smaller than the Hill limit $\approx$\,3.40\,\r{A}~\cite{Hill_1970_Earlyactinidesperiodic, Brodsky_1978_Magneticpropertiesactinide}. ${}^{13}$C NMR spectral data as well as spin echo decay measurements unravel two nonequivalent ${}^{13}$C sites, suggesting that the two nearest-neighbor C atoms sit in a slightly different local magnetic environment. Specific heat and magnetic susceptibility measurements find a moderately large value of the Sommerfeld coefficient, $\gamma$ = 45\,mJ\,mol$_{\mathrm{Pu}}^{-1}$\,K$^{-2}$, and enhanced Pauli paramagnetic behavior, suggestive of a weakly correlated metal. NMR Korringa analysis indicates a representative metallic property with moderate ferromagnetic spin fluctuations especially at lower temperature, demonstrating a delocalized 5$f$-electron character with weak electronic correlations. Band-structure calculations confirm a narrow 5$f$-band near $E_F$ in good agreement with these results. Taken together, the data suggest that Pu$_2$C$_3$ is a narrow-5$f$-band, weakly correlated metal with enhanced Pauli paramagnetism and moderate ferromagnetic fluctuations.

\section{Experimental and theoretical details}

\subsection{Synthesis and characterization}
Powder Pu$_2$C$_3$ was prepared  by arc melting ${}^{13}$C powder and $\alpha$-Pu together on a water-cooled copper hearth under an ultra-high-purity Ar atmosphere and a Zr getter. The Pu$_2$C$_3$ button was melted and flipped several times to improve homogeneity. The final Pu$_2$C$_3$ sample was wrapped in Ta foil, sealed in a quartz tube under vacuum of $\sim 10^{-4}$\,Torr, and annealed for two weeks at 1000\,${}^{\circ}$C.

All samples of Pu$_2$C$_3$ were isotopically enriched and labeled with ${}^{13}$C atoms by 99\,$\%$ for ${}^{13}$C NMR. The isotopic content of Pu is (by weight\,$\%$): 93.96\,$\%$ $^{239}$Pu, 5.908\,$\%$ $^{240}$Pu, 0.098\,$\%$ $^{241}$Pu, 0.025\,$\%$ $^{242}$Pu, and 0.013\,$\%$ $^{238}$Pu. 

Powder XRD data for the prepared Pu$_2$C$_3$ sample were collected at room temperature using a Bruker D8 Discover with DAVINCI design equipped with a Lynx-Eye detector and a Cu K$\alpha$ radiation source. Subsequent model fits to the data were performed using Bruker’s TOPAS v.7 software. A high-quality single Pu$_2$C$_3$ phase was verified.

\subsection{Magnetic susceptibility and heat capacity measurements}
Magnetic susceptibility and isothermal magnetization were measured using a Quantum Design physical properties measurement system (PPMS) Dynacool instrument equipped with the vibrating sample magnetometer option with a temperature between $T$ = 2 to 300\,K and magnetic fields to $B$\,($=\mu_{0}H$) = 6.0\,T. To safely measure Pu-containing samples and to prevent contamination hazards, the magnetization samples were mounted using small amounts of plastic wrap and GE varnish glued inside a plastic straw. The plastic straw was then sealed with custom caps that contain micron sized platinum frits which were glued to the end of the straw with
epoxy. The background signal from GE varnish was subtracted from the susceptibility data, in which we expect $\sim\pm15\,\%$ uncertainty due to the ambiguity in estimated mass of GE varnish.

Heat capacity data were also collected using the PPMS. The semiadiabatic pulse technique was used for data collection. Specific heat $C/T$ was measured between $T$ = 6.3 and 300~K.

\subsection{NMR measurements}
Powder Pu$_2$C$_3$ samples of a crystallite size larger than 20\,$\mu$m were loaded in a copper solenoid coil for NMR that is encapsulated in a Stycast 1266 epoxy cube to avoid radioactive contamination~\cite{Yasuoka_2019_Plutonium239NMR}. Two titanium frits with 2\,$\mu$m diameter pores were installed at both ends of coil’s axis to seal the containment and to allow thermal contact with ${}^4$He exchange gas in the cryostat. 

${}^{13}$C (nuclear spin $I$ = 1/2) NMR measurements were performed in Pu$_2$C$_3$ powder samples using commercially available spectrometers (REDSTONE, Tecmag Inc.) and standard Hahn-echo pulse sequences. A conventional superconducting magnet and $^{4}$He variable-temperature-insert cryostat were used to apply a magnetic field up to $B$ = 12.0\,T and control temperatures between $T$ = 1.8 and 300\,K. Typical pulse conditions consisted of a pulse length $t_{\pi/2}$ = 2 - 8\,$\mu$s and an inter-pulse delay $\tau$ = 15 - 40\,$\mu$s. The spin-lattice (longitudinal) relaxation time $T_1$ was determined from a single-exponential fit to the recovery of nuclear magnetization after inversion. To evaluate the nearest-neighbor C--C bond length, spin-echo decay (SED) oscillations were measured. Powder NMR spectrum and SED oscillations were analyzed and fit using SIMPSON version 4.2.1. \cite{Bak_2011_SIMPSONgeneralsimulation,Tosner_2014_Computerintensivesimulation}.

${}^{13}$C shifts were calculated in reference to the $^{63}$Cu NMR signal of the copper coil wrapped around the sample and the $^{23}$Na NMR signal from NaCl solution (0.1\,M in H$_2$O) measured in a separate coil. ${}^{13}$C shifts were defined relative to the shift in tetramethylsilane (TMS) using $^{23}$Na resonance and conventions defined in IUPAC Recommendations~\cite{Harris_2001_NMRnomenclatureNuclear,Harris_2008_FurtherconventionsNMR}. Temperatures below 4\,K were calibrated by $^{63}T_1$ measurements of Cu in the NMR coil based on the Korringa constant of metallic $^{63}$Cu spins ($^{63}T_{1}T$ = 1.27\,sec\,K)~\cite{Carter_1977_MetallicshiftsNMR}.

 \subsection{Band-structure calculations}
The electronic structure was calculated within density functional theory (DFT) under the generalized gradient approximation with the Perdew-Burke-Ernzerhof exchange correlation functional~\cite{Perdew_1996_GeneralizedGradientApproximation}  as implemented within the WIEN2K code~\cite{Blaha_2019_WIEN2kAPWloprogram}. Spin-orbit coupling was included via a second variational step with a relativistic local $p$-orbital on the Pu sites added to the set of basis functions.

 \begin{figure}[t]
	\includegraphics[width=6.5cm]{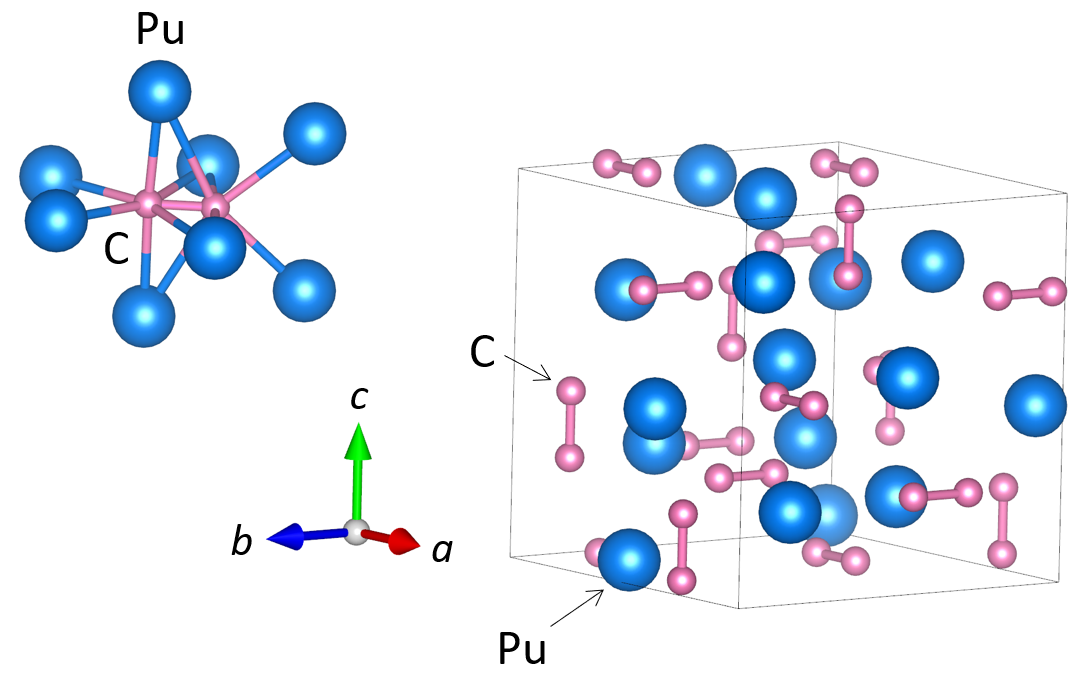}
	\caption{Crystallographic structure of Pu$_2$C$_3$. Pu atoms are denoted by large blue spheres and C atoms by small pink spheres. The local environment of a nearest neighbor C--C pair is shown on the left, and the full unit cell (gray box) is shown on the right.}
	\label{fig:1}
\end{figure}

\begin{figure}[t]
	\includegraphics[width=8.5cm]{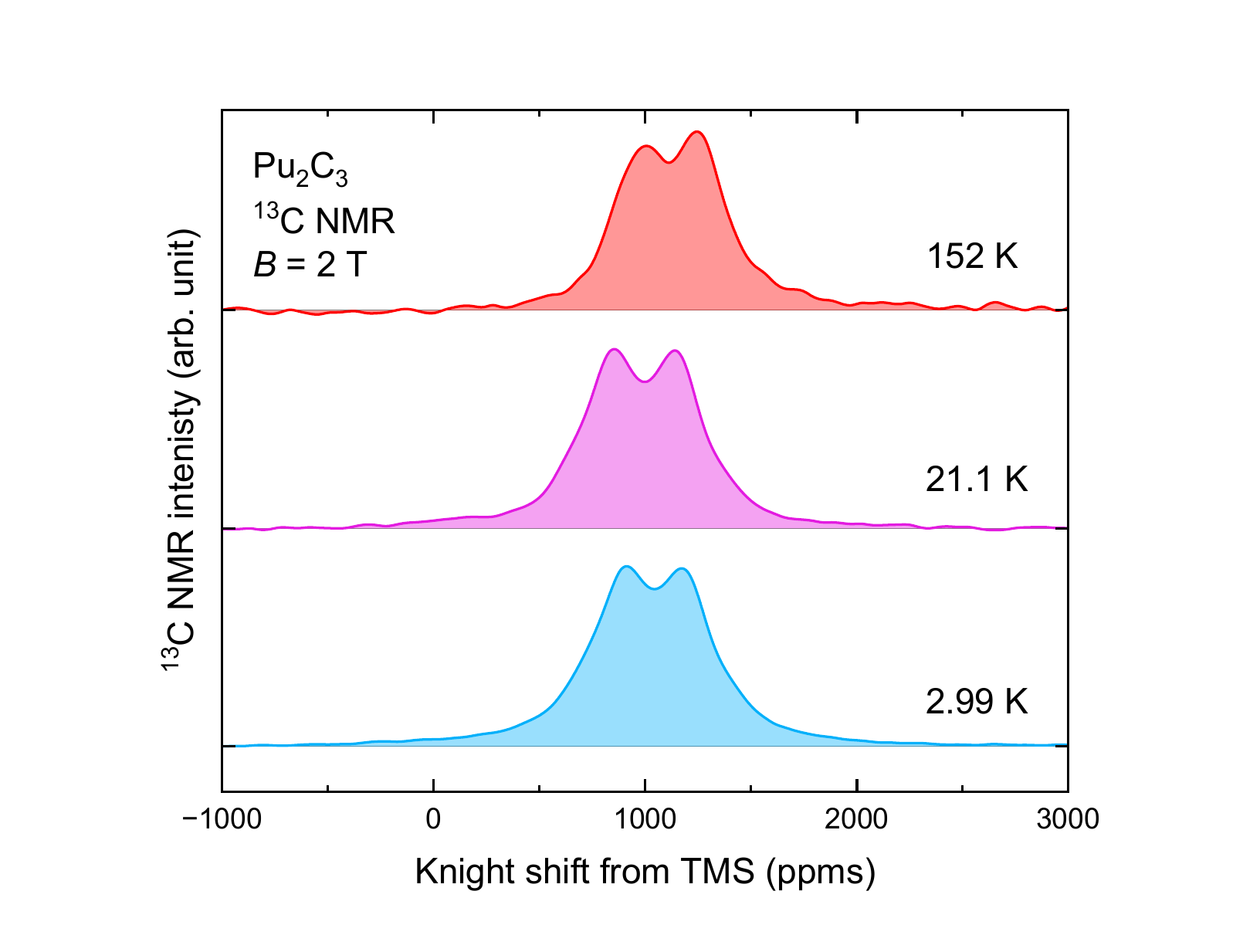}
	\caption{${}^{13}$C NMR spectra of Pu$_2$C$_3$ powder at selected temperatures at 2.0\,T.}
	\label{fig:5}
\end{figure}

\section{Results and Discussion}

\subsection{Structural analysis}

The powder samples used in all of our experiments were characterized with powder XRD. The Rietveld refinement reveals 99.8\,w$\%$ of Pu$_2$C$_3$ and 0.2\,w$\%$ of PuC. The unit cell of Pu$_2$C$_3$ with a body-centered-cubic (bcc) lattice contains eight formula units, with 16 plutonium atoms locating in 16($c$) (.3.) and 24 carbon atoms in 24($d$) (2..) spacial positions. The lattice has the space group $I\overline{4}3d$ (space group No.~220) with the lattice constant of $a$ = 8.129\,\AA, all in good agreement with previous XRD~\cite{Zachariasen_1952_Crystalchemicalstudies}.

Structural details were microscopically investigated using ${}^{13}$C NMR in Pu$_{2}$C$_{3}$ powder. Figure~\ref{fig:5} presents typical ${}^{13}$C spectra recorded at 2.0\,T, which show two overlapping lines with small temperature ($T$) dependence in its line shape and an average (center of gravity) isotropic shift value of $\sim$ 1000\,ppm at low $T$. 

\begin{figure}[!t]
	\includegraphics[width=6cm]{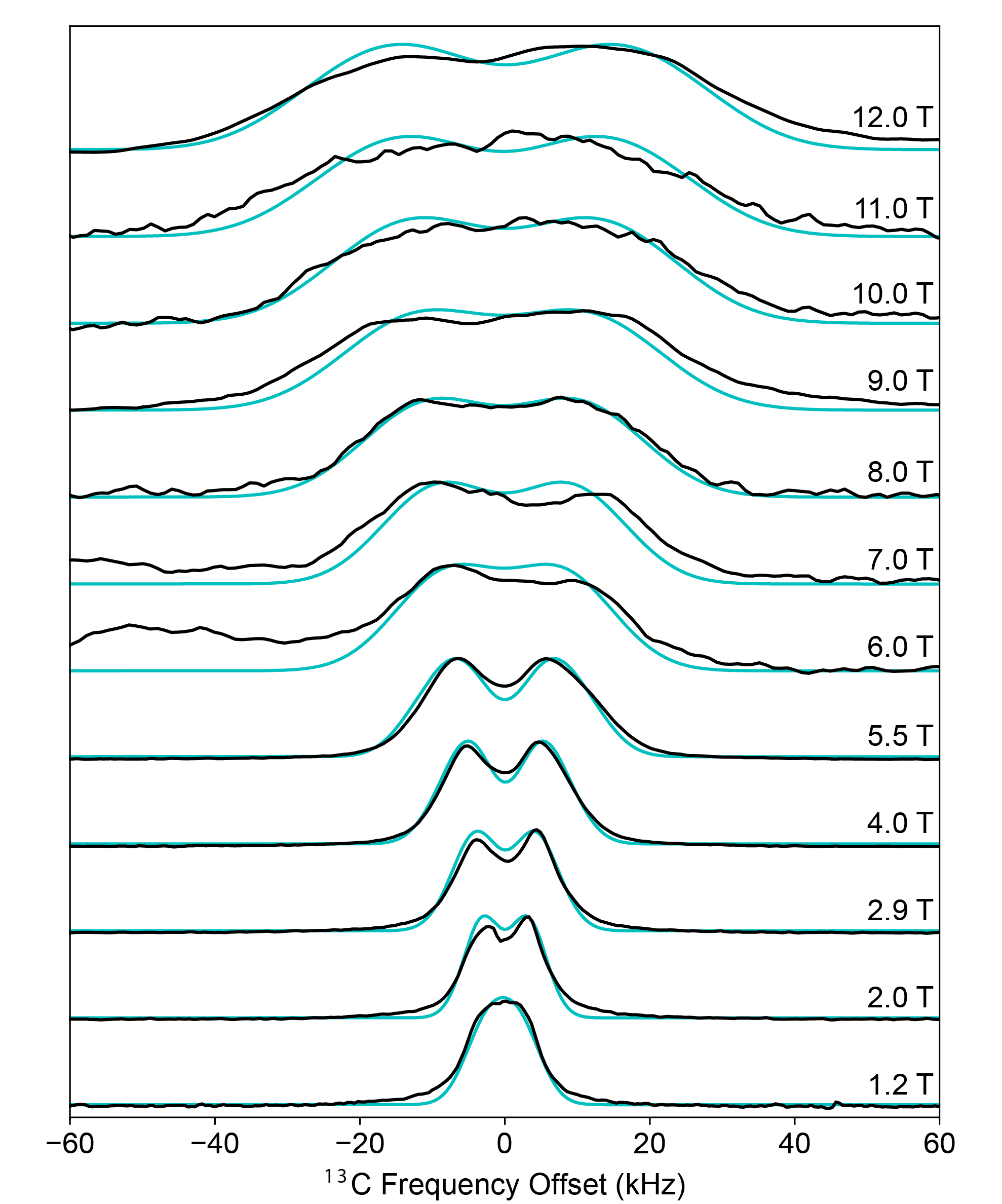}
	\caption{${}^{13}$C NMR spectra of Pu$_2$C$_3$ powder measured at different magnetic fields at 150~K. Experimental spectra are shown in black and simulated curves with SIMPSON in blue. The ${}^{13}$C frequency scale is shown as an offset relative to the site-averaged ${}^{13}$C isotropic shift $\overline{\delta_{\mathrm{iso}}}$ (center of gravity frequency of the spectrum). Spectra recorded between 1.2 and 5.5 T were with a "Fresh" sample and spectra between 6.0 and 12.0~T were with an "Aged" sample. The broad signal seen in the negative frequency end at 6.0 and 7.0~T originates from aged sample with some disorder effects. Note that for the data above 6.0~T, very short last delays were used in the pulse sequence to minimize the signal from the "Aged" sample that has a much longer $T_1$ relative to that of the "Fresh" sample (see the SI for details of disorder effects).}
	\label{fig:spectraH}
\end{figure}

In Fig.~\ref{fig:spectraH} the field dependence of the spectra at 150\,K is depicted with the magnetic field value varied from 1.2 to 12.0\,T. [Measurements from 1.2 to 5.5\,T were conducted on a “fresh” sample, whereas additional measurements from 6.0 to 12.0\,T occurred several months later after some aging effects took place, where the intensity of the ${}^{13}$C signal corresponding to  Pu$_2$C$_3$ was significantly reduced, and some structural disorder likely accrued but did not alter the NMR shift and relaxation rates from Pu$_2$C$_3$; see the Supplementary Information (SI) for details.] At 1.2\,T a single broad pattern appeared with little spectral details, while between 2.0 and 5.5\,T a well-resolved doublet splitting was observed. At and above 6.0\,T this doublet splitting pattern is still resolved despite the likely onset of structural disorder and the total pattern breadth continuously increases. 

\begin{figure}[t]
	\includegraphics[width=8cm]{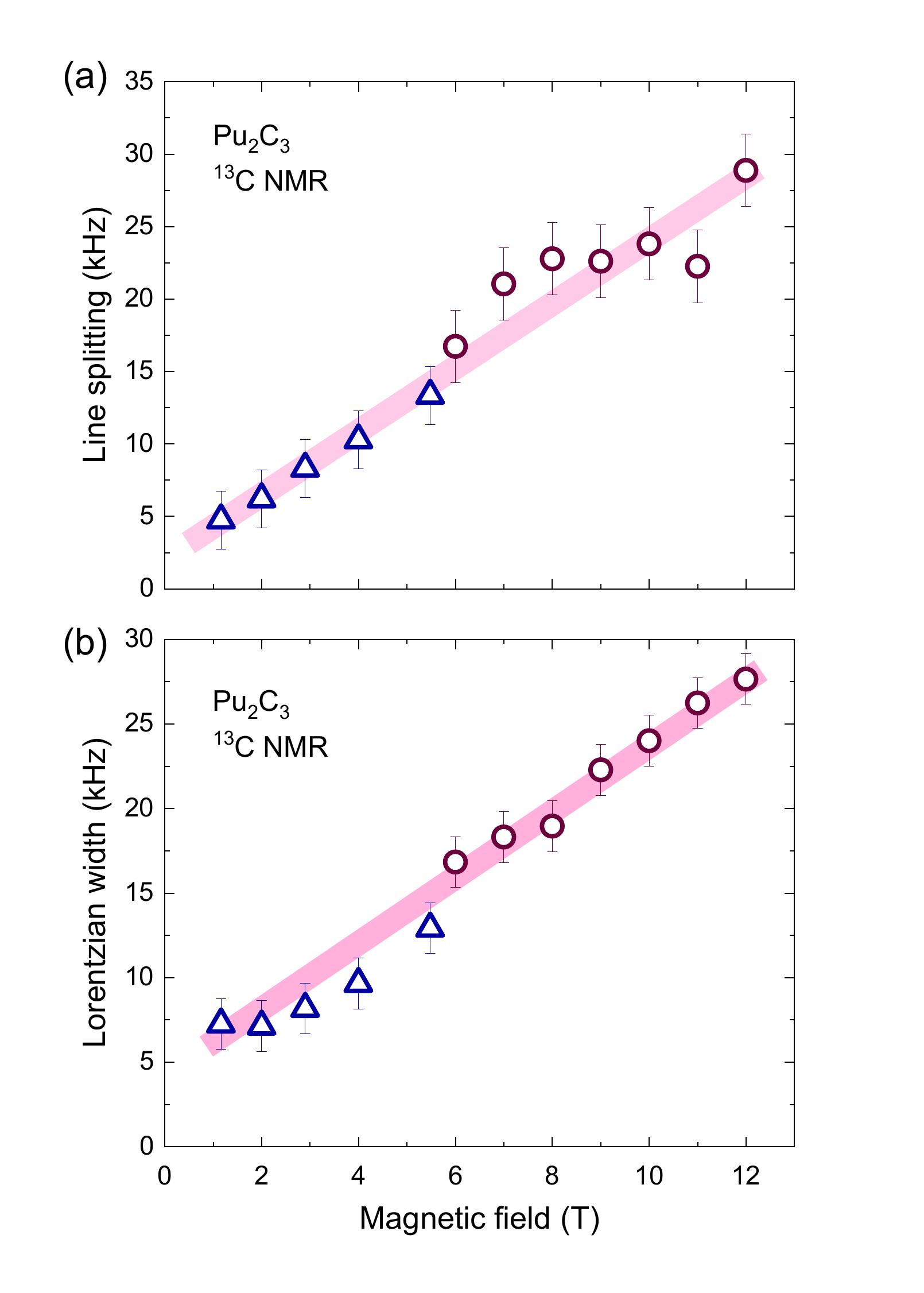}
	\caption{Magnetic field dependence of ${}^{13}$C NMR parameters of Pu$_{2}$C$_{3}$ powder. Spectra were fitted with two Lorentzian forms to deduce the Lorentzian splitting (a) and linewidth (b). Lower field data (triangles; below 5.7\,T) were taken at 150\,K in the "fresh" sample, whereas higher field data (circles; above 6\,T) were collected a few month later in an aged sample at 35\,K. Solid lines are guide to the eyes. Linewidth data at 150\,K were multiplied by a constant factor of 1.225 to account for the temperature dependence and match the low temperature data (see the SI). Error bars indicate estimated systematic uncertainties arising from differing pulse sequence parameters.}
	\label{fig:splitting_width}
\end{figure}

A Pake-doublet like pattern~\cite{Pake_1948_NuclearResonanceAbsorption} may be expected for the ${}^{13}$C--${}^{13}$C dipole-dipole coupling in this isotopically labeled material; however, both the splitting and the pattern breadth increase with $B$, which can only arise from magnetic shift (via the hyperfine interaction) 
rather than nuclear dipole-dipole coupling. Indeed, the spectra are reasonably fit with two Lorentzian forms to define the line splitting [Fig.~\ref{fig:splitting_width}(a)] and linewidth [Fig.~\ref{fig:splitting_width}(b)], both of which show a linear increase with increasing $B$. [Note that in Fig.~\ref{fig:splitting_width}(b), the saturation of the width for $B$ lower than 2.0\,T is possibly indicative of the underlying Pake-like powder spectrum.] Moreover, $T$ dependence of splitting and width approximately parallels that of the site-averaged isotropic shift $\overline{\delta_{\mathrm{iso}}}$ [see the SI and Fig.~\ref{fig:shiftT1}(a)].  All of these strongly suggest that there are two chemically nonequivalent ${}^{13}$C sites (dubbed C$_1$ and C$_2$ hereafter) corresponding to unique shift tensors.

To reveal the underlying C--C bonding nature, SED were measured. A clear oscillation was observed whose pattern was  analyzed using SIMPSON by assuming two nearest-neighbor ${}^{13}$C atoms with slightly different shift tensors and a direct nuclear dipole coupling. By fitting to the experimental pattern, we extracted a dipole coupling constant of $J=$~2.89\,kHz. This translates into a ${}^{13}$C--${}^{13}$C bond length of $r=$~1.38\,\r{A}, consistent with powder neutron diffraction~\cite{Green_1970_Crystallographicmagneticordering} and XRD~\cite{Jain_1993_EvaluationphasesPu} experiments. Further, we can determine the isotropic shift difference (between C$_1$ and C$_2$), $\delta_{\mathrm{iso}}^{\mathrm{C}_{1}} - \delta_{\mathrm{iso}}^{\mathrm{C}_{2}} = \Delta\delta_{\mathrm{iso}} \approx$ 240\,ppm, and the reduced anisotropy, $\delta \approx$~120\,ppm; then, the spectra are simulated using these constrained $J$, $\Delta\delta_{\mathrm{iso}}$, and $\delta$ parameters and $\eta=$~1.0 (where $\eta$ is an asymmetry parameter), as shown by blue curves in Fig.~\ref{fig:spectraH} (see the SI for details). A good agreement is seen between the experimental and simulated spectra, in which the doublet frequency splitting and pattern breadth are well captured, suggesting accurate $\Delta\delta_{\mathrm{iso}}$ and $\delta$ values; however, the agreement is lower for the higher-$B$ spectra, as they are likely further broadened by disorder and do not show the same level of resolution, which is approximated in simulations by increased line-broadening.

The shift and nuclear dipole-dipole coupling are further examined by fitting the $B$-dependent  spectra, with the shift tensors and $J$ as fit parameters. The $B$-dependent frequency separation of the splitting in Fig.~\ref{fig:spectraH} suggests a difference in the isotropic shift value between the two shift tensors (at C$_1$ and C$_2$), $ \Delta\delta_{\mathrm{iso}}$. Because the total pattern breadth varies with $B$, both $\delta$ and $\eta$ that affect the breadth must be included. Fitting the spectra becomes challenging when considering all three shift tensor parameters ($\Delta\delta_{\mathrm{iso}}$, $\delta$, and $\eta$) and within the low resolution of the experimental data. Therefore, the outcome of SED analysis was employed to isolate the $J$ solely originating from the ${}^{13}$C--${}^{13}$C dipole coupling (further details are given in the SI).

The spectral fits reveal a rather large site-averaged isotropic shift $\overline{\delta_{\mathrm{iso}}}~[= (\delta_{\mathrm{iso}}^{\mathrm{C}_{1}} + \delta_{\mathrm{iso}}^{\mathrm{C}_{2}})/2] \approx$ 1000\,ppm and a relatively small anisotropy $\delta = 120$\,ppm. This suggests that to the first approximation, we can safely omit anisotropy of the hyperfine interaction. Interestingly, the presence of a different hyperfine environment for the two carbon atoms (i.e., $\Delta\delta_{\mathrm{iso}} \neq 0$) suggests that the global bcc symmetry is broken from a local magnetic perspective, which may be due to a small shift and/or tilting of the C--C dimers off their perfect bcc lattice positions. Earlier studies~\cite{Zachariasen_1952_Crystalchemicalstudies, Jain_1993_EvaluationphasesPu} and our powder XRD might have missed these unique carbon environments due to the insensitivity of XRD to light elements such as carbon. In fact, the scattering cross section for a typical x-ray photon energy ($\sim10$\,keV) is at least two orders of magnitude bigger for Pu relative to C~\cite{Berger_2010_XCOMPhotonCross}, which means that the scattering events are dominated by Pu. In contrast, NMR is a local probe that is extremely sensitive to microscopic disorder via the hyperfine interaction, which in Pu$_2$C$_3$ seems to be rather sizable (see Sec.~\ref{shift_T1}) and could sharply amplify the underlying structural distortions. A similar discrepancy between XRD and NMR has been reported in $^{57}$Fe NMR of Fe$_3$O$_4$ nanocrystals~\cite{Lim_2018_MicroscopicStatesVerwey}, in which NMR detected disorder that XRD could not resolve.

(Cs, Rb)$_4$O$_6$ represents another example in which powder XRD~\cite{Helms_1939_UeberdieKristallstrukturen, Jansen_1991_NeueUntersuchungenzu} and neutron powder diffraction~\cite{Jansen_1999_Rb4O6studiedelastic} were originally unable to correctly identify a reduction of symmetry from $I\overline{4}$3d to $I\overline{4}$2d. In that case, NMR~\cite{Arcon_2013_InfluenceO2molecularorientation} was also able to provide evidence of a phase transition, and eventually single-crystal XRD~\cite{Sans_2014_StructuralImplicationsSpin} and careful powder neutron diffraction~\cite{Colman_2019_ElusiveValenceTransition} were able to resolve this distortion. Although the existing neutron-powder diffraction data from Pu$_2$C$_3$~\cite{Green_1970_Crystallographicmagneticordering} is inconsistent with the $I\overline{4}$2d structure specifically, it may be that a different reduction in symmetry resulting in two unique C sites with equal multiplicity has been missed.

\subsection{Specific heat and magnetic susceptibility}
\begin{figure}[t]
	\includegraphics[width=8cm]{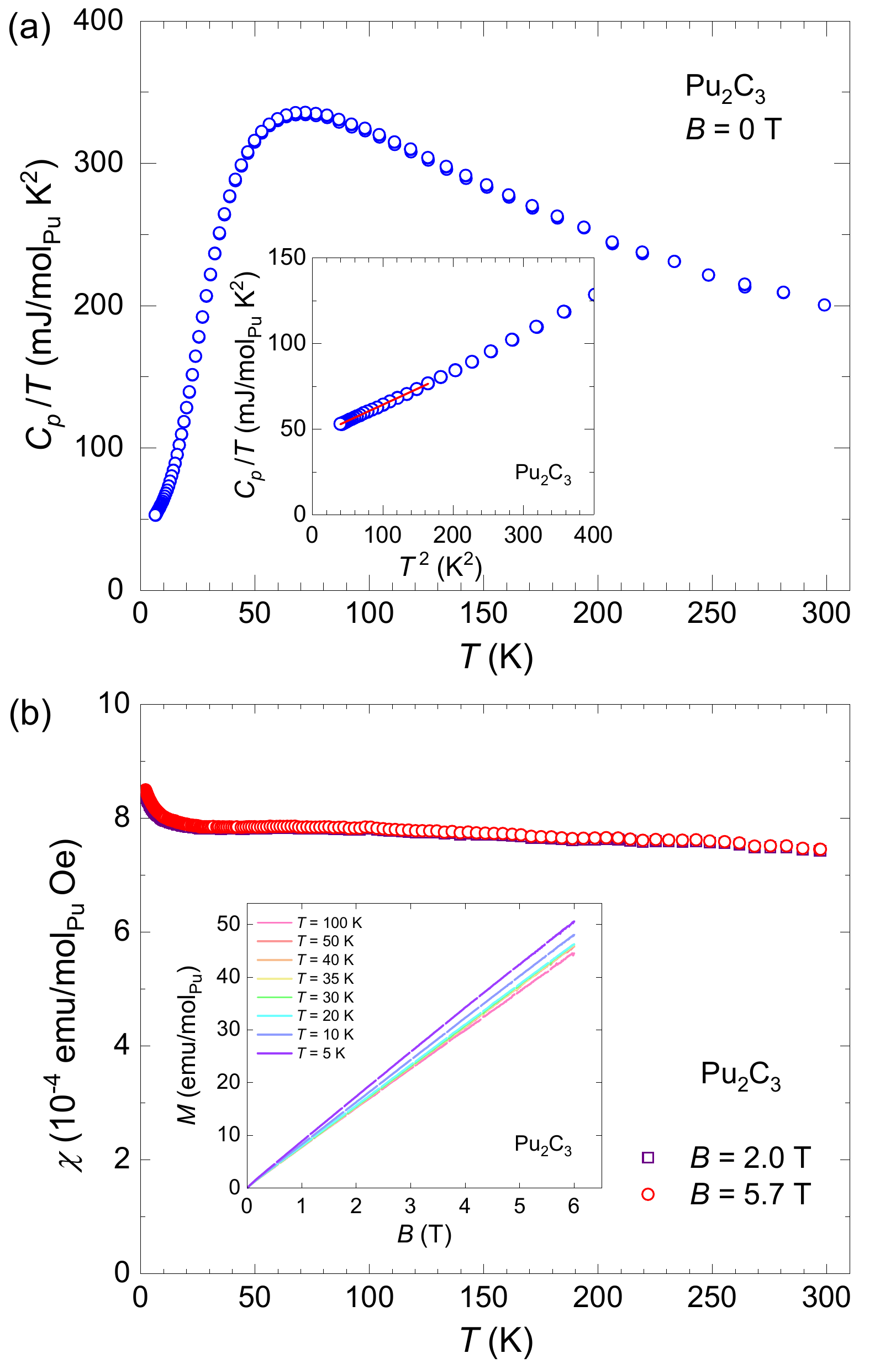}
	\caption{Specific heat and magnetic susceptibility of polycrystalline Pu$_2$C$_3$. (a) Temperature dependence of specific heat $C_p/T$ at zero field. The inset shows the low-temperature (6.3 $< T <$ 20~K) magnification plotted as $C_p/T$ against $T^{2}$. The solid line is the fit to the function $C_p/T = \gamma + \beta T^{2}$. (b) Temperature dependence of magnetic susceptibility $\chi$ at 2.0~T (black squares) and 5.7~T (red circles). The inset shows magnetization $M$ as a function of magnetic field $B~(=\mu_{0}H)$ at elevated temperatures from 5 to 100~K. 
    }
	\label{fig_heatcap_suscpt}
\end{figure}

Figure~\ref{fig_heatcap_suscpt}(a) shows specific heat data, $C_p/T$, collected from 6.3 to 300~K at zero field in a fresh sample. No anomaly was observed, and the data below 20~K is well fit with a $T^{2}$ form, $C_p/T = \gamma + \beta T^{2}$, where $\gamma$ is the linear electronic specific heat (Sommerfeld) coefficient reflecting the quasiparticle density of states at the Fermi level $E_\mathrm{F}$, $D(E_{\mathrm{F}})$, and $\beta T^{2}$ gives the phonon contribution at low $T$. The fit yields $\gamma$ = 45 mJ\,mol$_{\mathrm{Pu}}^{-1}$\,K$^{-2}$ and $\beta$ = 0.189 mJ\,mol$_{\mathrm{Pu}}^{-1}$\,K$^{-4}$. The Debye temperature is calculated to be $\Theta_{\mathrm{D}}$ = 371~K from $\beta = N(12/5)\pi^{4}R\Theta_{\mathrm{D}}^{-3}$, where $R$ = 8.314 J\,mol$^{-1}$\,K$^{-1}$ is the gas constant and $N$ is the number of atoms per formula unit ($N=2.5$ for Pu$_2$C$_3$, per Pu). Such a moderately enhanced value of $\gamma$ is often observed in itinerant uranium and plutonium compounds, with several examples summarized in Table~\ref{tab:Sommerfeld_coeffs}. The enhancement of $\gamma$ with respect to nonmagnetic analogs---e.g., La$_2$C$_3$ ($\gamma$ = 10.6 mJ\,mol$^{-1}$\,K$^{-2}$)~\cite{Kim_2007_Strongelectronphonon} and Y$_2$C$_3$ ($\gamma$ = 4.7 mJ\,mol$^{-1}$\,K$^{-2}$)~\cite{Akutagawa_2007_SuperconductivityY_2C_3Investigated}---indicates a reasonably large $D(E_{\mathrm{F}})$. Note that this $\gamma$ value in Pu$_2$C$_3$ is rather close to that in $\delta$-Pu, which is in line with the observation of photoemission spectra that found very similar characteristics in two compounds~\cite{Gouder_2007_Variability5fstates, Havela_2009_Magneticpropertiesplutonium}.

\begin{table}[t]
\centering
\caption{\label{tab:Sommerfeld_coeffs}The electronic Sommerfeld coefficient $\gamma$ of various related actinide (An) materials.}
\begin{tabular}{ccc}
Material	& $\gamma$ $(\mathrm{mJ}\,\mathrm{mol}_{\mathrm{An}}^{-1} \mathrm{K}^{-2}$) & Reference \\
\hline \\[-0.3cm]
Pu$_2$C$_3$ & 45                                                                        & this work \\
U$_2$C$_3$	& 42			                                                            & {}\cite{Eloirdi_2013_Evidencepersistentspin} \\
UAl$_3$     & 43			                                                            & {}\cite{Cornelius_1999_ElectronicpropertiesUX_3} \\
$\delta$-Pu	& 30-60		                                                                & {}\cite{Lashley_2003_ExperimentalElectronicHeat, Javorsky_2006_SpecificHeat} \\
PuCoGa$_5$  & 77                                                                        & {}\cite{Sarrao_2002_Plutoniumbasedsuperconductivity} \\
\end{tabular}
\end{table}

Magnetic susceptibility data as a function of $T$, $\chi(T)$, taken at 2.0 and 5.7~T are shown in Fig.~\ref{fig_heatcap_suscpt}(b). Above $\sim$~30~K, there is no evidence of Curie-Weiss behavior (at least up to 300~K), yet apparent Pauli paramagnetic behavior is observed with very small $T$ dependence, having an average value of $\chi \sim$ 8.0$\times$10$^{-4}$ emu\,mol$_{\mathrm{Pu}}^{-1}$. (Note that there is at least $\sim \pm 15 \%$ uncertainty in the size of $\chi$ due to the ambiguity we have in the background subtraction.) There is a little upturn in $\chi$ below $\sim$~30~K that could be most likely due to a small amount of impurity phases remained in our sample. At higher temperatures above 100~K, $\chi$ gradually decreases with increasing $T$ by a small amount, which might be ascribed to either small $T$ dependence of the Pauli paramagnetic susceptibility or a small localized component of Pu 5$f$ electrons reflecting their dual nature, as often discussed in Pu-based compounds~\cite{Joyce_2003_PhotoemissionElectronicStructure, Baek_2010_AnisotropicSpinFluctuations}. Inset shows $B$ dependence of magnetization, $M$, at temperatures from 5 to 100~K. Almost linear field dependence is observed in $M$ at above $\sim$~30~K from zero field up to 6~T, consistent with the Pauli paramagnetism picture.   

Similar $T$-insensitive susceptibility is observed in Ga-stabilized $\delta$-Pu below 300~K with an even smaller size of $\chi \sim$ 5.5$\times$10$^{-4}$ emu\,mol$_{\mathrm{Pu}}^{-1}$~\cite{MeotReymond_1996_Localization5felectrons}. A paramagnetic feature of this sort is widely reported in itinerant uranium compounds, such as UC~\cite{Eloirdi_2013_Evidencepersistentspin} and a range of UX$_{3}$ compounds (such as UAl$_3$, USi$_3$, UGe$_3$, and URh$_3$ as well as UGa$_3$ above its Néel temperature $T_\mathrm{N}$), also known as enhanced Pauli paramagnetism~\cite{Koelling_1985_felectronhybridization, Cornelius_1999_ElectronicpropertiesUX_3} and is understood by the itinerant nature of 5$f$-electron bands~\cite{Kaczorowski_1993_Magneticnonmagnetictransition}.

Additionally, from the approximately temperature-independent magnetic susceptibility and Sommerfeld coefficient, we can evaluate the Wilson ratio $R_W = \frac{1}{3} \pi^2 {k_B}^2 {\mu_B}^{-2} (\chi / \gamma)$, where $k_B$ is the Boltzmann constant and $\mu_B$ is the Bohr magneton. We find $R_W \approx 1.3$. This is in line with other weakly correlated materials~\cite{Lee_1986_Theoriesheavyelectron, Sheng_1995_NonKondoprediction}.

\subsection{NMR shift and spin-lattice relaxation rate}
\label{shift_T1}
\begin{figure}[t]
	\includegraphics[width=8.5cm]{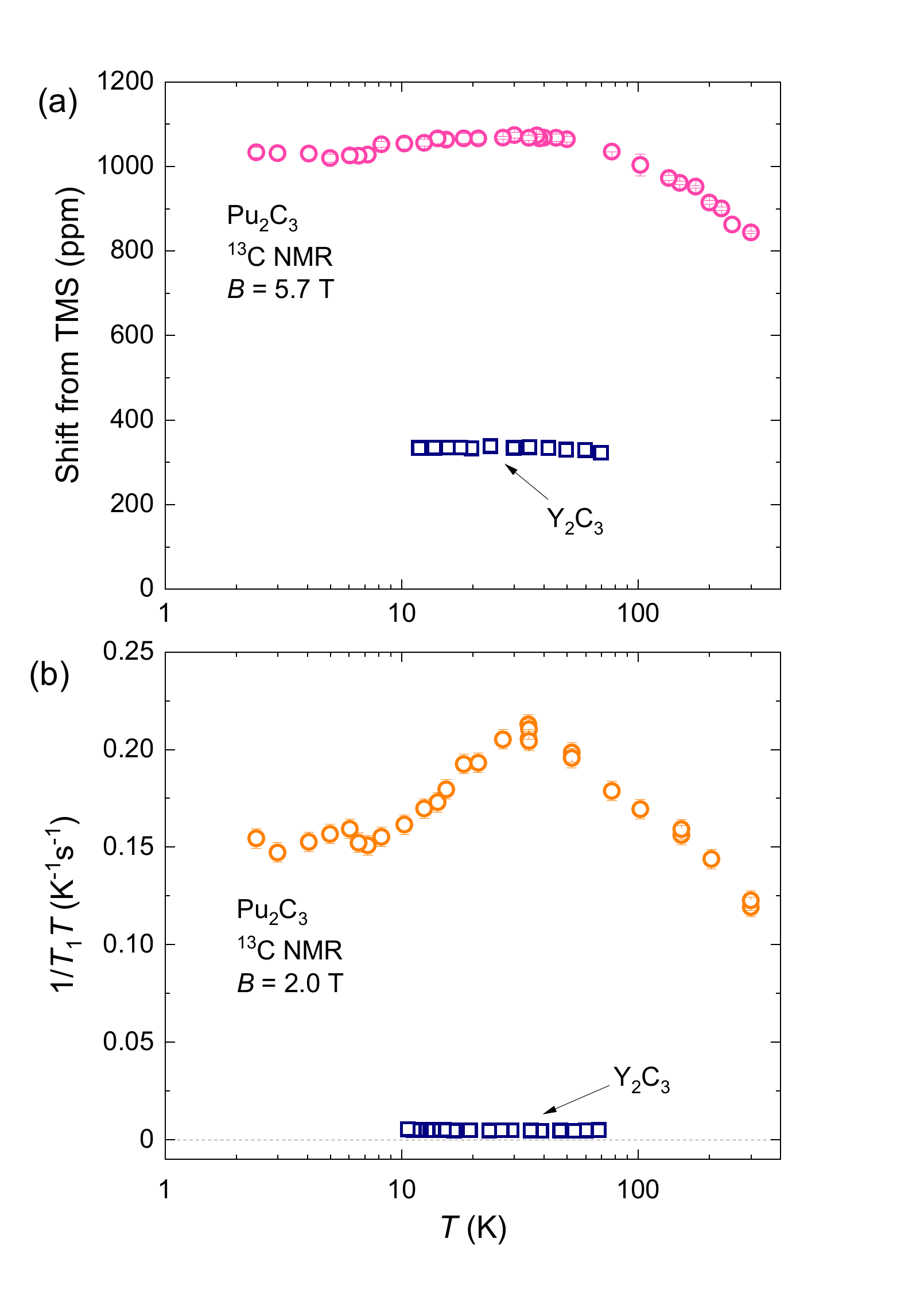}
	\caption{Temperature dependence of the average ${}^{13}$C isotropic shift $\overline{\delta_{\mathrm{iso}}}$ measured at 5.7~T (a) and spin-lattice relaxation rate divided by temperature $1/T_{1}T$ measured at 2.0~T (b) of fresh Pu$_{2}$C$_{3}$. Squared symbols represent the corresponding shift and $1/T_{1}T$ data in nonmagnetic analog Y$_2$C$_3$ in its metallic phase~\cite{Harada_2007_MultigapSuperconductivityY_2C_3}.}
	\label{fig:shiftT1}
\end{figure}

\begin{figure}[t]
	\includegraphics[width=8.5cm]{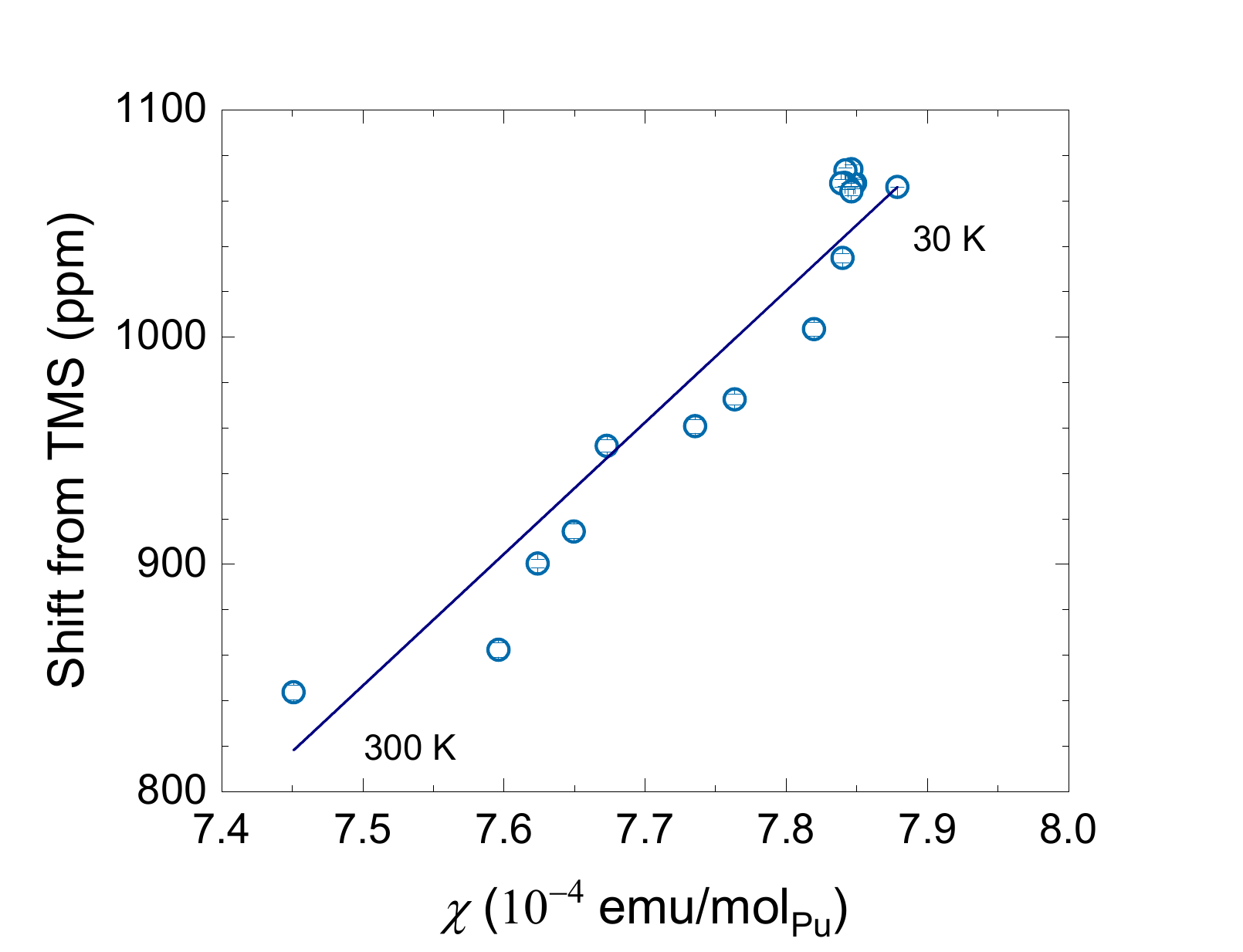}
	\caption{${}^{13}$C isotropic shift $\overline{\delta_{\mathrm{iso}}}$ plotted against susceptibility $\chi$ of Pu$_{2}$C$_{3}$ measured at 5.7~T. Temperature is used as an implicit parameter. The straight line is a least-square fit to the data.}
	\label{fig:K-chi}
\end{figure}

Figure~\ref{fig:shiftT1}(a) shows the site-averaged ${}^{13}$C NMR isotropic shift $\overline{\delta_{\mathrm{iso}}}$ as a function of $T$, measured at 5.7~T from 2 to 300~K. In a simple metallic system, the shift is given by a sum of $T$-independent orbital shift, $K_0$, and the electron spin shift, $K_\mathrm{spin}$, as: $\overline{\delta_{\mathrm{iso}}} = K_\mathrm{spin} + K_0$. Moderate $T$ dependence was observed above $T^*\approx$~30~K which scales with the susceptibility $\chi$. In fact, a good linear relationship is seen between $\overline{\delta_{\mathrm{iso}}}$ and $\chi$ from 30 to 300~K as shown in Fig.~\ref{fig:K-chi}. A linear fit to the $\overline{\delta_{\mathrm{iso}}}- \chi$ plot for $30 \leq T \leq 300$ K with a form $\overline{\delta_{\mathrm{iso}}} = (A/N_{\mathrm{A}}\mu_\mathrm{B}) \chi$ yields $A =$ 3.24~$\mathrm{T}/\mu_\mathrm{B}$ from the slope, where $A$ is the on-site hyperfine coupling constant and $N_{\mathrm{A}}$ is the Avogadro's number.  Below $T^*$, the shift exhibits a broad maximum at $\sim$\,30~K, levels off below 10~K, and approaches a value close to 1000\,ppm at 2~K. This low-$T$ value is three times bigger than in nonmagnetic analogs La$_2$C$_3$ (340\,ppm)~\cite{Potocnik_2014_Anomalouslocalspin} and Y$_2$C$_3$ (330\,ppm)~\cite{Harada_2007_MultigapSuperconductivityY_2C_3} that are simple metals, while three to four times smaller than in U$_2$C$_3$ (3000-4000\,ppm)~\cite{Boutard_1974_EtudeparRMN} that has a more Curie-Weiss like 5$f$-electron character at higher $T$~\cite{Eloirdi_2013_Evidencepersistentspin}. The moderate size of $\overline{\delta_{\mathrm{iso}}}$ found in Pu$_2$C$_3$ aligns well with the Wilson ratio and the picture of enhanced Pauli paramagnetism of delocalized 5$f$ electrons.

The $T$ dependence of ${}^{13}$C spin-lattice relaxation rate divided by temperature, $1/T_{1}T$, is plotted in Fig.~\ref{fig:shiftT1}(b), recorded at 2.0~T determined from the full echo intensity. (No difference was observed in $T_{1}$ at the two nonequivalent ${}^{13}$C sites corresponding to the two peaks in Fig.~\ref{fig:spectraH}, and there was no $B$ dependence up to 9.0~T). $1/T_{1}T$ well parallels the $T$ dependence of $\overline{\delta_{\mathrm{iso}}}$, forming a maximum at around $T^*$ and showing a flat feature below 10~K with a value of 0.150~K$^{-1}$\,sec$^{-1}$. This latter value is an order of magnitude bigger than in Y$_2$C$_3$ ($1/T_{1}T =$ 0.004~K$^{-1}$\,sec$^{-1}$)~\cite{Harada_2007_MultigapSuperconductivityY_2C_3}.

In weakly correlated metals, $1/T_{1}T$ is proportional to $[D(E_\mathrm{F})]^2 \times A^2$~\cite{Slichter_1990_PrinciplesMagneticResonance}. 
Within this framework  and omitting the hyperfine term $A$ (which is not known in Y$_2$C$_3$), the ratio of $1/T_{1}T$ in two systems scales to that of $[D(E_\mathrm{F})]^2$, i.e., $[D(E_\mathrm{F})]_{\mathrm{Pu_2C_3}}^2/[D(E_\mathrm{F})]_{\mathrm{Y_2C_3}}^2 \sim$~91, as estimated from the corresponding values of the Sommerfeld coefficient $\gamma$.
This is, however, a factor of $\sim 2.5$ too large in comparison with the experimental ratio: $(1/T_{1}T)_{\mathrm{Pu_2C_3}}/(1/T_{1}T)_{\mathrm{Y_2C_3}} \sim$~37. The discrepancy implies that either the hyperfine coupling constant is $\sqrt{2.5} \approx  1.6$ times larger in the Y compound than in Pu compound or the weakly correlated picture is not valid.

\begin{figure}[t]
	\includegraphics[width=8.5cm]{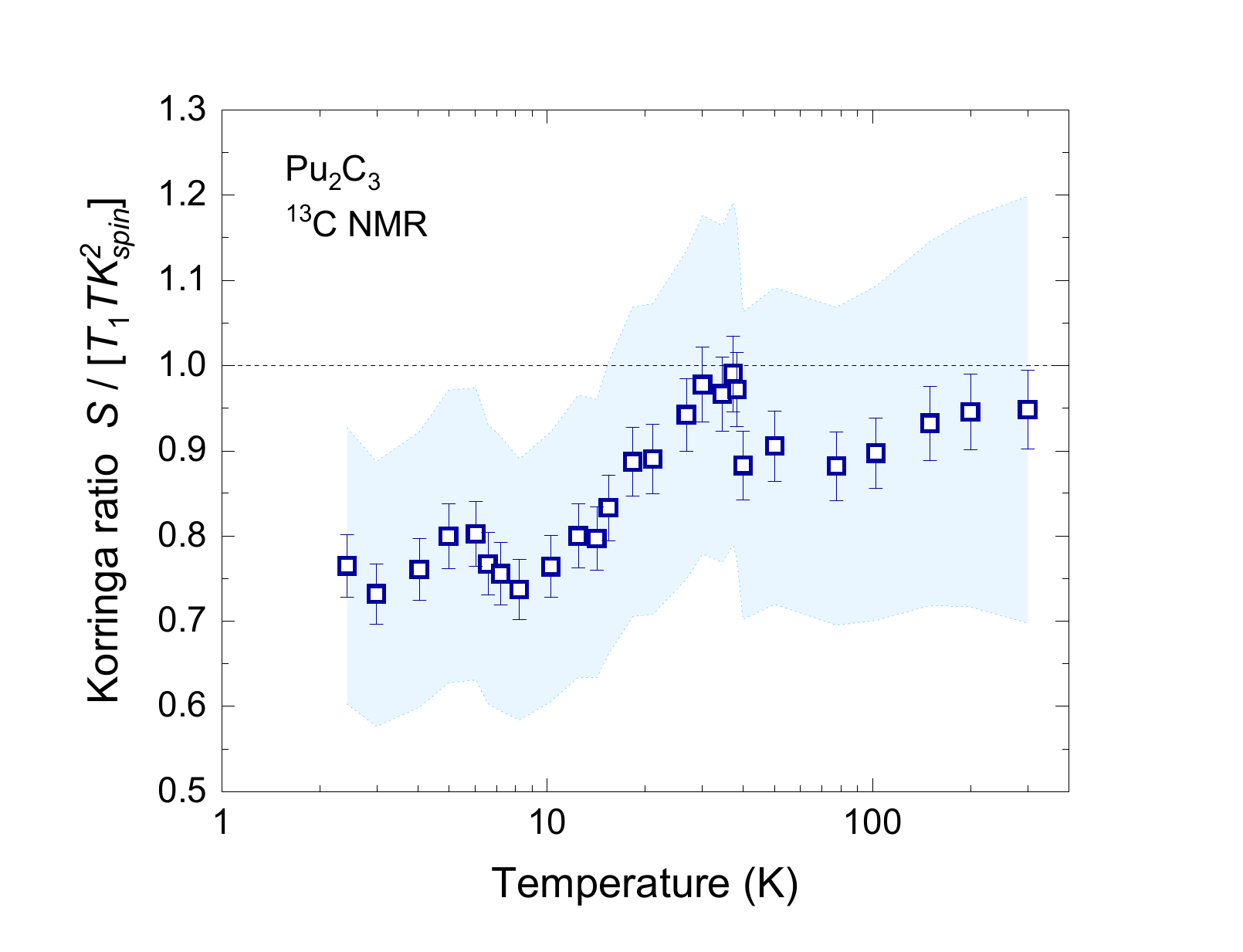}
	\caption{${}^{13}$C Korringa ratio $\mathcal{K} = S/(T_1TK^2_\mathrm{spin})$ in Pu$_2$C$_3$, plotted as a function of temperature with $S = \hbar \gamma_e^2/(4\pi k_\mathrm{B} \gamma_n^2)$. $\mathcal{K} = 1$ (dashed horizontal line) corresponds to the case for free electrons, while $\mathcal{K} < 1$ ($\mathcal{K} > 1$) suggests ferromagnetic (antiferromagnetic) spin fluctuations. The shaded band represents the range of $\mathcal{K}$ depending on the estimated size of orbital shift $K_0$. To derive the lower and upper bounds of this band, typical values of $K_0=$~0 and 200~ppm reported for ${}^{13}$C resonance in nonmagnetic insulators~\cite{Tossell_1992_NuclearMagneticShieldings}, respectively, are assumed.} 
	\label{fig:Korringa}
\end{figure}

To evaluate the electron correlation effects, we assume an isotropic hyperfine interaction and relate the spin part of the shift to the spin-lattice relaxation rate through $D(E_\mathrm{F})$ via the Korringa relation~\cite{Slichter_1990_PrinciplesMagneticResonance, Abragam_1961_PrinciplesNuclearMagnetism} 
\begin{equation}
 \frac{\hbar \gamma_e^2}{4\pi k_\mathrm{B} \gamma_n^2} \frac{1}{T_1TK^2_\mathrm{spin}} = 1,
\label{eq:1}
\end{equation}
\noindent where $\hbar$ is the reduced Planck constant and $\gamma_e$ ($\gamma_n$) is electronic (nuclear) gyromagnetic ratio. When electronic correlations are present, deviation from this form can be characterized by the Korringa ratio $\mathcal{K} = S/(T_1TK^2_\mathrm{spin})$ with $S = \hbar \gamma_e^2/(4\pi k_\mathrm{B} \gamma_n^2)$, as plotted in Fig.~\ref{fig:Korringa}. Although the size of $\mathcal{K}$ has a certain range originating from the uncertainty in the estimated size of $K_0$ (shaded region in Fig.~\ref{fig:Korringa}, given by assuming $K_0=$~0 and 200~ppm for lower and upper bounds, respectively, as typically reported for ${}^{13}$C resonance in nonmagnetic insulators~\cite{Tossell_1992_NuclearMagneticShieldings}), the overall trend of decreasing $\mathcal{K}$ towards lower $T$ can be clearly seen. Further, $\mathcal{K} < 1$ is observed at low $T$, which indicates the presence of ferromagnetic spin fluctuations at $E_\mathrm{F}$. Above and below $T^* \approx$~30~K, $\mathcal{K} = \mathrm{const.}$ is satisfied at around slightly different values of $\mathcal{K}$ ($\mathcal{K} \approx$~0.9 at $T>T^*$ and $\mathcal{K} \approx$~0.8 at $T<T^*$). This flat feature, together with $\mathcal{K} < 1$, provides strong evidence that Pu$_2$C$_3$ is a good metal with Stoner enhanced Pauli paramagnetism, in agreement with the observed finite value of the specific-heat Sommerfeld coefficient $\gamma$ and the associated picture of delocalized 5$f$ electrons also suggested by the Pu-Pu nearest neighbor distance being slightly smaller than the Hill limit~\cite{Hill_1970_Earlyactinidesperiodic, Brodsky_1978_Magneticpropertiesactinide}.
This is also consistent with previous DFT work, which predicted Pu$_2$C$_3$ to have a non-magnetic metallic ground state~\cite{Yang_2017_Densityfunctionalstudy}. 
Similar Korringa-type, constant-$\mathcal{K}$ behavior is observed in other Pu-based systems, e.g., Pu superconductors PuCoGa$_5$ (with $^{59}$Co NMR)~\cite{Baek_2010_AnisotropicSpinFluctuations} and PuRhGa$_5$ (with $^{69}$Ga NMR)~\cite{Sakai_2006_NMRShiftMeasurements} in the normal state (although $\mathcal{K} >1$ due to antiferromagnetic fluctuations); whereas, in Ga-stabilized $\delta$-Pu (with $^{69}$Ga NMR) $\mathcal{K}$ is not constant but $1/T_1T = \mathrm{const.}$ is found instead~\cite{Piskunov_2005_SpinsusceptibilityGa, Curro_2004_Scalingemergentbehavior}.

In the intermediate temperature range, however, a non-monotonic behavior is observed in $\mathcal{K}$ at around 30~K with a small peak-like structure. This nonmonotonicity might indicate underlying antiferromagnetic fluctuations that compete with ferromagnetic fluctuations and eventually cease with further cooling or a Kondo coherence development, as reported in heavy-fermion compounds CeAl$_3$~\cite{Lysak_1985_Nuclearmagneticresonance} and LiV$_2$O$_4$~\cite{Mahajan_1998_7Li51VNMR}. 

\begin{figure}[t]
	\includegraphics[width=8.5cm]{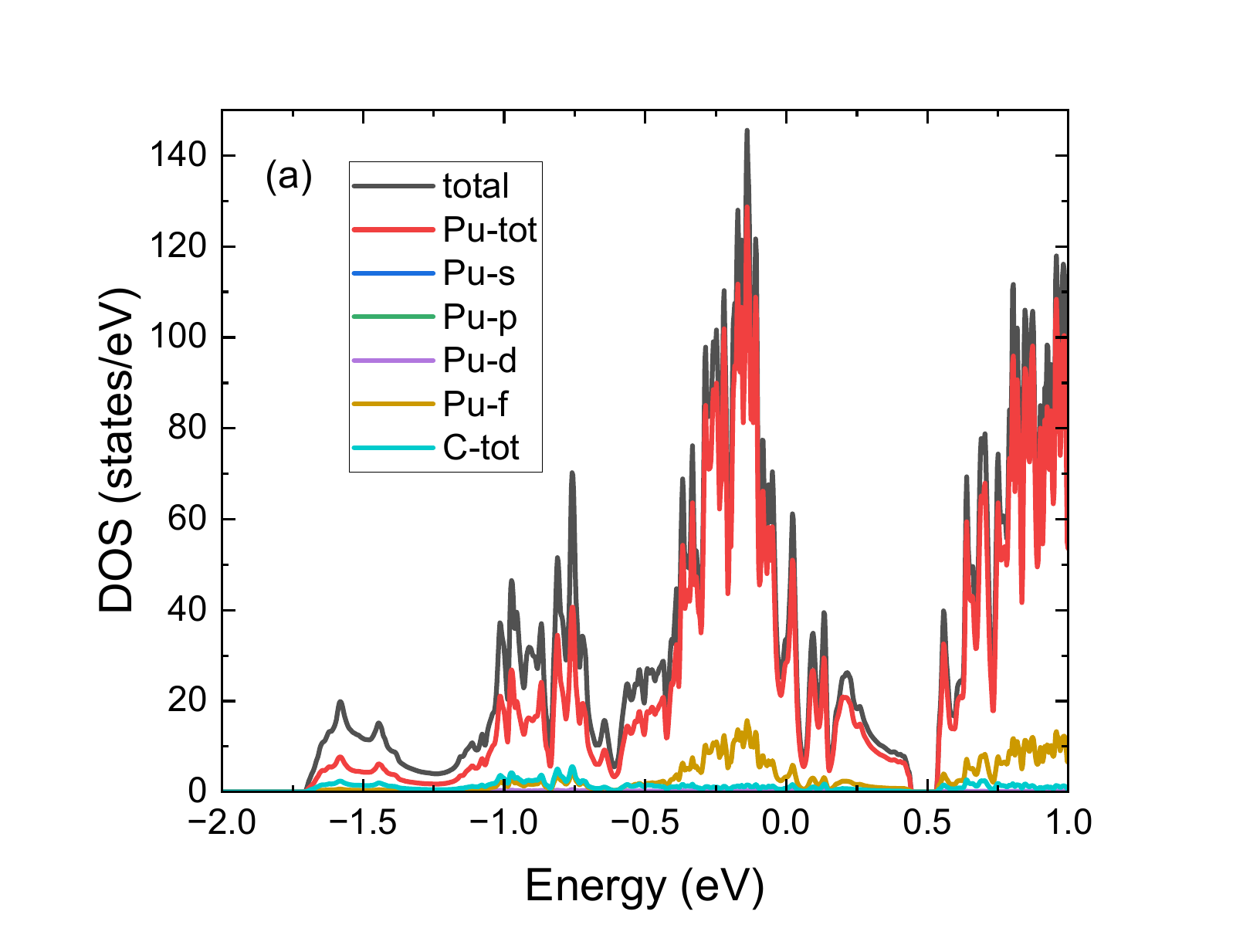}
        \includegraphics[width=8.5cm]{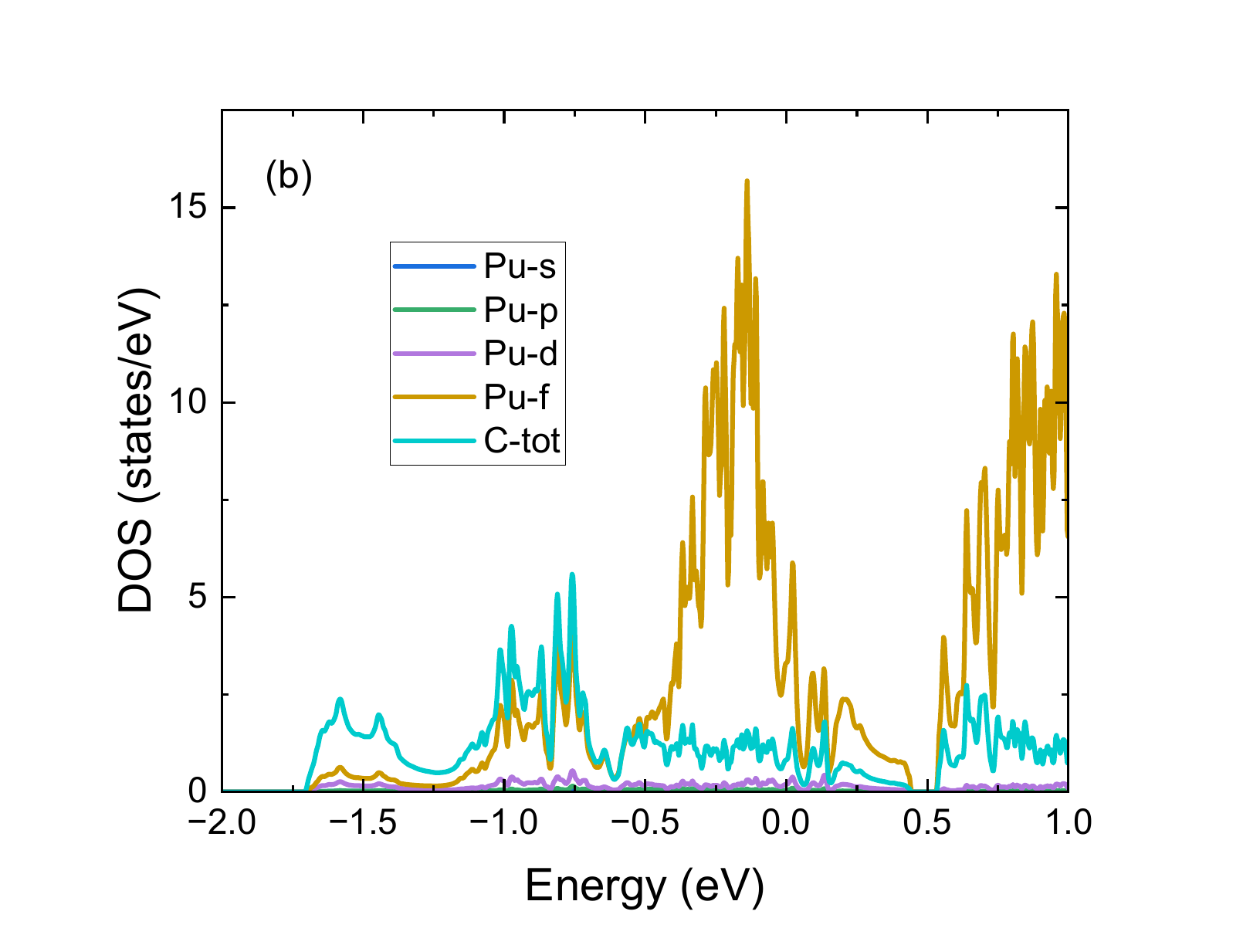}
	\caption{(a) Calculated electronic density of states (DOS) of Pu$_2$C$_3$ near the Fermi level ($E_\mathrm{F} = 0$) plotted as a function of energy. Note that the total DOS (dark gray) accounts for all of the electron density, whereas partial DOS (all other curves) do not consider the atomic multiplicity or the electron density outside the muffin tins, therefore, one does not expect the partial DOS to sum to the total in the figure. (b) Calculated partial DOS for Pu-s, -p, -d, and -f; and the total contribution from C.}
	\label{fig:band}
\end{figure}

\subsection{Band structure calculation}
The density of states as calculated within DFT is shown in Fig.~\ref{fig:band}. Close to the Fermi level $E_\mathrm{F}$ the density of states is dominated by the plutonium 5$f$-orbital weight. The expected Sommerfeld contribution to the heat capacity is 20 mJ mol$_{\mathrm{Pu}}^{-1}$ K$^{-2}$, based on the calculation, which is more than a factor of two smaller than the measured value of $\gamma$ (= 45 mJ mol$_{\mathrm{Pu}}^{-1}$ K$^{-2}$). This suggests the presence of moderate electronic correlations in Pu$_2$C$_3$.
\\
\\

\section{Conclusions}
We report a comprehensive study of structural and electronic properties of Pu-based cubic sesquicarbide Pu$_2$C$_3$. XRD finds a global cubic symmetry with the nearest-neighbor C--C bond length of $r$ = 1.38~\r{A}. ${}^{13}$C NMR powder spectrum has a clear line splitting that shows sizable field and temperature dependence, and the spin echo decay curve possesses a characteristic oscillator pattern. These findings suggest that the two carbon atoms locate in a unique local magnetic environment and the global cubic symmetry is locally broken, likely due to a small distortion of C--C bonds not resolved by XRD. Specific heat measurement finds metallic properties with a moderately enhanced low-temperature $C_p/T$ of $\gamma$ = 45 mJ mol$_{\mathrm{Pu}}^{-1}$ K$^{-2}$, and magnetic susceptibility reveals enhanced Pauli paramagnetism. The Wilson ratio $R_W \approx 1.3$ indicates moderately enhanced electronic correlations. NMR shift and spin-lattice relaxation rate find Korringa behavior characteristic of a metallic ground state accompanied by moderate ferromagnetic spin fluctuations setting in below 30~K. The DFT-calculated band structure confirms the presence of a narrow 5$f$-electron band near the Fermi level, consistent with the experimental finding of itinerant 5$f$ electrons. Our results offer firm evidence for a weakly correlated metallic ground state in Pu$_2$C$_3$, which serves as a good reference for a delocalized 5$f$-electrons system and becomes important for understanding the complex interplay of localization and itinerancy in plutonium compounds.

\begin{acknowledgments}
We thank C. S. Kengle, A. Scheie, P. F. S. Rosa, H. Sakai, and H. Yasuoka for fruitful discussions. Crystal growth, NMR measurements, and DFT were supported by the U.S. Department of Energy, Office of Basic Energy Sciences "Quantum Fluctuations in Narrow Band Systems" program. Thermodynamic measurements were supported by the Laboratory Directed Research and Development program.
\end{acknowledgments}

\bibliography{Pu2C3}

\begin{thebibliography}{68}%
\makeatletter
\providecommand \@ifxundefined [1]{%
 \@ifx{#1\undefined}
}%
\providecommand \@ifnum [1]{%
 \ifnum #1\expandafter \@firstoftwo
 \else \expandafter \@secondoftwo
 \fi
}%
\providecommand \@ifx [1]{%
 \ifx #1\expandafter \@firstoftwo
 \else \expandafter \@secondoftwo
 \fi
}%
\providecommand \natexlab [1]{#1}%
\providecommand \enquote  [1]{``#1''}%
\providecommand \bibnamefont  [1]{#1}%
\providecommand \bibfnamefont [1]{#1}%
\providecommand \citenamefont [1]{#1}%
\providecommand \href@noop [0]{\@secondoftwo}%
\providecommand \href [0]{\begingroup \@sanitize@url \@href}%
\providecommand \@href[1]{\@@startlink{#1}\@@href}%
\providecommand \@@href[1]{\endgroup#1\@@endlink}%
\providecommand \@sanitize@url [0]{\catcode `\\12\catcode `\$12\catcode `\&12\catcode `\#12\catcode `\^12\catcode `\_12\catcode `\%12\relax}%
\providecommand \@@startlink[1]{}%
\providecommand \@@endlink[0]{}%
\providecommand \url  [0]{\begingroup\@sanitize@url \@url }%
\providecommand \@url [1]{\endgroup\@href {#1}{\urlprefix }}%
\providecommand \urlprefix  [0]{URL }%
\providecommand \Eprint [0]{\href }%
\providecommand \doibase [0]{https://doi.org/}%
\providecommand \selectlanguage [0]{\@gobble}%
\providecommand \bibinfo  [0]{\@secondoftwo}%
\providecommand \bibfield  [0]{\@secondoftwo}%
\providecommand \translation [1]{[#1]}%
\providecommand \BibitemOpen [0]{}%
\providecommand \bibitemStop [0]{}%
\providecommand \bibitemNoStop [0]{.\EOS\space}%
\providecommand \EOS [0]{\spacefactor3000\relax}%
\providecommand \BibitemShut  [1]{\csname bibitem#1\endcsname}%
\let\auto@bib@innerbib\@empty
\bibitem [{\citenamefont {Bauer}\ and\ \citenamefont {Thompson}(2015)}]{Bauer_2015_PlutoniumBasedHeavy}%
  \BibitemOpen
  \bibfield  {author} {\bibinfo {author} {\bibfnamefont {E.~D.}\ \bibnamefont {Bauer}}\ and\ \bibinfo {author} {\bibfnamefont {J.~D.}\ \bibnamefont {Thompson}},\ }\bibfield  {title} {\bibinfo {title} {{Plutonium-Based Heavy-Fermion Systems}},\ }\href {https://doi.org/10.1146/annurev-conmatphys-031214-014508} {\bibfield  {journal} {\bibinfo  {journal} {Annual Review of Condensed Matter Physics}\ }\textbf {\bibinfo {volume} {6}},\ \bibinfo {pages} {137} (\bibinfo {year} {2015})}\BibitemShut {NoStop}%
\bibitem [{\citenamefont {Sarrao}\ \emph {et~al.}(2002)\citenamefont {Sarrao}, \citenamefont {Morales}, \citenamefont {Thompson}, \citenamefont {Scott}, \citenamefont {Stewart}, \citenamefont {Wastin}, \citenamefont {Rebizant}, \citenamefont {Boulet}, \citenamefont {Colineau},\ and\ \citenamefont {Lander}}]{Sarrao_2002_Plutoniumbasedsuperconductivity}%
  \BibitemOpen
  \bibfield  {author} {\bibinfo {author} {\bibfnamefont {J.~L.}\ \bibnamefont {Sarrao}}, \bibinfo {author} {\bibfnamefont {L.~A.}\ \bibnamefont {Morales}}, \bibinfo {author} {\bibfnamefont {J.~D.}\ \bibnamefont {Thompson}}, \bibinfo {author} {\bibfnamefont {B.~L.}\ \bibnamefont {Scott}}, \bibinfo {author} {\bibfnamefont {G.~R.}\ \bibnamefont {Stewart}}, \bibinfo {author} {\bibfnamefont {F.}~\bibnamefont {Wastin}}, \bibinfo {author} {\bibfnamefont {J.}~\bibnamefont {Rebizant}}, \bibinfo {author} {\bibfnamefont {P.}~\bibnamefont {Boulet}}, \bibinfo {author} {\bibfnamefont {E.}~\bibnamefont {Colineau}},\ and\ \bibinfo {author} {\bibfnamefont {G.~H.}\ \bibnamefont {Lander}},\ }\bibfield  {title} {\bibinfo {title} {{Plutonium-based superconductivity with a transition temperature above 18~K}},\ }\href {https://doi.org/10.1038/nature01212} {\bibfield  {journal} {\bibinfo  {journal} {Nature}\ }\textbf {\bibinfo {volume} {420}},\ \bibinfo {pages} {297} (\bibinfo {year} {2002})}\BibitemShut {NoStop}%
\bibitem [{\citenamefont {Shim}\ \emph {et~al.}(2007)\citenamefont {Shim}, \citenamefont {Haule},\ and\ \citenamefont {Kotliar}}]{Shim_2007_Fluctuatingvalencecorrelated}%
  \BibitemOpen
  \bibfield  {author} {\bibinfo {author} {\bibfnamefont {J.~H.}\ \bibnamefont {Shim}}, \bibinfo {author} {\bibfnamefont {K.}~\bibnamefont {Haule}},\ and\ \bibinfo {author} {\bibfnamefont {G.}~\bibnamefont {Kotliar}},\ }\bibfield  {title} {\bibinfo {title} {{Fluctuating valence in a correlated solid and the anomalous properties of $\delta$-plutonium}},\ }\href {https://doi.org/10.1038/nature05647} {\bibfield  {journal} {\bibinfo  {journal} {Nature}\ }\textbf {\bibinfo {volume} {446}},\ \bibinfo {pages} {513} (\bibinfo {year} {2007})}\BibitemShut {NoStop}%
\bibitem [{\citenamefont {Harrison}\ \emph {et~al.}(2024)\citenamefont {Harrison}, \citenamefont {Chappell},\ and\ \citenamefont {Tobash}}]{Harrison_2024_Indicationsflatbands}%
  \BibitemOpen
  \bibfield  {author} {\bibinfo {author} {\bibfnamefont {N.}~\bibnamefont {Harrison}}, \bibinfo {author} {\bibfnamefont {G.~L.}\ \bibnamefont {Chappell}},\ and\ \bibinfo {author} {\bibfnamefont {P.~H.}\ \bibnamefont {Tobash}},\ }\bibfield  {title} {\bibinfo {title} {{Indications of flat bands driving the $\delta$ to $\alpha$ volume collapse of plutonium}},\ }\href {https://doi.org/10.1073/pnas.2308729121} {\bibfield  {journal} {\bibinfo  {journal} {Proceedings of the National Academy of Sciences}\ }\textbf {\bibinfo {volume} {121}},\ \bibinfo {pages} {e2308729121} (\bibinfo {year} {2024})}\BibitemShut {NoStop}%
\bibitem [{\citenamefont {Lashley}\ \emph {et~al.}(2003)\citenamefont {Lashley}, \citenamefont {Singleton}, \citenamefont {Migliori}, \citenamefont {Betts}, \citenamefont {Fisher}, \citenamefont {Smith},\ and\ \citenamefont {McQueeney}}]{Lashley_2003_ExperimentalElectronicHeat}%
  \BibitemOpen
  \bibfield  {author} {\bibinfo {author} {\bibfnamefont {J.~C.}\ \bibnamefont {Lashley}}, \bibinfo {author} {\bibfnamefont {J.}~\bibnamefont {Singleton}}, \bibinfo {author} {\bibfnamefont {A.}~\bibnamefont {Migliori}}, \bibinfo {author} {\bibfnamefont {J.~B.}\ \bibnamefont {Betts}}, \bibinfo {author} {\bibfnamefont {R.~A.}\ \bibnamefont {Fisher}}, \bibinfo {author} {\bibfnamefont {J.~L.}\ \bibnamefont {Smith}},\ and\ \bibinfo {author} {\bibfnamefont {R.~J.}\ \bibnamefont {McQueeney}},\ }\bibfield  {title} {\bibinfo {title} {{Experimental Electronic Heat Capacities of $\alpha$- and $\delta$-Plutonium: Heavy-Fermion Physics in an Element}},\ }\href {https://doi.org/10.1103/physrevlett.91.205901} {\bibfield  {journal} {\bibinfo  {journal} {Physical Review Letters}\ }\textbf {\bibinfo {volume} {91}},\ \bibinfo {pages} {205901} (\bibinfo {year} {2003})}\BibitemShut {NoStop}%
\bibitem [{\citenamefont {Javorský}\ \emph {et~al.}(2006)\citenamefont {Javorský}, \citenamefont {Havela}, \citenamefont {Wastin}, \citenamefont {Colineau}, ,\ and\ \citenamefont {Bouëxière}}]{Javorsky_2006_SpecificHeat}%
  \BibitemOpen
  \bibfield  {author} {\bibinfo {author} {\bibfnamefont {P.}~\bibnamefont {Javorský}}, \bibinfo {author} {\bibfnamefont {L.}~\bibnamefont {Havela}}, \bibinfo {author} {\bibfnamefont {F.}~\bibnamefont {Wastin}}, \bibinfo {author} {\bibfnamefont {E.}~\bibnamefont {Colineau}}, ,\ and\ \bibinfo {author} {\bibfnamefont {D.}~\bibnamefont {Bouëxière}},\ }\bibfield  {title} {\bibinfo {title} {{Specific Heat of $\delta$-Pu Stabilized by Am}},\ }\href {https://doi.org/10.1103/physrevlett.96.156404} {\bibfield  {journal} {\bibinfo  {journal} {Physical Review Letters}\ }\textbf {\bibinfo {volume} {96}},\ \bibinfo {pages} {156404} (\bibinfo {year} {2006})}\BibitemShut {NoStop}%
\bibitem [{\citenamefont {Janoschek}\ \emph {et~al.}(2015)\citenamefont {Janoschek}, \citenamefont {Das}, \citenamefont {Chakrabarti}, \citenamefont {Abernathy}, \citenamefont {Lumsden}, \citenamefont {Lawrence}, \citenamefont {Thompson}, \citenamefont {Lander}, \citenamefont {Mitchell}, \citenamefont {Richmond}, \citenamefont {Ramos}, \citenamefont {Trouw}, \citenamefont {Zhu}, \citenamefont {Haule}, \citenamefont {Kotliar},\ and\ \citenamefont {Bauer}}]{Janoschek_2015_valencefluctuatingground}%
  \BibitemOpen
  \bibfield  {author} {\bibinfo {author} {\bibfnamefont {M.}~\bibnamefont {Janoschek}}, \bibinfo {author} {\bibfnamefont {P.}~\bibnamefont {Das}}, \bibinfo {author} {\bibfnamefont {B.}~\bibnamefont {Chakrabarti}}, \bibinfo {author} {\bibfnamefont {D.~L.}\ \bibnamefont {Abernathy}}, \bibinfo {author} {\bibfnamefont {M.~D.}\ \bibnamefont {Lumsden}}, \bibinfo {author} {\bibfnamefont {J.~M.}\ \bibnamefont {Lawrence}}, \bibinfo {author} {\bibfnamefont {J.~D.}\ \bibnamefont {Thompson}}, \bibinfo {author} {\bibfnamefont {G.~H.}\ \bibnamefont {Lander}}, \bibinfo {author} {\bibfnamefont {J.~N.}\ \bibnamefont {Mitchell}}, \bibinfo {author} {\bibfnamefont {S.}~\bibnamefont {Richmond}}, \bibinfo {author} {\bibfnamefont {M.}~\bibnamefont {Ramos}}, \bibinfo {author} {\bibfnamefont {F.}~\bibnamefont {Trouw}}, \bibinfo {author} {\bibfnamefont {J.-X.}\ \bibnamefont {Zhu}}, \bibinfo {author} {\bibfnamefont {K.}~\bibnamefont {Haule}}, \bibinfo {author} {\bibfnamefont {G.}~\bibnamefont {Kotliar}},\ and\ \bibinfo {author}
  {\bibfnamefont {E.~D.}\ \bibnamefont {Bauer}},\ }\bibfield  {title} {\bibinfo {title} {{The valence-fluctuating ground state of plutonium}},\ }\href {https://doi.org/10.1126/sciadv.1500188} {\bibfield  {journal} {\bibinfo  {journal} {Science Advances}\ }\textbf {\bibinfo {volume} {1}},\ \bibinfo {pages} {e1500188} (\bibinfo {year} {2015})}\BibitemShut {NoStop}%
\bibitem [{\citenamefont {Ramshaw}\ \emph {et~al.}(2015)\citenamefont {Ramshaw}, \citenamefont {Shekhter}, \citenamefont {McDonald}, \citenamefont {Betts}, \citenamefont {Mitchell}, \citenamefont {Tobash}, \citenamefont {Mielke}, \citenamefont {Bauer},\ and\ \citenamefont {Migliori}}]{Ramshaw_2015_Avoidedvalencetransition}%
  \BibitemOpen
  \bibfield  {author} {\bibinfo {author} {\bibfnamefont {B.~J.}\ \bibnamefont {Ramshaw}}, \bibinfo {author} {\bibfnamefont {A.}~\bibnamefont {Shekhter}}, \bibinfo {author} {\bibfnamefont {R.~D.}\ \bibnamefont {McDonald}}, \bibinfo {author} {\bibfnamefont {J.~B.}\ \bibnamefont {Betts}}, \bibinfo {author} {\bibfnamefont {J.~N.}\ \bibnamefont {Mitchell}}, \bibinfo {author} {\bibfnamefont {P.~H.}\ \bibnamefont {Tobash}}, \bibinfo {author} {\bibfnamefont {C.~H.}\ \bibnamefont {Mielke}}, \bibinfo {author} {\bibfnamefont {E.~D.}\ \bibnamefont {Bauer}},\ and\ \bibinfo {author} {\bibfnamefont {A.}~\bibnamefont {Migliori}},\ }\bibfield  {title} {\bibinfo {title} {{Avoided valence transition in a plutonium superconductor}},\ }\href {https://doi.org/10.1073/pnas.1421174112} {\bibfield  {journal} {\bibinfo  {journal} {Proceedings of the National Academy of Sciences}\ }\textbf {\bibinfo {volume} {112}},\ \bibinfo {pages} {3285} (\bibinfo {year} {2015})}\BibitemShut {NoStop}%
\bibitem [{\citenamefont {Morss}\ \emph {et~al.}(2011)\citenamefont {Morss}, \citenamefont {Edelstein}, \citenamefont {Fuger},\ and\ \citenamefont {Katz}}]{Morss_2011_ChemistryActinideTransactinide}%
  \BibitemOpen
  \bibfield  {author} {\bibinfo {author} {\bibfnamefont {L.~R.}\ \bibnamefont {Morss}}, \bibinfo {author} {\bibfnamefont {N.~M.}\ \bibnamefont {Edelstein}}, \bibinfo {author} {\bibfnamefont {J.}~\bibnamefont {Fuger}},\ and\ \bibinfo {author} {\bibfnamefont {J.~J.}\ \bibnamefont {Katz}},\ }\href {https://doi.org/10.1007/978-94-007-0211-0} {\emph {\bibinfo {title} {{The Chemistry of the Actinide and Transactinide Elements}}}}\ (\bibinfo  {publisher} {Springer Netherlands},\ \bibinfo {address} {Dordrecht},\ \bibinfo {year} {2011})\BibitemShut {NoStop}%
\bibitem [{\citenamefont {Clark}\ \emph {et~al.}(2006)\citenamefont {Clark}, \citenamefont {Hecker}, \citenamefont {Jarvinen},\ and\ \citenamefont {Neu}}]{Clark_2006_PlutoniumPlutoniumCompounds}%
  \BibitemOpen
  \bibfield  {author} {\bibinfo {author} {\bibfnamefont {D.~L.}\ \bibnamefont {Clark}}, \bibinfo {author} {\bibfnamefont {S.~S.}\ \bibnamefont {Hecker}}, \bibinfo {author} {\bibfnamefont {G.~D.}\ \bibnamefont {Jarvinen}},\ and\ \bibinfo {author} {\bibfnamefont {M.~P.}\ \bibnamefont {Neu}},\ }\bibinfo {title} {{Plutonium and Plutonium Compounds}},\ in\ \href {https://doi.org/10.1002/0471238961.1612212013151819.a01.pub3} {\emph {\bibinfo {booktitle} {Kirk‐Othmer Encyclopedia of Chemical Technology}}}\ (\bibinfo  {publisher} {John Wiley \& Sons, Ltd},\ \bibinfo {year} {2006})\BibitemShut {NoStop}%
\bibitem [{\citenamefont {Green}\ \emph {et~al.}(1970)\citenamefont {Green}, \citenamefont {Arnold}, \citenamefont {Leary},\ and\ \citenamefont {Nereson}}]{Green_1970_Crystallographicmagneticordering}%
  \BibitemOpen
  \bibfield  {author} {\bibinfo {author} {\bibfnamefont {J.~L.}\ \bibnamefont {Green}}, \bibinfo {author} {\bibfnamefont {G.~P.}\ \bibnamefont {Arnold}}, \bibinfo {author} {\bibfnamefont {J.~A.}\ \bibnamefont {Leary}},\ and\ \bibinfo {author} {\bibfnamefont {N.~G.}\ \bibnamefont {Nereson}},\ }\bibfield  {title} {\bibinfo {title} {Crystallographic and magnetic ordering studies of plutonium carbides using neutron diffraction},\ }\href {https://doi.org/10.1016/0022-3115(70)90194-7} {\bibfield  {journal} {\bibinfo  {journal} {Journal of Nuclear Materials}\ }\textbf {\bibinfo {volume} {34}},\ \bibinfo {pages} {281} (\bibinfo {year} {1970})}\BibitemShut {NoStop}%
\bibitem [{\citenamefont {Ellinger}\ \emph {et~al.}(1968)\citenamefont {Ellinger}, \citenamefont {Miner}, \citenamefont {O'Boyle},\ and\ \citenamefont {Schonfeld}}]{Ellinger_1968_ConstitutionPlutoniumAlloys}%
  \BibitemOpen
  \bibfield  {author} {\bibinfo {author} {\bibfnamefont {F.~H.}\ \bibnamefont {Ellinger}}, \bibinfo {author} {\bibfnamefont {W.~N.}\ \bibnamefont {Miner}}, \bibinfo {author} {\bibfnamefont {D.~R.}\ \bibnamefont {O'Boyle}},\ and\ \bibinfo {author} {\bibfnamefont {F.~W.}\ \bibnamefont {Schonfeld}},\ }\href {https://doi.org/10.2172/4800722} {\emph {\bibinfo {title} {{Constitution of Plutonium Alloys}}}},\ \bibinfo {type} {Tech. Rep.}\ (\bibinfo {year} {1968})\BibitemShut {NoStop}%
\bibitem [{\citenamefont {Raphael}\ and\ \citenamefont {de~Novion}(1969)}]{Raphael_1969_Susceptibilitiesmagnetiquesdes}%
  \BibitemOpen
  \bibfield  {author} {\bibinfo {author} {\bibfnamefont {G.}~\bibnamefont {Raphael}}\ and\ \bibinfo {author} {\bibfnamefont {C.}~\bibnamefont {de~Novion}},\ }\bibfield  {title} {\bibinfo {title} {{Susceptibilities magnetiques des mononitrures et sesquicarbures de thorium, uranium et plutonium}},\ }\href {https://doi.org/10.1016/0038-1098(69)90668-1} {\bibfield  {journal} {\bibinfo  {journal} {Solid State Communications}\ }\textbf {\bibinfo {volume} {7}},\ \bibinfo {pages} {791} (\bibinfo {year} {1969})}\BibitemShut {NoStop}%
\bibitem [{\citenamefont {Gouder}\ \emph {et~al.}(2007)\citenamefont {Gouder}, \citenamefont {Havela}, \citenamefont {Shick}, \citenamefont {Huber}, \citenamefont {Wastin},\ and\ \citenamefont {Rebizant}}]{Gouder_2007_Variability5fstates}%
  \BibitemOpen
  \bibfield  {author} {\bibinfo {author} {\bibfnamefont {T.}~\bibnamefont {Gouder}}, \bibinfo {author} {\bibfnamefont {L.}~\bibnamefont {Havela}}, \bibinfo {author} {\bibfnamefont {A.~B.}\ \bibnamefont {Shick}}, \bibinfo {author} {\bibfnamefont {F.}~\bibnamefont {Huber}}, \bibinfo {author} {\bibfnamefont {F.}~\bibnamefont {Wastin}},\ and\ \bibinfo {author} {\bibfnamefont {J.}~\bibnamefont {Rebizant}},\ }\bibfield  {title} {\bibinfo {title} {{Variability of 5$f$ states in plutonium carbides}},\ }\href {https://doi.org/10.1088/0953-8984/19/47/476201} {\bibfield  {journal} {\bibinfo  {journal} {Journal of Physics: Condensed Matter}\ }\textbf {\bibinfo {volume} {19}},\ \bibinfo {pages} {476201} (\bibinfo {year} {2007})}\BibitemShut {NoStop}%
\bibitem [{\citenamefont {Joyce}\ \emph {et~al.}(2003)\citenamefont {Joyce}, \citenamefont {Wills}, \citenamefont {Durakiewicz}, \citenamefont {Butterfield}, \citenamefont {Guziewicz}, \citenamefont {Sarrao}, \citenamefont {Morales}, \citenamefont {Arko},\ and\ \citenamefont {Eriksson}}]{Joyce_2003_PhotoemissionElectronicStructure}%
  \BibitemOpen
  \bibfield  {author} {\bibinfo {author} {\bibfnamefont {J.~J.}\ \bibnamefont {Joyce}}, \bibinfo {author} {\bibfnamefont {J.~M.}\ \bibnamefont {Wills}}, \bibinfo {author} {\bibfnamefont {T.}~\bibnamefont {Durakiewicz}}, \bibinfo {author} {\bibfnamefont {M.~T.}\ \bibnamefont {Butterfield}}, \bibinfo {author} {\bibfnamefont {E.}~\bibnamefont {Guziewicz}}, \bibinfo {author} {\bibfnamefont {J.~L.}\ \bibnamefont {Sarrao}}, \bibinfo {author} {\bibfnamefont {L.~A.}\ \bibnamefont {Morales}}, \bibinfo {author} {\bibfnamefont {A.~J.}\ \bibnamefont {Arko}},\ and\ \bibinfo {author} {\bibfnamefont {O.}~\bibnamefont {Eriksson}},\ }\bibfield  {title} {\bibinfo {title} {{Photoemission and the Electronic Structure of PuCoGa$_5$}},\ }\href {https://doi.org/10.1103/physrevlett.91.176401} {\bibfield  {journal} {\bibinfo  {journal} {Physical Review Letters}\ }\textbf {\bibinfo {volume} {91}},\ \bibinfo {pages} {176401} (\bibinfo {year} {2003})}\BibitemShut {NoStop}%
\bibitem [{\citenamefont {Havela}\ \emph {et~al.}(2009)\citenamefont {Havela}, \citenamefont {Shick},\ and\ \citenamefont {Gouder}}]{Havela_2009_Magneticpropertiesplutonium}%
  \BibitemOpen
  \bibfield  {author} {\bibinfo {author} {\bibfnamefont {L.}~\bibnamefont {Havela}}, \bibinfo {author} {\bibfnamefont {A.}~\bibnamefont {Shick}},\ and\ \bibinfo {author} {\bibfnamefont {T.}~\bibnamefont {Gouder}},\ }\bibfield  {title} {\bibinfo {title} {{Magnetic properties of plutonium and Pu compounds}},\ }\href {https://doi.org/10.1063/1.3062946} {\bibfield  {journal} {\bibinfo  {journal} {Journal of Applied Physics}\ }\textbf {\bibinfo {volume} {105}},\ \bibinfo {pages} {07E130} (\bibinfo {year} {2009})}\BibitemShut {NoStop}%
\bibitem [{\citenamefont {Atoji}\ and\ \citenamefont {Williams}(1961)}]{Atoji_1961_Neutron‐DiffractionStudiesLa_2C_3}%
  \BibitemOpen
  \bibfield  {author} {\bibinfo {author} {\bibfnamefont {M.}~\bibnamefont {Atoji}}\ and\ \bibinfo {author} {\bibfnamefont {D.~E.}\ \bibnamefont {Williams}},\ }\bibfield  {title} {\bibinfo {title} {{Neutron‐Diffraction Studies of La$_2$C$_3$, Ce$_2$C$_3$, Pr$_2$C$_3$, and Tb$_2$C$_3$}},\ }\href {https://doi.org/10.1063/1.1732193} {\bibfield  {journal} {\bibinfo  {journal} {Journal of Chemical Physics}\ }\textbf {\bibinfo {volume} {35}},\ \bibinfo {pages} {1960} (\bibinfo {year} {1961})}\BibitemShut {NoStop}%
\bibitem [{\citenamefont {Jin}\ \emph {et~al.}(2021)\citenamefont {Jin}, \citenamefont {Liu}, \citenamefont {Zhang}, \citenamefont {Dai},\ and\ \citenamefont {Liu}}]{Jin_2021_Sixfoldfourfoldthreefold}%
  \BibitemOpen
  \bibfield  {author} {\bibinfo {author} {\bibfnamefont {L.}~\bibnamefont {Jin}}, \bibinfo {author} {\bibfnamefont {Y.}~\bibnamefont {Liu}}, \bibinfo {author} {\bibfnamefont {X.}~\bibnamefont {Zhang}}, \bibinfo {author} {\bibfnamefont {X.}~\bibnamefont {Dai}},\ and\ \bibinfo {author} {\bibfnamefont {G.}~\bibnamefont {Liu}},\ }\bibfield  {title} {\bibinfo {title} {{Sixfold, fourfold, and threefold excitations in the rare-earth metal carbide R$_2$C$_3$}},\ }\href {https://doi.org/10.1103/physrevb.104.045111} {\bibfield  {journal} {\bibinfo  {journal} {Physical Review B}\ }\textbf {\bibinfo {volume} {104}},\ \bibinfo {pages} {045111} (\bibinfo {year} {2021})}\BibitemShut {NoStop}%
\bibitem [{\citenamefont {Kim}\ \emph {et~al.}(2007)\citenamefont {Kim}, \citenamefont {Xie}, \citenamefont {Kremer}, \citenamefont {Babizhetskyy}, \citenamefont {Jepsen}, \citenamefont {Simon}, \citenamefont {Ahn}, \citenamefont {Raquet}, \citenamefont {Rakoto}, \citenamefont {Broto},\ and\ \citenamefont {Ouladdiaf}}]{Kim_2007_Strongelectronphonon}%
  \BibitemOpen
  \bibfield  {author} {\bibinfo {author} {\bibfnamefont {J.~S.}\ \bibnamefont {Kim}}, \bibinfo {author} {\bibfnamefont {W.}~\bibnamefont {Xie}}, \bibinfo {author} {\bibfnamefont {R.~K.}\ \bibnamefont {Kremer}}, \bibinfo {author} {\bibfnamefont {V.}~\bibnamefont {Babizhetskyy}}, \bibinfo {author} {\bibfnamefont {O.}~\bibnamefont {Jepsen}}, \bibinfo {author} {\bibfnamefont {A.}~\bibnamefont {Simon}}, \bibinfo {author} {\bibfnamefont {K.~S.}\ \bibnamefont {Ahn}}, \bibinfo {author} {\bibfnamefont {B.}~\bibnamefont {Raquet}}, \bibinfo {author} {\bibfnamefont {H.}~\bibnamefont {Rakoto}}, \bibinfo {author} {\bibfnamefont {J.-M.}\ \bibnamefont {Broto}},\ and\ \bibinfo {author} {\bibfnamefont {B.}~\bibnamefont {Ouladdiaf}},\ }\bibfield  {title} {\bibinfo {title} {{Strong electron-phonon coupling in the rare-earth carbide superconductor La$_2$C$_3$}},\ }\href {https://doi.org/10.1103/physrevb.76.014516} {\bibfield  {journal} {\bibinfo  {journal} {Physical Review B}\ }\textbf {\bibinfo {volume} {76}},\ \bibinfo {pages}
  {014516} (\bibinfo {year} {2007})}\BibitemShut {NoStop}%
\bibitem [{\citenamefont {Sugawara}\ \emph {et~al.}(2007)\citenamefont {Sugawara}, \citenamefont {Sato}, \citenamefont {Souma}, \citenamefont {Takahashi},\ and\ \citenamefont {Ochiai}}]{Sugawara_2007_Anomaloussuperconductinggap}%
  \BibitemOpen
  \bibfield  {author} {\bibinfo {author} {\bibfnamefont {K.}~\bibnamefont {Sugawara}}, \bibinfo {author} {\bibfnamefont {T.}~\bibnamefont {Sato}}, \bibinfo {author} {\bibfnamefont {S.}~\bibnamefont {Souma}}, \bibinfo {author} {\bibfnamefont {T.}~\bibnamefont {Takahashi}},\ and\ \bibinfo {author} {\bibfnamefont {A.}~\bibnamefont {Ochiai}},\ }\bibfield  {title} {\bibinfo {title} {{Anomalous superconducting-gap symmetry of noncentrosymmetric La$_2$C$_3$ observed by ultrahigh-resolution photoemission spectroscopy}},\ }\href {https://doi.org/10.1103/physrevb.76.132512} {\bibfield  {journal} {\bibinfo  {journal} {Physical Review B}\ }\textbf {\bibinfo {volume} {76}},\ \bibinfo {pages} {132512} (\bibinfo {year} {2007})}\BibitemShut {NoStop}%
\bibitem [{\citenamefont {Akutagawa}\ and\ \citenamefont {Akimitsu}(2007)}]{Akutagawa_2007_SuperconductivityY_2C_3Investigated}%
  \BibitemOpen
  \bibfield  {author} {\bibinfo {author} {\bibfnamefont {S.}~\bibnamefont {Akutagawa}}\ and\ \bibinfo {author} {\bibfnamefont {J.}~\bibnamefont {Akimitsu}},\ }\bibfield  {title} {\bibinfo {title} {{Superconductivity of Y$_2$C$_3$ Investigated by Specific Heat Measurement}},\ }\href {https://doi.org/10.1143/jpsj.76.024713} {\bibfield  {journal} {\bibinfo  {journal} {Journal of the Physical Society of Japan}\ }\textbf {\bibinfo {volume} {76}},\ \bibinfo {pages} {024713} (\bibinfo {year} {2007})}\BibitemShut {NoStop}%
\bibitem [{\citenamefont {Chen}\ \emph {et~al.}(2011)\citenamefont {Chen}, \citenamefont {Salamon}, \citenamefont {Akutagawa}, \citenamefont {Akimitsu}, \citenamefont {Singleton}, \citenamefont {Zhang}, \citenamefont {Jiao},\ and\ \citenamefont {Yuan}}]{Chen_2011_Evidencenodalgap}%
  \BibitemOpen
  \bibfield  {author} {\bibinfo {author} {\bibfnamefont {J.}~\bibnamefont {Chen}}, \bibinfo {author} {\bibfnamefont {M.~B.}\ \bibnamefont {Salamon}}, \bibinfo {author} {\bibfnamefont {S.}~\bibnamefont {Akutagawa}}, \bibinfo {author} {\bibfnamefont {J.}~\bibnamefont {Akimitsu}}, \bibinfo {author} {\bibfnamefont {J.}~\bibnamefont {Singleton}}, \bibinfo {author} {\bibfnamefont {J.~L.}\ \bibnamefont {Zhang}}, \bibinfo {author} {\bibfnamefont {L.}~\bibnamefont {Jiao}},\ and\ \bibinfo {author} {\bibfnamefont {H.~Q.}\ \bibnamefont {Yuan}},\ }\bibfield  {title} {\bibinfo {title} {{Evidence of nodal gap structure in the noncentrosymmetric superconductor Y$_2$C$_3$}},\ }\href {https://doi.org/10.1103/physrevb.83.144529} {\bibfield  {journal} {\bibinfo  {journal} {Physical Review B}\ }\textbf {\bibinfo {volume} {83}},\ \bibinfo {pages} {144529} (\bibinfo {year} {2011})}\BibitemShut {NoStop}%
\bibitem [{\citenamefont {Smidman}\ \emph {et~al.}(2017)\citenamefont {Smidman}, \citenamefont {Salamon}, \citenamefont {Yuan},\ and\ \citenamefont {Agterberg}}]{Smidman_2017_Superconductivityspin–orbitcoupling}%
  \BibitemOpen
  \bibfield  {author} {\bibinfo {author} {\bibfnamefont {M.}~\bibnamefont {Smidman}}, \bibinfo {author} {\bibfnamefont {M.~B.}\ \bibnamefont {Salamon}}, \bibinfo {author} {\bibfnamefont {H.~Q.}\ \bibnamefont {Yuan}},\ and\ \bibinfo {author} {\bibfnamefont {D.~F.}\ \bibnamefont {Agterberg}},\ }\bibfield  {title} {\bibinfo {title} {Superconductivity and spin–orbit coupling in non-centrosymmetric materials: a review},\ }\href {https://doi.org/10.1088/1361-6633/80/3/036501} {\bibfield  {journal} {\bibinfo  {journal} {Reports on Progress in Physics}\ }\textbf {\bibinfo {volume} {80}},\ \bibinfo {pages} {036501} (\bibinfo {year} {2017})}\BibitemShut {NoStop}%
\bibitem [{\citenamefont {De~Novion}\ \emph {et~al.}(1965)\citenamefont {De~Novion}, \citenamefont {Costa},\ and\ \citenamefont {Dean}}]{DeNovion_1965_Existencedunetransition}%
  \BibitemOpen
  \bibfield  {author} {\bibinfo {author} {\bibfnamefont {C.}~\bibnamefont {De~Novion}}, \bibinfo {author} {\bibfnamefont {P.}~\bibnamefont {Costa}},\ and\ \bibinfo {author} {\bibfnamefont {G.}~\bibnamefont {Dean}},\ }\bibfield  {title} {\bibinfo {title} {{Existence d'une transition magnetique dans le sesquicarbure d'uranium U$_2$C$_3$}},\ }\href {https://doi.org/10.1016/0031-9163(65)90097-1} {\bibfield  {journal} {\bibinfo  {journal} {Physics Letters}\ }\textbf {\bibinfo {volume} {19}},\ \bibinfo {pages} {455} (\bibinfo {year} {1965})}\BibitemShut {NoStop}%
\bibitem [{\citenamefont {Koelling}\ \emph {et~al.}(1985)\citenamefont {Koelling}, \citenamefont {Dunlap},\ and\ \citenamefont {Crabtree}}]{Koelling_1985_felectronhybridization}%
  \BibitemOpen
  \bibfield  {author} {\bibinfo {author} {\bibfnamefont {D.~D.}\ \bibnamefont {Koelling}}, \bibinfo {author} {\bibfnamefont {B.~D.}\ \bibnamefont {Dunlap}},\ and\ \bibinfo {author} {\bibfnamefont {G.~W.}\ \bibnamefont {Crabtree}},\ }\bibfield  {title} {\bibinfo {title} {{$f$-electron hybridization and heavy-fermion compounds}},\ }\href {https://doi.org/10.1103/physrevb.31.4966} {\bibfield  {journal} {\bibinfo  {journal} {Physical Review B}\ }\textbf {\bibinfo {volume} {31}},\ \bibinfo {pages} {4966} (\bibinfo {year} {1985})}\BibitemShut {NoStop}%
\bibitem [{\citenamefont {Cornelius}\ \emph {et~al.}(1999)\citenamefont {Cornelius}, \citenamefont {Arko}, \citenamefont {Sarrao}, \citenamefont {Thompson}, \citenamefont {Hundley}, \citenamefont {Booth}, \citenamefont {Harrison},\ and\ \citenamefont {Oppeneer}}]{Cornelius_1999_ElectronicpropertiesUX_3}%
  \BibitemOpen
  \bibfield  {author} {\bibinfo {author} {\bibfnamefont {A.~L.}\ \bibnamefont {Cornelius}}, \bibinfo {author} {\bibfnamefont {A.~J.}\ \bibnamefont {Arko}}, \bibinfo {author} {\bibfnamefont {J.~L.}\ \bibnamefont {Sarrao}}, \bibinfo {author} {\bibfnamefont {J.~D.}\ \bibnamefont {Thompson}}, \bibinfo {author} {\bibfnamefont {M.~F.}\ \bibnamefont {Hundley}}, \bibinfo {author} {\bibfnamefont {C.~H.}\ \bibnamefont {Booth}}, \bibinfo {author} {\bibfnamefont {N.}~\bibnamefont {Harrison}},\ and\ \bibinfo {author} {\bibfnamefont {P.~M.}\ \bibnamefont {Oppeneer}},\ }\bibfield  {title} {\bibinfo {title} {{Electronic properties of UX$_3$ (X = Ga, Al, and Sn) compounds in high magnetic fields: Transport, specific heat, magnetization, and quantum oscillations}},\ }\href {https://doi.org/10.1103/physrevb.59.14473} {\bibfield  {journal} {\bibinfo  {journal} {Physical Review B}\ }\textbf {\bibinfo {volume} {59}},\ \bibinfo {pages} {14473} (\bibinfo {year} {1999})}\BibitemShut {NoStop}%
\bibitem [{\citenamefont {Jullien}\ and\ \citenamefont {Coqblin}(1976)}]{Jullien_1976_Existencedunetransition}%
  \BibitemOpen
  \bibfield  {author} {\bibinfo {author} {\bibfnamefont {R.}~\bibnamefont {Jullien}}\ and\ \bibinfo {author} {\bibfnamefont {B.}~\bibnamefont {Coqblin}},\ }\bibfield  {title} {\bibinfo {title} {{Existence d'une transition magnetique dans le sesquicarbure d'uranium U$_2$C$_3$}},\ }\href {https://doi.org/10.1007/bf00654717} {\bibfield  {journal} {\bibinfo  {journal} {Journal of Low Temperature Physics}\ }\textbf {\bibinfo {volume} {22}},\ \bibinfo {pages} {437} (\bibinfo {year} {1976})}\BibitemShut {NoStop}%
\bibitem [{\citenamefont {Béal-Monod}\ \emph {et~al.}(1968)\citenamefont {Béal-Monod}, \citenamefont {Ma},\ and\ \citenamefont {Fredkin}}]{BealMonod_1968_TemperatureDependenceSpin}%
  \BibitemOpen
  \bibfield  {author} {\bibinfo {author} {\bibfnamefont {M.~T.}\ \bibnamefont {Béal-Monod}}, \bibinfo {author} {\bibfnamefont {S.-K.}\ \bibnamefont {Ma}},\ and\ \bibinfo {author} {\bibfnamefont {D.~R.}\ \bibnamefont {Fredkin}},\ }\bibfield  {title} {\bibinfo {title} {{Temperature Dependence of the Spin Susceptibility of a Nearly Ferromagnetic Fermi Liquid}},\ }\href {https://doi.org/10.1103/physrevlett.20.929} {\bibfield  {journal} {\bibinfo  {journal} {Physical Review Letters}\ }\textbf {\bibinfo {volume} {20}},\ \bibinfo {pages} {929} (\bibinfo {year} {1968})}\BibitemShut {NoStop}%
\bibitem [{\citenamefont {Trainor}\ \emph {et~al.}(1975)\citenamefont {Trainor}, \citenamefont {Brodsky},\ and\ \citenamefont {Culbert}}]{Trainor_1975_SpecificHeatSpin}%
  \BibitemOpen
  \bibfield  {author} {\bibinfo {author} {\bibfnamefont {R.~J.}\ \bibnamefont {Trainor}}, \bibinfo {author} {\bibfnamefont {M.~B.}\ \bibnamefont {Brodsky}},\ and\ \bibinfo {author} {\bibfnamefont {H.~V.}\ \bibnamefont {Culbert}},\ }\bibfield  {title} {\bibinfo {title} {{Specific Heat of the Spin-Fluctuation System UAl$_2$}},\ }\href {https://doi.org/10.1103/physrevlett.34.1019} {\bibfield  {journal} {\bibinfo  {journal} {Physical Review Letters}\ }\textbf {\bibinfo {volume} {34}},\ \bibinfo {pages} {1019} (\bibinfo {year} {1975})}\BibitemShut {NoStop}%
\bibitem [{\citenamefont {Hill}(1970)}]{Hill_1970_Earlyactinidesperiodic}%
  \BibitemOpen
  \bibfield  {author} {\bibinfo {author} {\bibfnamefont {H.~H.}\ \bibnamefont {Hill}},\ }\bibfield  {title} {\bibinfo {title} {{Early actinides: the periodic system's $f$ electron transition metal series}},\ }\href {https://www.osti.gov/biblio/4065028} {\bibfield  {journal} {\bibinfo  {journal} {Nucl. Met., Met. Soc. AIME 17}\ ,\ \bibinfo {pages} {2}} (\bibinfo {year} {1970})}\BibitemShut {NoStop}%
\bibitem [{\citenamefont {Brodsky}(1978)}]{Brodsky_1978_Magneticpropertiesactinide}%
  \BibitemOpen
  \bibfield  {author} {\bibinfo {author} {\bibfnamefont {M.~B.}\ \bibnamefont {Brodsky}},\ }\bibfield  {title} {\bibinfo {title} {Magnetic properties of the actinide elements and their metallic compounds},\ }\href {https://doi.org/10.1088/0034-4885/41/10/001} {\bibfield  {journal} {\bibinfo  {journal} {Reports on Progress in Physics}\ }\textbf {\bibinfo {volume} {41}},\ \bibinfo {pages} {1547} (\bibinfo {year} {1978})}\BibitemShut {NoStop}%
\bibitem [{\citenamefont {Yasuoka}(2019)}]{Yasuoka_2019_Plutonium239NMR}%
  \BibitemOpen
  \bibfield  {author} {\bibinfo {author} {\bibfnamefont {H.}~\bibnamefont {Yasuoka}},\ }\bibinfo {title} {{Plutonium Handbook: Plutonium-239 NMR of Plutonium Compounds in the Solid State}}\ (\bibinfo  {publisher} {American Nuclear Society},\ \bibinfo {year} {2019})\ Chap.~\bibinfo {chapter} {42}, p.\ \bibinfo {pages} {2938},\ \bibinfo {edition} {7th}\ ed.\BibitemShut {Stop}%
\bibitem [{\citenamefont {Bak}\ \emph {et~al.}(2011)\citenamefont {Bak}, \citenamefont {Rasmussen},\ and\ \citenamefont {Nielsen}}]{Bak_2011_SIMPSONgeneralsimulation}%
  \BibitemOpen
  \bibfield  {author} {\bibinfo {author} {\bibfnamefont {M.}~\bibnamefont {Bak}}, \bibinfo {author} {\bibfnamefont {J.~T.}\ \bibnamefont {Rasmussen}},\ and\ \bibinfo {author} {\bibfnamefont {N.~C.}\ \bibnamefont {Nielsen}},\ }\bibfield  {title} {\bibinfo {title} {{SIMPSON: A general simulation program for solid-state NMR spectroscopy}},\ }\href {https://doi.org/10.1016/j.jmr.2011.09.008} {\bibfield  {journal} {\bibinfo  {journal} {Journal of Magnetic Resonance}\ }\textbf {\bibinfo {volume} {213}},\ \bibinfo {pages} {366} (\bibinfo {year} {2011})}\BibitemShut {NoStop}%
\bibitem [{\citenamefont {Tošner}\ \emph {et~al.}(2014)\citenamefont {Tošner}, \citenamefont {Andersen}, \citenamefont {Stevensson}, \citenamefont {Edén}, \citenamefont {Nielsen},\ and\ \citenamefont {Vosegaard}}]{Tosner_2014_Computerintensivesimulation}%
  \BibitemOpen
  \bibfield  {author} {\bibinfo {author} {\bibfnamefont {Z.}~\bibnamefont {Tošner}}, \bibinfo {author} {\bibfnamefont {R.}~\bibnamefont {Andersen}}, \bibinfo {author} {\bibfnamefont {B.}~\bibnamefont {Stevensson}}, \bibinfo {author} {\bibfnamefont {M.}~\bibnamefont {Edén}}, \bibinfo {author} {\bibfnamefont {N.~C.}\ \bibnamefont {Nielsen}},\ and\ \bibinfo {author} {\bibfnamefont {T.}~\bibnamefont {Vosegaard}},\ }\bibfield  {title} {\bibinfo {title} {{Computer-intensive simulation of solid-state NMR experiments using SIMPSON}},\ }\href {https://doi.org/10.1016/j.jmr.2014.07.002} {\bibfield  {journal} {\bibinfo  {journal} {Journal of Magnetic Resonance}\ }\textbf {\bibinfo {volume} {246}},\ \bibinfo {pages} {79} (\bibinfo {year} {2014})}\BibitemShut {NoStop}%
\bibitem [{\citenamefont {Harris}\ \emph {et~al.}(2001)\citenamefont {Harris}, \citenamefont {Becker}, \citenamefont {Cabral~de Menezes}, \citenamefont {Goodfellow},\ and\ \citenamefont {Granger}}]{Harris_2001_NMRnomenclatureNuclear}%
  \BibitemOpen
  \bibfield  {author} {\bibinfo {author} {\bibfnamefont {R.~K.}\ \bibnamefont {Harris}}, \bibinfo {author} {\bibfnamefont {E.~D.}\ \bibnamefont {Becker}}, \bibinfo {author} {\bibfnamefont {S.~M.}\ \bibnamefont {Cabral~de Menezes}}, \bibinfo {author} {\bibfnamefont {R.}~\bibnamefont {Goodfellow}},\ and\ \bibinfo {author} {\bibfnamefont {P.}~\bibnamefont {Granger}},\ }\bibfield  {title} {\bibinfo {title} {{{NMR} nomenclature. Nuclear spin properties and conventions for chemical shifts ({IUPAC} Recommendations 2001)}},\ }\href {https://doi.org/10.1351/pac200173111795} {\bibfield  {journal} {\bibinfo  {journal} {Pure and Applied Chemistry}\ }\textbf {\bibinfo {volume} {73}},\ \bibinfo {pages} {1795} (\bibinfo {year} {2001})}\BibitemShut {NoStop}%
\bibitem [{\citenamefont {Harris}\ \emph {et~al.}(2008)\citenamefont {Harris}, \citenamefont {Becker}, \citenamefont {Cabral~de Menezes}, \citenamefont {Granger}, \citenamefont {Hoffman},\ and\ \citenamefont {Zilm}}]{Harris_2008_FurtherconventionsNMR}%
  \BibitemOpen
  \bibfield  {author} {\bibinfo {author} {\bibfnamefont {R.~K.}\ \bibnamefont {Harris}}, \bibinfo {author} {\bibfnamefont {E.~D.}\ \bibnamefont {Becker}}, \bibinfo {author} {\bibfnamefont {S.~M.}\ \bibnamefont {Cabral~de Menezes}}, \bibinfo {author} {\bibfnamefont {P.}~\bibnamefont {Granger}}, \bibinfo {author} {\bibfnamefont {R.~E.}\ \bibnamefont {Hoffman}},\ and\ \bibinfo {author} {\bibfnamefont {K.~W.}\ \bibnamefont {Zilm}},\ }\bibfield  {title} {\bibinfo {title} {{Further conventions for NMR shielding and chemical shifts (IUPAC Recommendations 2008)}},\ }\href {https://doi.org/10.1351/pac200880010059} {\bibfield  {journal} {\bibinfo  {journal} {Pure and Applied Chemistry}\ }\textbf {\bibinfo {volume} {80}},\ \bibinfo {pages} {59} (\bibinfo {year} {2008})}\BibitemShut {NoStop}%
\bibitem [{\citenamefont {Carter}\ \emph {et~al.}(1977)\citenamefont {Carter}, \citenamefont {Bennett},\ and\ \citenamefont {Kahan}}]{Carter_1977_MetallicshiftsNMR}%
  \BibitemOpen
  \bibfield  {author} {\bibinfo {author} {\bibfnamefont {G.}~\bibnamefont {Carter}}, \bibinfo {author} {\bibfnamefont {L.~H.}\ \bibnamefont {Bennett}},\ and\ \bibinfo {author} {\bibfnamefont {D.~J.}\ \bibnamefont {Kahan}},\ }\href@noop {} {\emph {\bibinfo {title} {Metallic Shifts in NMR}}},\ \bibinfo {series} {Progress in materials science}, Vol.~\bibinfo {volume} {3}\ (\bibinfo  {publisher} {Pergamon Press},\ \bibinfo {address} {Oxford},\ \bibinfo {year} {1977})\BibitemShut {NoStop}%
\bibitem [{\citenamefont {Perdew}\ \emph {et~al.}(1996)\citenamefont {Perdew}, \citenamefont {Burke},\ and\ \citenamefont {Ernzerhof}}]{Perdew_1996_GeneralizedGradientApproximation}%
  \BibitemOpen
  \bibfield  {author} {\bibinfo {author} {\bibfnamefont {J.~P.}\ \bibnamefont {Perdew}}, \bibinfo {author} {\bibfnamefont {K.}~\bibnamefont {Burke}},\ and\ \bibinfo {author} {\bibfnamefont {M.}~\bibnamefont {Ernzerhof}},\ }\bibfield  {title} {\bibinfo {title} {{Generalized Gradient Approximation Made Simple}},\ }\href {https://doi.org/10.1103/physrevlett.77.3865} {\bibfield  {journal} {\bibinfo  {journal} {Physical Review Letters}\ }\textbf {\bibinfo {volume} {77}},\ \bibinfo {pages} {3865} (\bibinfo {year} {1996})}\BibitemShut {NoStop}%
\bibitem [{\citenamefont {Blaha}\ \emph {et~al.}(2019)\citenamefont {Blaha}, \citenamefont {Schwarz}, \citenamefont {Tran}, \citenamefont {Laskowski}, \citenamefont {Madsen},\ and\ \citenamefont {Marks}}]{Blaha_2019_WIEN2kAPWloprogram}%
  \BibitemOpen
  \bibfield  {author} {\bibinfo {author} {\bibfnamefont {P.}~\bibnamefont {Blaha}}, \bibinfo {author} {\bibfnamefont {K.}~\bibnamefont {Schwarz}}, \bibinfo {author} {\bibfnamefont {F.}~\bibnamefont {Tran}}, \bibinfo {author} {\bibfnamefont {R.}~\bibnamefont {Laskowski}}, \bibinfo {author} {\bibfnamefont {G.~K.~H.}\ \bibnamefont {Madsen}},\ and\ \bibinfo {author} {\bibfnamefont {L.~D.}\ \bibnamefont {Marks}},\ }\bibfield  {title} {\bibinfo {title} {{WIEN2k: An APW+lo program for calculating the properties of solids}},\ }\href {https://doi.org/10.1063/1.5143061} {\bibfield  {journal} {\bibinfo  {journal} {The Journal of Chemical Physics}\ }\textbf {\bibinfo {volume} {152}},\ \bibinfo {pages} {074101} (\bibinfo {year} {2019})}\BibitemShut {NoStop}%
\bibitem [{\citenamefont {Zachariasen}(1952)}]{Zachariasen_1952_Crystalchemicalstudies}%
  \BibitemOpen
  \bibfield  {author} {\bibinfo {author} {\bibfnamefont {W.~H.}\ \bibnamefont {Zachariasen}},\ }\bibfield  {title} {{\bibinfo {title} {{Crystal chemical studies of the 5f-series of elements. XV. The crystal structure of plutonium sesquicarbide}}},\ }\href {https://doi.org/10.1107/s0365110x52000058} {\bibfield  {journal} {\bibinfo  {journal} {Acta Crystallographica}\ }\textbf {\bibinfo {volume} {5}},\ \bibinfo {pages} {17} (\bibinfo {year} {1952})}\BibitemShut {NoStop}%
\bibitem [{\citenamefont {Pake}(1948)}]{Pake_1948_NuclearResonanceAbsorption}%
  \BibitemOpen
  \bibfield  {author} {\bibinfo {author} {\bibfnamefont {G.~E.}\ \bibnamefont {Pake}},\ }\bibfield  {title} {\bibinfo {title} {{Nuclear Resonance Absorption in Hydrated Crystals: Fine Structure of the Proton Line}},\ }\href {https://doi.org/10.1063/1.1746878} {\bibfield  {journal} {\bibinfo  {journal} {The Journal of Chemical Physics}\ }\textbf {\bibinfo {volume} {16}},\ \bibinfo {pages} {327} (\bibinfo {year} {1948})}\BibitemShut {NoStop}%
\bibitem [{\citenamefont {Jain}\ and\ \citenamefont {Ganguly}(1993)}]{Jain_1993_EvaluationphasesPu}%
  \BibitemOpen
  \bibfield  {author} {\bibinfo {author} {\bibfnamefont {G.}~\bibnamefont {Jain}}\ and\ \bibinfo {author} {\bibfnamefont {C.}~\bibnamefont {Ganguly}},\ }\bibfield  {title} {\bibinfo {title} {{Evaluation of phases in Pu-C-O and (U, Pu)-C-O systems by X-ray diffractometry and chemical analysis}},\ }\href {https://doi.org/10.1016/0022-3115(93)90259-2} {\bibfield  {journal} {\bibinfo  {journal} {Journal of Nuclear Materials}\ }\textbf {\bibinfo {volume} {207}},\ \bibinfo {pages} {169} (\bibinfo {year} {1993})}\BibitemShut {NoStop}%
\bibitem [{\citenamefont {Berger}\ \emph {et~al.}(2010)\citenamefont {Berger}, \citenamefont {Hubbell}, \citenamefont {Seltzer}, \citenamefont {Chang}, \citenamefont {Coursey}, \citenamefont {Sukumar}, \citenamefont {Zucker},\ and\ \citenamefont {Olsen}}]{Berger_2010_XCOMPhotonCross}%
  \BibitemOpen
  \bibfield  {author} {\bibinfo {author} {\bibfnamefont {M.~J.}\ \bibnamefont {Berger}}, \bibinfo {author} {\bibfnamefont {J.~H.}\ \bibnamefont {Hubbell}}, \bibinfo {author} {\bibfnamefont {S.~M.}\ \bibnamefont {Seltzer}}, \bibinfo {author} {\bibfnamefont {J.}~\bibnamefont {Chang}}, \bibinfo {author} {\bibfnamefont {J.~S.}\ \bibnamefont {Coursey}}, \bibinfo {author} {\bibfnamefont {R.}~\bibnamefont {Sukumar}}, \bibinfo {author} {\bibfnamefont {D.~S.}\ \bibnamefont {Zucker}},\ and\ \bibinfo {author} {\bibfnamefont {K.}~\bibnamefont {Olsen}},\ }\href {https://doi.org/10.18434/T48G6X} {\bibinfo {title} {{XCOM-Photon Cross Sections Database, NIST Standard Reference Database 8 (XGAM)}}} (\bibinfo {year} {2010})\BibitemShut {NoStop}%
\bibitem [{\citenamefont {Lim}\ \emph {et~al.}(2018)\citenamefont {Lim}, \citenamefont {Choi}, \citenamefont {Lee}, \citenamefont {Lee}, \citenamefont {Nahm}, \citenamefont {Kim}, \citenamefont {Kim}, \citenamefont {Park}, \citenamefont {Lee}, \citenamefont {Hong}, \citenamefont {Kwon},\ and\ \citenamefont {Hyeon}}]{Lim_2018_MicroscopicStatesVerwey}%
  \BibitemOpen
  \bibfield  {author} {\bibinfo {author} {\bibfnamefont {S.}~\bibnamefont {Lim}}, \bibinfo {author} {\bibfnamefont {B.}~\bibnamefont {Choi}}, \bibinfo {author} {\bibfnamefont {S.~Y.}\ \bibnamefont {Lee}}, \bibinfo {author} {\bibfnamefont {S.}~\bibnamefont {Lee}}, \bibinfo {author} {\bibfnamefont {H.-H.}\ \bibnamefont {Nahm}}, \bibinfo {author} {\bibfnamefont {Y.-H.}\ \bibnamefont {Kim}}, \bibinfo {author} {\bibfnamefont {T.}~\bibnamefont {Kim}}, \bibinfo {author} {\bibfnamefont {J.-G.}\ \bibnamefont {Park}}, \bibinfo {author} {\bibfnamefont {J.}~\bibnamefont {Lee}}, \bibinfo {author} {\bibfnamefont {J.}~\bibnamefont {Hong}}, \bibinfo {author} {\bibfnamefont {S.~G.}\ \bibnamefont {Kwon}},\ and\ \bibinfo {author} {\bibfnamefont {T.}~\bibnamefont {Hyeon}},\ }\bibfield  {title} {\bibinfo {title} {{Microscopic States and the Verwey Transition of Magnetite Nanocrystals Investigated by Nuclear Magnetic Resonance}},\ }\href {https://doi.org/10.1021/acs.nanolett.7b04866} {\bibfield  {journal} {\bibinfo  {journal}
  {Nano Letters}\ }\textbf {\bibinfo {volume} {18}},\ \bibinfo {pages} {1745} (\bibinfo {year} {2018})}\BibitemShut {NoStop}%
\bibitem [{\citenamefont {Helms}\ and\ \citenamefont {Klemm}(1939)}]{Helms_1939_UeberdieKristallstrukturen}%
  \BibitemOpen
  \bibfield  {author} {\bibinfo {author} {\bibfnamefont {A.}~\bibnamefont {Helms}}\ and\ \bibinfo {author} {\bibfnamefont {W.}~\bibnamefont {Klemm}},\ }\bibfield  {title} {\bibinfo {title} {Über die kristallstrukturen der rubidium‐ und cäsiumsesquioxyde},\ }\href {https://doi.org/10.1002/zaac.19392420210} {\bibfield  {journal} {\bibinfo  {journal} {Zeitschrift für anorganische und allgemeine Chemie}\ }\textbf {\bibinfo {volume} {242}},\ \bibinfo {pages} {201} (\bibinfo {year} {1939})}\BibitemShut {NoStop}%
\bibitem [{\citenamefont {Jansen}\ and\ \citenamefont {Korber}(1991)}]{Jansen_1991_NeueUntersuchungenzu}%
  \BibitemOpen
  \bibfield  {author} {\bibinfo {author} {\bibfnamefont {M.}~\bibnamefont {Jansen}}\ and\ \bibinfo {author} {\bibfnamefont {N.}~\bibnamefont {Korber}},\ }\bibfield  {title} {\bibinfo {title} {{Neue Untersuchungen zu Präparation und Struktur von Rb$_4$O$_6$}},\ }\href {https://doi.org/10.1002/zaac.19915980116} {\bibfield  {journal} {\bibinfo  {journal} {Zeitschrift für anorganische und allgemeine Chemie}\ }\textbf {\bibinfo {volume} {598}},\ \bibinfo {pages} {163} (\bibinfo {year} {1991})}\BibitemShut {NoStop}%
\bibitem [{\citenamefont {Jansen}\ \emph {et~al.}(1999)\citenamefont {Jansen}, \citenamefont {Hagenmayer},\ and\ \citenamefont {Korber}}]{Jansen_1999_Rb4O6studiedelastic}%
  \BibitemOpen
  \bibfield  {author} {\bibinfo {author} {\bibfnamefont {M.}~\bibnamefont {Jansen}}, \bibinfo {author} {\bibfnamefont {R.}~\bibnamefont {Hagenmayer}},\ and\ \bibinfo {author} {\bibfnamefont {N.}~\bibnamefont {Korber}},\ }\bibfield  {title} {\bibinfo {title} {{Rb$_4$O$_6$ studied by elastic and inelastic neutron scattering}},\ }\href {https://doi.org/10.1016/s1387-1609(00)88570-9} {\bibfield  {journal} {\bibinfo  {journal} {Comptes Rendus de l’Académie des Sciences - Series IIC - Chemistry}\ }\textbf {\bibinfo {volume} {2}},\ \bibinfo {pages} {591} (\bibinfo {year} {1999})}\BibitemShut {NoStop}%
\bibitem [{\citenamefont {Arčon}\ \emph {et~al.}(2013)\citenamefont {Arčon}, \citenamefont {Anderle}, \citenamefont {Klanjšek}, \citenamefont {Sans}, \citenamefont {Mühle}, \citenamefont {Adler}, \citenamefont {Schnelle}, \citenamefont {Jansen},\ and\ \citenamefont {Felser}}]{Arcon_2013_InfluenceO2molecularorientation}%
  \BibitemOpen
  \bibfield  {author} {\bibinfo {author} {\bibfnamefont {D.}~\bibnamefont {Arčon}}, \bibinfo {author} {\bibfnamefont {K.}~\bibnamefont {Anderle}}, \bibinfo {author} {\bibfnamefont {M.}~\bibnamefont {Klanjšek}}, \bibinfo {author} {\bibfnamefont {A.}~\bibnamefont {Sans}}, \bibinfo {author} {\bibfnamefont {C.}~\bibnamefont {Mühle}}, \bibinfo {author} {\bibfnamefont {P.}~\bibnamefont {Adler}}, \bibinfo {author} {\bibfnamefont {W.}~\bibnamefont {Schnelle}}, \bibinfo {author} {\bibfnamefont {M.}~\bibnamefont {Jansen}},\ and\ \bibinfo {author} {\bibfnamefont {C.}~\bibnamefont {Felser}},\ }\bibfield  {title} {\bibinfo {title} {Influence of {O$_2$} molecular orientation on {$p$}-orbital ordering and exchange pathways in {Cs$_4$O$_6$}},\ }\href {https://doi.org/10.1103/physrevb.88.224409} {\bibfield  {journal} {\bibinfo  {journal} {Physical Review B}\ }\textbf {\bibinfo {volume} {88}},\ \bibinfo {pages} {224409} (\bibinfo {year} {2013})}\BibitemShut {NoStop}%
\bibitem [{\citenamefont {Sans}\ \emph {et~al.}(2014)\citenamefont {Sans}, \citenamefont {Nuss}, \citenamefont {Fecher}, \citenamefont {Mühle}, \citenamefont {Felser},\ and\ \citenamefont {Jansen}}]{Sans_2014_StructuralImplicationsSpin}%
  \BibitemOpen
  \bibfield  {author} {\bibinfo {author} {\bibfnamefont {A.}~\bibnamefont {Sans}}, \bibinfo {author} {\bibfnamefont {J.}~\bibnamefont {Nuss}}, \bibinfo {author} {\bibfnamefont {G.~H.}\ \bibnamefont {Fecher}}, \bibinfo {author} {\bibfnamefont {C.}~\bibnamefont {Mühle}}, \bibinfo {author} {\bibfnamefont {C.}~\bibnamefont {Felser}},\ and\ \bibinfo {author} {\bibfnamefont {M.}~\bibnamefont {Jansen}},\ }\bibfield  {title} {\bibinfo {title} {{Structural Implications of Spin, Charge, and Orbital Ordering in Rubidium Sesquioxide, Rb$_4$O$_6$}},\ }\href {https://doi.org/10.1002/zaac.201400125} {\bibfield  {journal} {\bibinfo  {journal} {Zeitschrift für anorganische und allgemeine Chemie}\ }\textbf {\bibinfo {volume} {640}},\ \bibinfo {pages} {1239} (\bibinfo {year} {2014})}\BibitemShut {NoStop}%
\bibitem [{\citenamefont {Colman}\ \emph {et~al.}(2019)\citenamefont {Colman}, \citenamefont {Okur}, \citenamefont {Kockelmann}, \citenamefont {Brown}, \citenamefont {Sans}, \citenamefont {Felser}, \citenamefont {Jansen},\ and\ \citenamefont {Prassides}}]{Colman_2019_ElusiveValenceTransition}%
  \BibitemOpen
  \bibfield  {author} {\bibinfo {author} {\bibfnamefont {R.~H.}\ \bibnamefont {Colman}}, \bibinfo {author} {\bibfnamefont {H.~E.}\ \bibnamefont {Okur}}, \bibinfo {author} {\bibfnamefont {W.}~\bibnamefont {Kockelmann}}, \bibinfo {author} {\bibfnamefont {C.~M.}\ \bibnamefont {Brown}}, \bibinfo {author} {\bibfnamefont {A.}~\bibnamefont {Sans}}, \bibinfo {author} {\bibfnamefont {C.}~\bibnamefont {Felser}}, \bibinfo {author} {\bibfnamefont {M.}~\bibnamefont {Jansen}},\ and\ \bibinfo {author} {\bibfnamefont {K.}~\bibnamefont {Prassides}},\ }\bibfield  {title} {\bibinfo {title} {{Elusive Valence Transition in Mixed-Valence Sesquioxide Cs$_4$O$_6$}},\ }\href {https://doi.org/10.1021/acs.inorgchem.9b02122} {\bibfield  {journal} {\bibinfo  {journal} {Inorganic Chemistry}\ }\textbf {\bibinfo {volume} {58}},\ \bibinfo {pages} {14532} (\bibinfo {year} {2019})}\BibitemShut {NoStop}%
\bibitem [{\citenamefont {Eloirdi}\ \emph {et~al.}(2013)\citenamefont {Eloirdi}, \citenamefont {Fuchs}, \citenamefont {Griveau}, \citenamefont {Colineau}, \citenamefont {Shick}, \citenamefont {Manara},\ and\ \citenamefont {Caciuffo}}]{Eloirdi_2013_Evidencepersistentspin}%
  \BibitemOpen
  \bibfield  {author} {\bibinfo {author} {\bibfnamefont {R.}~\bibnamefont {Eloirdi}}, \bibinfo {author} {\bibfnamefont {A.~J.}\ \bibnamefont {Fuchs}}, \bibinfo {author} {\bibfnamefont {J.-C.}\ \bibnamefont {Griveau}}, \bibinfo {author} {\bibfnamefont {E.}~\bibnamefont {Colineau}}, \bibinfo {author} {\bibfnamefont {A.~B.}\ \bibnamefont {Shick}}, \bibinfo {author} {\bibfnamefont {D.}~\bibnamefont {Manara}},\ and\ \bibinfo {author} {\bibfnamefont {R.}~\bibnamefont {Caciuffo}},\ }\bibfield  {title} {\bibinfo {title} {{Evidence for persistent spin fluctuations in uranium sesquicarbide}},\ }\href {https://doi.org/10.1103/physrevb.87.214414} {\bibfield  {journal} {\bibinfo  {journal} {Physical Review B}\ }\textbf {\bibinfo {volume} {87}},\ \bibinfo {pages} {214414} (\bibinfo {year} {2013})}\BibitemShut {NoStop}%
\bibitem [{\citenamefont {Baek}\ \emph {et~al.}(2010)\citenamefont {Baek}, \citenamefont {Sakai}, \citenamefont {Bauer}, \citenamefont {Mitchell}, \citenamefont {Kennison}, \citenamefont {Ronning},\ and\ \citenamefont {Thompson}}]{Baek_2010_AnisotropicSpinFluctuations}%
  \BibitemOpen
  \bibfield  {author} {\bibinfo {author} {\bibfnamefont {S.-H.}\ \bibnamefont {Baek}}, \bibinfo {author} {\bibfnamefont {H.}~\bibnamefont {Sakai}}, \bibinfo {author} {\bibfnamefont {E.~D.}\ \bibnamefont {Bauer}}, \bibinfo {author} {\bibfnamefont {J.~N.}\ \bibnamefont {Mitchell}}, \bibinfo {author} {\bibfnamefont {J.~A.}\ \bibnamefont {Kennison}}, \bibinfo {author} {\bibfnamefont {F.}~\bibnamefont {Ronning}},\ and\ \bibinfo {author} {\bibfnamefont {J.~D.}\ \bibnamefont {Thompson}},\ }\bibfield  {title} {\bibinfo {title} {{Anisotropic Spin Fluctuations and Superconductivity in ``115'' Heavy Fermion Compounds: ${}^{59}$Co NMR Study in PuCoGa$_5$}},\ }\href {https://doi.org/10.1103/physrevlett.105.217002} {\bibfield  {journal} {\bibinfo  {journal} {Physical Review Letters}\ }\textbf {\bibinfo {volume} {105}},\ \bibinfo {pages} {217002} (\bibinfo {year} {2010})}\BibitemShut {NoStop}%
\bibitem [{\citenamefont {Méot-Reymond}\ and\ \citenamefont {Fournier}(1996)}]{MeotReymond_1996_Localization5felectrons}%
  \BibitemOpen
  \bibfield  {author} {\bibinfo {author} {\bibfnamefont {S.}~\bibnamefont {Méot-Reymond}}\ and\ \bibinfo {author} {\bibfnamefont {J.}~\bibnamefont {Fournier}},\ }\bibfield  {title} {\bibinfo {title} {{Localization of 5$f$ electrons in $\delta$-plutonium: Evidence for the Kondo effect}},\ }\href {https://doi.org/10.1016/0925-8388(95)01937-5} {\bibfield  {journal} {\bibinfo  {journal} {Journal of Alloys and Compounds}\ }\textbf {\bibinfo {volume} {232}},\ \bibinfo {pages} {119} (\bibinfo {year} {1996})}\BibitemShut {NoStop}%
\bibitem [{\citenamefont {Kaczorowski}\ \emph {et~al.}(1993)\citenamefont {Kaczorowski}, \citenamefont {Troć}, \citenamefont {Badurski}, \citenamefont {Böhm}, \citenamefont {Shlyk},\ and\ \citenamefont {Steglich}}]{Kaczorowski_1993_Magneticnonmagnetictransition}%
  \BibitemOpen
  \bibfield  {author} {\bibinfo {author} {\bibfnamefont {D.}~\bibnamefont {Kaczorowski}}, \bibinfo {author} {\bibfnamefont {R.}~\bibnamefont {Troć}}, \bibinfo {author} {\bibfnamefont {D.}~\bibnamefont {Badurski}}, \bibinfo {author} {\bibfnamefont {A.}~\bibnamefont {Böhm}}, \bibinfo {author} {\bibfnamefont {L.}~\bibnamefont {Shlyk}},\ and\ \bibinfo {author} {\bibfnamefont {F.}~\bibnamefont {Steglich}},\ }\bibfield  {title} {\bibinfo {title} {{Magnetic-to-nonmagnetic transition in the pseudobinary system U(Ga$_{1-x}$Sn$_x$)$_3$}},\ }\href {https://doi.org/10.1103/physrevb.48.16425} {\bibfield  {journal} {\bibinfo  {journal} {Physical Review B}\ }\textbf {\bibinfo {volume} {48}},\ \bibinfo {pages} {16425} (\bibinfo {year} {1993})}\BibitemShut {NoStop}%
\bibitem [{\citenamefont {Lee}\ \emph {et~al.}(1986)\citenamefont {Lee}, \citenamefont {Rice}, \citenamefont {Serene}, \citenamefont {Sham},\ and\ \citenamefont {Wilkins}}]{Lee_1986_Theoriesheavyelectron}%
  \BibitemOpen
  \bibfield  {author} {\bibinfo {author} {\bibfnamefont {P.~A.}\ \bibnamefont {Lee}}, \bibinfo {author} {\bibfnamefont {T.~M.}\ \bibnamefont {Rice}}, \bibinfo {author} {\bibfnamefont {J.~W.}\ \bibnamefont {Serene}}, \bibinfo {author} {\bibfnamefont {L.~J.}\ \bibnamefont {Sham}},\ and\ \bibinfo {author} {\bibfnamefont {J.~W.}\ \bibnamefont {Wilkins}},\ }\bibfield  {title} {\bibinfo {title} {{Theories of Heavy-Electron Systems}},\ }\href@noop {} {\bibfield  {journal} {\bibinfo  {journal} {Comments on Condensed Matter Physics}\ }\textbf {\bibinfo {volume} {12}},\ \bibinfo {pages} {99} (\bibinfo {year} {1986})}\BibitemShut {NoStop}%
\bibitem [{\citenamefont {Sheng}\ and\ \citenamefont {Cooper}(1995)}]{Sheng_1995_NonKondoprediction}%
  \BibitemOpen
  \bibfield  {author} {\bibinfo {author} {\bibfnamefont {Q.~G.}\ \bibnamefont {Sheng}}\ and\ \bibinfo {author} {\bibfnamefont {B.~R.}\ \bibnamefont {Cooper}},\ }\bibfield  {title} {\bibinfo {title} {{Non-Kondo prediction of Wilson ratio for heavy fermion systems}},\ }\href {https://doi.org/10.1080/09500839508241624} {\bibfield  {journal} {\bibinfo  {journal} {Philosophical Magazine Letters}\ }\textbf {\bibinfo {volume} {72}},\ \bibinfo {pages} {123} (\bibinfo {year} {1995})}\BibitemShut {NoStop}%
\bibitem [{\citenamefont {Potočnik}\ \emph {et~al.}(2014)\citenamefont {Potočnik}, \citenamefont {Jeglič}, \citenamefont {Kobayashi}, \citenamefont {Kawashima}, \citenamefont {Kuchida}, \citenamefont {Akimitsu},\ and\ \citenamefont {Arčon}}]{Potocnik_2014_Anomalouslocalspin}%
  \BibitemOpen
  \bibfield  {author} {\bibinfo {author} {\bibfnamefont {A.}~\bibnamefont {Potočnik}}, \bibinfo {author} {\bibfnamefont {P.}~\bibnamefont {Jeglič}}, \bibinfo {author} {\bibfnamefont {K.}~\bibnamefont {Kobayashi}}, \bibinfo {author} {\bibfnamefont {K.}~\bibnamefont {Kawashima}}, \bibinfo {author} {\bibfnamefont {S.}~\bibnamefont {Kuchida}}, \bibinfo {author} {\bibfnamefont {J.}~\bibnamefont {Akimitsu}},\ and\ \bibinfo {author} {\bibfnamefont {D.}~\bibnamefont {Arčon}},\ }\bibfield  {title} {\bibinfo {title} {{Anomalous local spin susceptibilities in noncentrosymmetric La$_2$C$_3$ superconductor}},\ }\href {https://doi.org/10.1103/physrevb.90.104507} {\bibfield  {journal} {\bibinfo  {journal} {Physical Review B}\ }\textbf {\bibinfo {volume} {90}},\ \bibinfo {pages} {104507} (\bibinfo {year} {2014})}\BibitemShut {NoStop}%
\bibitem [{\citenamefont {Harada}\ \emph {et~al.}(2007)\citenamefont {Harada}, \citenamefont {Akutagawa}, \citenamefont {Miyamichi}, \citenamefont {Mukuda}, \citenamefont {Kitaoka},\ and\ \citenamefont {Akimitsu}}]{Harada_2007_MultigapSuperconductivityY_2C_3}%
  \BibitemOpen
  \bibfield  {author} {\bibinfo {author} {\bibfnamefont {A.}~\bibnamefont {Harada}}, \bibinfo {author} {\bibfnamefont {S.}~\bibnamefont {Akutagawa}}, \bibinfo {author} {\bibfnamefont {Y.}~\bibnamefont {Miyamichi}}, \bibinfo {author} {\bibfnamefont {H.}~\bibnamefont {Mukuda}}, \bibinfo {author} {\bibfnamefont {Y.}~\bibnamefont {Kitaoka}},\ and\ \bibinfo {author} {\bibfnamefont {J.}~\bibnamefont {Akimitsu}},\ }\bibfield  {title} {\bibinfo {title} {{Multigap Superconductivity in Y$_2$C$_3$: A $^{13}$C-NMR Study}},\ }\href {https://doi.org/10.1143/jpsj.76.023704} {\bibfield  {journal} {\bibinfo  {journal} {Journal of the Physical Society of Japan}\ }\textbf {\bibinfo {volume} {76}},\ \bibinfo {pages} {023704} (\bibinfo {year} {2007})}\BibitemShut {NoStop}%
\bibitem [{\citenamefont {Boutard}\ and\ \citenamefont {de~Novion}(1974)}]{Boutard_1974_EtudeparRMN}%
  \BibitemOpen
  \bibfield  {author} {\bibinfo {author} {\bibfnamefont {J.-L.}\ \bibnamefont {Boutard}}\ and\ \bibinfo {author} {\bibfnamefont {C.-H.}\ \bibnamefont {de~Novion}},\ }\bibfield  {title} {\bibinfo {title} {{Etude par RMN et rayons X de sesquicarbure d'uranium}},\ }\href {https://doi.org/10.1016/0038-1098(74)90212-9} {\bibfield  {journal} {\bibinfo  {journal} {Solid State Communications}\ }\textbf {\bibinfo {volume} {14}},\ \bibinfo {pages} {181} (\bibinfo {year} {1974})}\BibitemShut {NoStop}%
\bibitem [{\citenamefont {Slichter}(1990)}]{Slichter_1990_PrinciplesMagneticResonance}%
  \BibitemOpen
  \bibfield  {author} {\bibinfo {author} {\bibfnamefont {C.~P.}\ \bibnamefont {Slichter}},\ }\href {https://doi.org/10.1007/978-3-662-09441-9} {\emph {\bibinfo {title} {{Principles of Magnetic Resonance}}}}\ (\bibinfo  {publisher} {Springer Berlin Heidelberg},\ \bibinfo {year} {1990})\BibitemShut {NoStop}%
\bibitem [{\citenamefont {Tossell}(1992)}]{Tossell_1992_NuclearMagneticShieldings}%
  \BibitemOpen
  \bibfield  {author} {\bibinfo {author} {\bibfnamefont {J.~A.}\ \bibnamefont {Tossell}},\ }\href {https://doi.org/10.1007/978-94-011-1652-7} {\emph {\bibinfo {title} {{Nuclear Magnetic Shieldings and Molecular Structure}}}}\ (\bibinfo  {publisher} {Springer Science+Business Media Dordrecht},\ \bibinfo {year} {1992})\BibitemShut {NoStop}%
\bibitem [{\citenamefont {Abragam}(1961)}]{Abragam_1961_PrinciplesNuclearMagnetism}%
  \BibitemOpen
  \bibfield  {author} {\bibinfo {author} {\bibfnamefont {A.}~\bibnamefont {Abragam}},\ }\href {https://global.oup.com/academic/product/principles-of-nuclear-magnetism-9780198520146?cc=us&lang=en&} {\emph {\bibinfo {title} {{Principles of Nuclear Magnetism}}}}\ (\bibinfo  {publisher} {Oxford University Press},\ \bibinfo {address} {New York},\ \bibinfo {year} {1961})\BibitemShut {NoStop}%
\bibitem [{\citenamefont {Yang}\ \emph {et~al.}(2017)\citenamefont {Yang}, \citenamefont {Tang}, \citenamefont {Gao},\ and\ \citenamefont {Ao}}]{Yang_2017_Densityfunctionalstudy}%
  \BibitemOpen
  \bibfield  {author} {\bibinfo {author} {\bibfnamefont {R.}~\bibnamefont {Yang}}, \bibinfo {author} {\bibfnamefont {B.}~\bibnamefont {Tang}}, \bibinfo {author} {\bibfnamefont {T.}~\bibnamefont {Gao}},\ and\ \bibinfo {author} {\bibfnamefont {B.~Y.}\ \bibnamefont {Ao}},\ }\bibfield  {title} {\bibinfo {title} {{Density functional study of Pu$_2$C$_3$}},\ }\href {https://doi.org/10.1140/epjb/e2017-70749-8} {\bibfield  {journal} {\bibinfo  {journal} {The European Physical Journal B}\ }\textbf {\bibinfo {volume} {90}},\ \bibinfo {pages} {145} (\bibinfo {year} {2017})}\BibitemShut {NoStop}%
\bibitem [{\citenamefont {Sakai}\ \emph {et~al.}(2006)\citenamefont {Sakai}, \citenamefont {Tokunaga}, \citenamefont {Fujimoto}, \citenamefont {Kambe}, \citenamefont {E.~Walstedt}, \citenamefont {Yasuoka}, \citenamefont {Aoki}, \citenamefont {Homma}, \citenamefont {Yamamoto}, \citenamefont {Nakamura}, \citenamefont {Shiokawa}, \citenamefont {Nakajima}, \citenamefont {Arai}, \citenamefont {D.~Matsuda}, \citenamefont {Haga},\ and\ \citenamefont {Ōnuki}}]{Sakai_2006_NMRShiftMeasurements}%
  \BibitemOpen
  \bibfield  {author} {\bibinfo {author} {\bibfnamefont {H.}~\bibnamefont {Sakai}}, \bibinfo {author} {\bibfnamefont {Y.}~\bibnamefont {Tokunaga}}, \bibinfo {author} {\bibfnamefont {T.}~\bibnamefont {Fujimoto}}, \bibinfo {author} {\bibfnamefont {S.}~\bibnamefont {Kambe}}, \bibinfo {author} {\bibfnamefont {R.}~\bibnamefont {E.~Walstedt}}, \bibinfo {author} {\bibfnamefont {H.}~\bibnamefont {Yasuoka}}, \bibinfo {author} {\bibfnamefont {D.}~\bibnamefont {Aoki}}, \bibinfo {author} {\bibfnamefont {Y.}~\bibnamefont {Homma}}, \bibinfo {author} {\bibfnamefont {E.}~\bibnamefont {Yamamoto}}, \bibinfo {author} {\bibfnamefont {A.}~\bibnamefont {Nakamura}}, \bibinfo {author} {\bibfnamefont {Y.}~\bibnamefont {Shiokawa}}, \bibinfo {author} {\bibfnamefont {K.}~\bibnamefont {Nakajima}}, \bibinfo {author} {\bibfnamefont {Y.}~\bibnamefont {Arai}}, \bibinfo {author} {\bibfnamefont {T.}~\bibnamefont {D.~Matsuda}}, \bibinfo {author} {\bibfnamefont {Y.}~\bibnamefont {Haga}},\ and\ \bibinfo {author} {\bibfnamefont {Y.}~\bibnamefont
  {Ōnuki}},\ }\bibfield  {title} {\bibinfo {title} {{NMR Shift Measurements of ${}^{69}$Ga in Unconventional Superconductor PuRhGa$_5$}},\ }\href {https://doi.org/10.1143/jpsjs.75s.50} {\bibfield  {journal} {\bibinfo  {journal} {Journal of the Physical Society of Japan}\ }\textbf {\bibinfo {volume} {75}},\ \bibinfo {pages} {50} (\bibinfo {year} {2006})}\BibitemShut {NoStop}%
\bibitem [{\citenamefont {Piskunov}\ \emph {et~al.}(2005)\citenamefont {Piskunov}, \citenamefont {Mikhalev}, \citenamefont {Gerashenko}, \citenamefont {Pogudin}, \citenamefont {Ogloblichev}, \citenamefont {Verkhovskii}, \citenamefont {Tankeyev}, \citenamefont {Arkhipov}, \citenamefont {Zouev},\ and\ \citenamefont {Lekomtsev}}]{Piskunov_2005_SpinsusceptibilityGa}%
  \BibitemOpen
  \bibfield  {author} {\bibinfo {author} {\bibfnamefont {Y.}~\bibnamefont {Piskunov}}, \bibinfo {author} {\bibfnamefont {K.}~\bibnamefont {Mikhalev}}, \bibinfo {author} {\bibfnamefont {A.}~\bibnamefont {Gerashenko}}, \bibinfo {author} {\bibfnamefont {A.}~\bibnamefont {Pogudin}}, \bibinfo {author} {\bibfnamefont {V.}~\bibnamefont {Ogloblichev}}, \bibinfo {author} {\bibfnamefont {S.}~\bibnamefont {Verkhovskii}}, \bibinfo {author} {\bibfnamefont {A.}~\bibnamefont {Tankeyev}}, \bibinfo {author} {\bibfnamefont {V.}~\bibnamefont {Arkhipov}}, \bibinfo {author} {\bibfnamefont {Y.}~\bibnamefont {Zouev}},\ and\ \bibinfo {author} {\bibfnamefont {S.}~\bibnamefont {Lekomtsev}},\ }\bibfield  {title} {\bibinfo {title} {{Spin susceptibility of Ga-stabilized $\delta$-Pu probed by ${}^{69}$Ga NMR}},\ }\href {https://doi.org/10.1103/physrevb.71.174410} {\bibfield  {journal} {\bibinfo  {journal} {Physical Review B}\ }\textbf {\bibinfo {volume} {71}},\ \bibinfo {pages} {174410} (\bibinfo {year} {2005})}\BibitemShut {NoStop}%
\bibitem [{\citenamefont {Curro}\ \emph {et~al.}(2004)\citenamefont {Curro}, \citenamefont {Young}, \citenamefont {Schmalian},\ and\ \citenamefont {Pines}}]{Curro_2004_Scalingemergentbehavior}%
  \BibitemOpen
  \bibfield  {author} {\bibinfo {author} {\bibfnamefont {N.~J.}\ \bibnamefont {Curro}}, \bibinfo {author} {\bibfnamefont {B.-L.}\ \bibnamefont {Young}}, \bibinfo {author} {\bibfnamefont {J.}~\bibnamefont {Schmalian}},\ and\ \bibinfo {author} {\bibfnamefont {D.}~\bibnamefont {Pines}},\ }\bibfield  {title} {\bibinfo {title} {Scaling in the emergent behavior of heavy-electron materials},\ }\href {https://doi.org/10.1103/physrevb.70.235117} {\bibfield  {journal} {\bibinfo  {journal} {Physical Review B}\ }\textbf {\bibinfo {volume} {70}},\ \bibinfo {pages} {235117} (\bibinfo {year} {2004})}\BibitemShut {NoStop}%
\bibitem [{\citenamefont {Lysak}\ and\ \citenamefont {MacLaughlin}(1985)}]{Lysak_1985_Nuclearmagneticresonance}%
  \BibitemOpen
  \bibfield  {author} {\bibinfo {author} {\bibfnamefont {M.~J.}\ \bibnamefont {Lysak}}\ and\ \bibinfo {author} {\bibfnamefont {D.~E.}\ \bibnamefont {MacLaughlin}},\ }\bibfield  {title} {\bibinfo {title} {{Nuclear magnetic resonance and unstable rare-earth magnetism in CeAl$_3$}},\ }\href {https://doi.org/10.1103/physrevb.31.6963} {\bibfield  {journal} {\bibinfo  {journal} {Physical Review B}\ }\textbf {\bibinfo {volume} {31}},\ \bibinfo {pages} {6963} (\bibinfo {year} {1985})}\BibitemShut {NoStop}%
\bibitem [{\citenamefont {Mahajan}\ \emph {et~al.}(1998)\citenamefont {Mahajan}, \citenamefont {Sala}, \citenamefont {Lee}, \citenamefont {Borsa}, \citenamefont {Kondo},\ and\ \citenamefont {Johnston}}]{Mahajan_1998_7Li51VNMR}%
  \BibitemOpen
  \bibfield  {author} {\bibinfo {author} {\bibfnamefont {A.~V.}\ \bibnamefont {Mahajan}}, \bibinfo {author} {\bibfnamefont {R.}~\bibnamefont {Sala}}, \bibinfo {author} {\bibfnamefont {E.}~\bibnamefont {Lee}}, \bibinfo {author} {\bibfnamefont {F.}~\bibnamefont {Borsa}}, \bibinfo {author} {\bibfnamefont {S.}~\bibnamefont {Kondo}},\ and\ \bibinfo {author} {\bibfnamefont {D.~C.}\ \bibnamefont {Johnston}},\ }\bibfield  {title} {\bibinfo {title} {{${}^7$Li and ${}^{51}$V NMR study of the heavy-fermion compound LiV$_2$O$_4$}},\ }\href {https://doi.org/10.1103/physrevb.57.8890} {\bibfield  {journal} {\bibinfo  {journal} {Physical Review B}\ }\textbf {\bibinfo {volume} {57}},\ \bibinfo {pages} {8890} (\bibinfo {year} {1998})}\BibitemShut {NoStop}%
\end{thebibliography}%


\begin{thebibliography}{20}%
\makeatletter
\providecommand \@ifxundefined [1]{%
 \@ifx{#1\undefined}
}%
\providecommand \@ifnum [1]{%
 \ifnum #1\expandafter \@firstoftwo
 \else \expandafter \@secondoftwo
 \fi
}%
\providecommand \@ifx [1]{%
 \ifx #1\expandafter \@firstoftwo
 \else \expandafter \@secondoftwo
 \fi
}%
\providecommand \natexlab [1]{#1}%
\providecommand \enquote  [1]{``#1''}%
\providecommand \bibnamefont  [1]{#1}%
\providecommand \bibfnamefont [1]{#1}%
\providecommand \citenamefont [1]{#1}%
\providecommand \href@noop [0]{\@secondoftwo}%
\providecommand \href [0]{\begingroup \@sanitize@url \@href}%
\providecommand \@href[1]{\@@startlink{#1}\@@href}%
\providecommand \@@href[1]{\endgroup#1\@@endlink}%
\providecommand \@sanitize@url [0]{\catcode `\\12\catcode `\$12\catcode `\&12\catcode `\#12\catcode `\^12\catcode `\_12\catcode `\%12\relax}%
\providecommand \@@startlink[1]{}%
\providecommand \@@endlink[0]{}%
\providecommand \url  [0]{\begingroup\@sanitize@url \@url }%
\providecommand \@url [1]{\endgroup\@href {#1}{\urlprefix }}%
\providecommand \urlprefix  [0]{URL }%
\providecommand \Eprint [0]{\href }%
\providecommand \doibase [0]{https://doi.org/}%
\providecommand \selectlanguage [0]{\@gobble}%
\providecommand \bibinfo  [0]{\@secondoftwo}%
\providecommand \bibfield  [0]{\@secondoftwo}%
\providecommand \translation [1]{[#1]}%
\providecommand \BibitemOpen [0]{}%
\providecommand \bibitemStop [0]{}%
\providecommand \bibitemNoStop [0]{.\EOS\space}%
\providecommand \EOS [0]{\spacefactor3000\relax}%
\providecommand \BibitemShut  [1]{\csname bibitem#1\endcsname}%
\let\auto@bib@innerbib\@empty
\bibitem [{\citenamefont {Gupta}\ \emph {et~al.}(2021)\citenamefont {Gupta}, \citenamefont {Stait-Gardner},\ and\ \citenamefont {Price}}]{Gupta_2021_IsItTime}%
  \BibitemOpen
  \bibfield  {author} {\bibinfo {author} {\bibfnamefont {A.}~\bibnamefont {Gupta}}, \bibinfo {author} {\bibfnamefont {T.}~\bibnamefont {Stait-Gardner}},\ and\ \bibinfo {author} {\bibfnamefont {W.~S.}\ \bibnamefont {Price}},\ }\bibfield  {title} {\bibinfo {title} {{Is It Time to Forgo the Use of the Terms “Spin–Lattice” and “Spin–Spin” Relaxation in NMR and MRI?}},\ }\href {https://doi.org/10.1021/acs.jpclett.1c00945} {\bibfield  {journal} {\bibinfo  {journal} {The Journal of Physical Chemistry Letters}\ }\textbf {\bibinfo {volume} {12}},\ \bibinfo {pages} {6305} (\bibinfo {year} {2021})}\BibitemShut {NoStop}%
\bibitem [{\citenamefont {Dementyev}\ \emph {et~al.}(2003)\citenamefont {Dementyev}, \citenamefont {Li}, \citenamefont {MacLean},\ and\ \citenamefont {Barrett}}]{Dementyev_2003_AnomaliesNMRsilicon}%
  \BibitemOpen
  \bibfield  {author} {\bibinfo {author} {\bibfnamefont {A.~E.}\ \bibnamefont {Dementyev}}, \bibinfo {author} {\bibfnamefont {D.}~\bibnamefont {Li}}, \bibinfo {author} {\bibfnamefont {K.}~\bibnamefont {MacLean}},\ and\ \bibinfo {author} {\bibfnamefont {S.~E.}\ \bibnamefont {Barrett}},\ }\bibfield  {title} {\bibinfo {title} {{Anomalies in the NMR of silicon: Unexpected spin echoes in a dilute dipolar solid}},\ }\href {https://doi.org/10.1103/physrevb.68.153302} {\bibfield  {journal} {\bibinfo  {journal} {Physical Review B}\ }\textbf {\bibinfo {volume} {68}},\ \bibinfo {pages} {153302} (\bibinfo {year} {2003})}\BibitemShut {NoStop}%
\bibitem [{\citenamefont {Franzoni}\ and\ \citenamefont {Levstein}(2005)}]{Franzoni_2005_Manifestationsabsencespin}%
  \BibitemOpen
  \bibfield  {author} {\bibinfo {author} {\bibfnamefont {M.~B.}\ \bibnamefont {Franzoni}}\ and\ \bibinfo {author} {\bibfnamefont {P.~R.}\ \bibnamefont {Levstein}},\ }\bibfield  {title} {\bibinfo {title} {{Manifestations of the absence of spin diffusion in multipulse NMR experiments on diluted dipolar solids}},\ }\href {https://doi.org/10.1103/physrevb.72.235410} {\bibfield  {journal} {\bibinfo  {journal} {Physical Review B}\ }\textbf {\bibinfo {volume} {72}},\ \bibinfo {pages} {235410} (\bibinfo {year} {2005})}\BibitemShut {NoStop}%
\bibitem [{\citenamefont {Li}\ \emph {et~al.}(2007)\citenamefont {Li}, \citenamefont {Dementyev}, \citenamefont {Dong}, \citenamefont {Ramos},\ and\ \citenamefont {Barrett}}]{Li_2007_GeneratingUnexpectedSpin}%
  \BibitemOpen
  \bibfield  {author} {\bibinfo {author} {\bibfnamefont {D.}~\bibnamefont {Li}}, \bibinfo {author} {\bibfnamefont {A.~E.}\ \bibnamefont {Dementyev}}, \bibinfo {author} {\bibfnamefont {Y.}~\bibnamefont {Dong}}, \bibinfo {author} {\bibfnamefont {R.~G.}\ \bibnamefont {Ramos}},\ and\ \bibinfo {author} {\bibfnamefont {S.~E.}\ \bibnamefont {Barrett}},\ }\bibfield  {title} {\bibinfo {title} {Generating unexpected spin echoes in dipolar solids with $\ensuremath{\pi}$ pulses},\ }\href {https://doi.org/10.1103/physrevlett.98.190401} {\bibfield  {journal} {\bibinfo  {journal} {Physical Review Letters}\ }\textbf {\bibinfo {volume} {98}},\ \bibinfo {pages} {190401} (\bibinfo {year} {2007})}\BibitemShut {NoStop}%
\bibitem [{\citenamefont {Li}\ \emph {et~al.}(2008)\citenamefont {Li}, \citenamefont {Dong}, \citenamefont {Ramos}, \citenamefont {Murray}, \citenamefont {MacLean}, \citenamefont {Dementyev},\ and\ \citenamefont {Barrett}}]{Li_2008_Intrinsicoriginspin}%
  \BibitemOpen
  \bibfield  {author} {\bibinfo {author} {\bibfnamefont {D.}~\bibnamefont {Li}}, \bibinfo {author} {\bibfnamefont {Y.}~\bibnamefont {Dong}}, \bibinfo {author} {\bibfnamefont {R.~G.}\ \bibnamefont {Ramos}}, \bibinfo {author} {\bibfnamefont {J.~D.}\ \bibnamefont {Murray}}, \bibinfo {author} {\bibfnamefont {K.}~\bibnamefont {MacLean}}, \bibinfo {author} {\bibfnamefont {A.~E.}\ \bibnamefont {Dementyev}},\ and\ \bibinfo {author} {\bibfnamefont {S.~E.}\ \bibnamefont {Barrett}},\ }\bibfield  {title} {\bibinfo {title} {Intrinsic origin of spin echoes in dipolar solids generated by strong $\ensuremath{\pi}$ pulses},\ }\href {https://doi.org/10.1103/physrevb.77.214306} {\bibfield  {journal} {\bibinfo  {journal} {Physical Review B}\ }\textbf {\bibinfo {volume} {77}},\ \bibinfo {pages} {214306} (\bibinfo {year} {2008})}\BibitemShut {NoStop}%
\bibitem [{\citenamefont {Siegel}\ \emph {et~al.}(2004)\citenamefont {Siegel}, \citenamefont {Nakashima},\ and\ \citenamefont {Wasylishen}}]{Siegel_2004_ApplicationMultiplePulse}%
  \BibitemOpen
  \bibfield  {author} {\bibinfo {author} {\bibfnamefont {R.}~\bibnamefont {Siegel}}, \bibinfo {author} {\bibfnamefont {T.~T.}\ \bibnamefont {Nakashima}},\ and\ \bibinfo {author} {\bibfnamefont {R.~E.}\ \bibnamefont {Wasylishen}},\ }\bibfield  {title} {\bibinfo {title} {{Application of Multiple-Pulse Experiments to Characterize Broad NMR Chemical-Shift Powder Patterns from Spin-1/2 Nuclei in the Solid State}},\ }\href {https://doi.org/10.1021/jp031048c} {\bibfield  {journal} {\bibinfo  {journal} {The Journal of Physical Chemistry B}\ }\textbf {\bibinfo {volume} {108}},\ \bibinfo {pages} {2218} (\bibinfo {year} {2004})}\BibitemShut {NoStop}%
\bibitem [{\citenamefont {Altenhof}\ \emph {et~al.}(2022)\citenamefont {Altenhof}, \citenamefont {Gan},\ and\ \citenamefont {Schurko}}]{Altenhof_2022_Reducingeffectsweak}%
  \BibitemOpen
  \bibfield  {author} {\bibinfo {author} {\bibfnamefont {A.~R.}\ \bibnamefont {Altenhof}}, \bibinfo {author} {\bibfnamefont {Z.}~\bibnamefont {Gan}},\ and\ \bibinfo {author} {\bibfnamefont {R.~W.}\ \bibnamefont {Schurko}},\ }\bibfield  {title} {\bibinfo {title} {Reducing the effects of weak homonuclear dipolar coupling with {CPMG} pulse sequences for static and spinning solids},\ }\href {https://doi.org/10.1016/j.jmr.2022.107174} {\bibfield  {journal} {\bibinfo  {journal} {Journal of Magnetic Resonance}\ }\textbf {\bibinfo {volume} {337}},\ \bibinfo {pages} {107174} (\bibinfo {year} {2022})}\BibitemShut {NoStop}%
\bibitem [{\citenamefont {Yannoni}\ \emph {et~al.}(1991)\citenamefont {Yannoni}, \citenamefont {Bernier}, \citenamefont {Bethune}, \citenamefont {Meijer},\ and\ \citenamefont {Salem}}]{Yannoni_1991_NMRdeterminationbond}%
  \BibitemOpen
  \bibfield  {author} {\bibinfo {author} {\bibfnamefont {C.~S.}\ \bibnamefont {Yannoni}}, \bibinfo {author} {\bibfnamefont {P.~P.}\ \bibnamefont {Bernier}}, \bibinfo {author} {\bibfnamefont {D.~S.}\ \bibnamefont {Bethune}}, \bibinfo {author} {\bibfnamefont {G.}~\bibnamefont {Meijer}},\ and\ \bibinfo {author} {\bibfnamefont {J.~R.}\ \bibnamefont {Salem}},\ }\bibfield  {title} {\bibinfo {title} {{NMR determination of the bond lengths in C60}},\ }\href {https://doi.org/10.1021/ja00008a068} {\bibfield  {journal} {\bibinfo  {journal} {Journal of the American Chemical Society}\ }\textbf {\bibinfo {volume} {113}},\ \bibinfo {pages} {3190} (\bibinfo {year} {1991})}\BibitemShut {NoStop}%
\bibitem [{\citenamefont {Baltisberger}\ \emph {et~al.}(2012)\citenamefont {Baltisberger}, \citenamefont {Walder}, \citenamefont {Keeler}, \citenamefont {Kaseman}, \citenamefont {Sanders},\ and\ \citenamefont {Grandinetti}}]{Baltisberger_2012_CommunicationPhaseincremented}%
  \BibitemOpen
  \bibfield  {author} {\bibinfo {author} {\bibfnamefont {J.~H.}\ \bibnamefont {Baltisberger}}, \bibinfo {author} {\bibfnamefont {B.~J.}\ \bibnamefont {Walder}}, \bibinfo {author} {\bibfnamefont {E.~G.}\ \bibnamefont {Keeler}}, \bibinfo {author} {\bibfnamefont {D.~C.}\ \bibnamefont {Kaseman}}, \bibinfo {author} {\bibfnamefont {K.~J.}\ \bibnamefont {Sanders}},\ and\ \bibinfo {author} {\bibfnamefont {P.~J.}\ \bibnamefont {Grandinetti}},\ }\bibfield  {title} {\bibinfo {title} {{Communication: Phase incremented echo train acquisition in NMR spectroscopy}},\ }\href {https://doi.org/10.1063/1.4728105} {\bibfield  {journal} {\bibinfo  {journal} {The Journal of Chemical Physics}\ }\textbf {\bibinfo {volume} {136}},\ \bibinfo {pages} {211104} (\bibinfo {year} {2012})}\BibitemShut {NoStop}%
\bibitem [{\citenamefont {Srivastava}\ \emph {et~al.}(2018{\natexlab{a}})\citenamefont {Srivastava}, \citenamefont {Florian}, \citenamefont {Baltisberger},\ and\ \citenamefont {Grandinetti}}]{Srivastava_2018_Correlatinggeminal2JSi-O-Si}%
  \BibitemOpen
  \bibfield  {author} {\bibinfo {author} {\bibfnamefont {D.~J.}\ \bibnamefont {Srivastava}}, \bibinfo {author} {\bibfnamefont {P.}~\bibnamefont {Florian}}, \bibinfo {author} {\bibfnamefont {J.~H.}\ \bibnamefont {Baltisberger}},\ and\ \bibinfo {author} {\bibfnamefont {P.~J.}\ \bibnamefont {Grandinetti}},\ }\bibfield  {title} {\bibinfo {title} {{Correlating geminal 2JSi–O–Si couplings to structure in framework silicates}},\ }\href {https://doi.org/10.1039/c7cp06486a} {\bibfield  {journal} {\bibinfo  {journal} {Physical Chemistry Chemical Physics}\ }\textbf {\bibinfo {volume} {20}},\ \bibinfo {pages} {562} (\bibinfo {year} {2018}{\natexlab{a}})}\BibitemShut {NoStop}%
\bibitem [{\citenamefont {Srivastava}\ \emph {et~al.}(2018{\natexlab{b}})\citenamefont {Srivastava}, \citenamefont {Baltisberger}, \citenamefont {Florian}, \citenamefont {Fayon}, \citenamefont {Shakhovoy}, \citenamefont {Deschamps}, \citenamefont {Sadiki},\ and\ \citenamefont {Grandinetti}}]{Srivastava_2018_Correlatingstructuraldistributions}%
  \BibitemOpen
  \bibfield  {author} {\bibinfo {author} {\bibfnamefont {D.~J.}\ \bibnamefont {Srivastava}}, \bibinfo {author} {\bibfnamefont {J.~H.}\ \bibnamefont {Baltisberger}}, \bibinfo {author} {\bibfnamefont {P.}~\bibnamefont {Florian}}, \bibinfo {author} {\bibfnamefont {F.}~\bibnamefont {Fayon}}, \bibinfo {author} {\bibfnamefont {R.~A.}\ \bibnamefont {Shakhovoy}}, \bibinfo {author} {\bibfnamefont {M.}~\bibnamefont {Deschamps}}, \bibinfo {author} {\bibfnamefont {N.}~\bibnamefont {Sadiki}},\ and\ \bibinfo {author} {\bibfnamefont {P.~J.}\ \bibnamefont {Grandinetti}},\ }\bibfield  {title} {\bibinfo {title} {{Correlating structural distributions in silica glass with two-dimensional $J$-resolved spectroscopy}},\ }\href {https://doi.org/10.1103/physrevb.98.134202} {\bibfield  {journal} {\bibinfo  {journal} {Physical Review B}\ }\textbf {\bibinfo {volume} {98}},\ \bibinfo {pages} {134202} (\bibinfo {year} {2018}{\natexlab{b}})}\BibitemShut {NoStop}%
\bibitem [{\citenamefont {Jardón-Álvarez}\ \emph {et~al.}(2021)\citenamefont {Jardón-Álvarez}, \citenamefont {Bovee},\ and\ \citenamefont {Grandinetti}}]{JardonAlvarez_2021_Silicon29echo}%
  \BibitemOpen
  \bibfield  {author} {\bibinfo {author} {\bibfnamefont {D.}~\bibnamefont {Jardón-Álvarez}}, \bibinfo {author} {\bibfnamefont {M.~O.}\ \bibnamefont {Bovee}},\ and\ \bibinfo {author} {\bibfnamefont {P.~J.}\ \bibnamefont {Grandinetti}},\ }\bibfield  {title} {\bibinfo {title} {{Silicon-29 echo train coherence lifetimes and geminal 2J-couplings in network modified silicate glasses}},\ }\href {https://doi.org/10.1016/j.jmr.2021.107097} {\bibfield  {journal} {\bibinfo  {journal} {Journal of Magnetic Resonance}\ }\textbf {\bibinfo {volume} {333}},\ \bibinfo {pages} {107097} (\bibinfo {year} {2021})}\BibitemShut {NoStop}%
\bibitem [{\citenamefont {Bak}\ \emph {et~al.}(2011)\citenamefont {Bak}, \citenamefont {Rasmussen},\ and\ \citenamefont {Nielsen}}]{Bak_2011_SIMPSONgeneralsimulation}%
  \BibitemOpen
  \bibfield  {author} {\bibinfo {author} {\bibfnamefont {M.}~\bibnamefont {Bak}}, \bibinfo {author} {\bibfnamefont {J.~T.}\ \bibnamefont {Rasmussen}},\ and\ \bibinfo {author} {\bibfnamefont {N.~C.}\ \bibnamefont {Nielsen}},\ }\bibfield  {title} {\bibinfo {title} {{SIMPSON: A general simulation program for solid-state NMR spectroscopy}},\ }\href {https://doi.org/10.1016/j.jmr.2011.09.008} {\bibfield  {journal} {\bibinfo  {journal} {Journal of Magnetic Resonance}\ }\textbf {\bibinfo {volume} {213}},\ \bibinfo {pages} {366} (\bibinfo {year} {2011})}\BibitemShut {NoStop}%
\bibitem [{\citenamefont {Tošner}\ \emph {et~al.}(2014)\citenamefont {Tošner}, \citenamefont {Andersen}, \citenamefont {Stevensson}, \citenamefont {Edén}, \citenamefont {Nielsen},\ and\ \citenamefont {Vosegaard}}]{Tosner_2014_Computerintensivesimulation}%
  \BibitemOpen
  \bibfield  {author} {\bibinfo {author} {\bibfnamefont {Z.}~\bibnamefont {Tošner}}, \bibinfo {author} {\bibfnamefont {R.}~\bibnamefont {Andersen}}, \bibinfo {author} {\bibfnamefont {B.}~\bibnamefont {Stevensson}}, \bibinfo {author} {\bibfnamefont {M.}~\bibnamefont {Edén}}, \bibinfo {author} {\bibfnamefont {N.~C.}\ \bibnamefont {Nielsen}},\ and\ \bibinfo {author} {\bibfnamefont {T.}~\bibnamefont {Vosegaard}},\ }\bibfield  {title} {\bibinfo {title} {{Computer-intensive simulation of solid-state NMR experiments using SIMPSON}},\ }\href {https://doi.org/10.1016/j.jmr.2014.07.002} {\bibfield  {journal} {\bibinfo  {journal} {Journal of Magnetic Resonance}\ }\textbf {\bibinfo {volume} {246}},\ \bibinfo {pages} {79} (\bibinfo {year} {2014})}\BibitemShut {NoStop}%
\bibitem [{\citenamefont {Haeberlen}(1976)}]{Haeberlen_1976_Advancesmagneticresonance}%
  \BibitemOpen
  \bibfield  {author} {\bibinfo {author} {\bibfnamefont {U.}~\bibnamefont {Haeberlen}},\ }\bibinfo {title} {Advances in magnetic resonance},\ in\ \href {https://doi.org/10.1016/b978-0-12-025561-0.50001-0} {\emph {\bibinfo {booktitle} {{High Resolution NMR in Solids Selective Averaging}}}}\ (\bibinfo  {publisher} {Elsevier},\ \bibinfo {address} {New York},\ \bibinfo {year} {1976})\ p.~\bibinfo {pages} {ii}\BibitemShut {NoStop}%
\bibitem [{\citenamefont {Abragam}(1961)}]{Abragam_1961_PrinciplesNuclearMagnetism}%
  \BibitemOpen
  \bibfield  {author} {\bibinfo {author} {\bibfnamefont {A.}~\bibnamefont {Abragam}},\ }\href {https://global.oup.com/academic/product/principles-of-nuclear-magnetism-9780198520146?cc=us&lang=en&} {\emph {\bibinfo {title} {{Principles of Nuclear Magnetism}}}}\ (\bibinfo  {publisher} {Oxford University Press},\ \bibinfo {address} {New York},\ \bibinfo {year} {1961})\BibitemShut {NoStop}%
\bibitem [{\citenamefont {Pake}(1948)}]{Pake_1948_NuclearResonanceAbsorption}%
  \BibitemOpen
  \bibfield  {author} {\bibinfo {author} {\bibfnamefont {G.~E.}\ \bibnamefont {Pake}},\ }\bibfield  {title} {\bibinfo {title} {{Nuclear Resonance Absorption in Hydrated Crystals: Fine Structure of the Proton Line}},\ }\href {https://doi.org/10.1063/1.1746878} {\bibfield  {journal} {\bibinfo  {journal} {The Journal of Chemical Physics}\ }\textbf {\bibinfo {volume} {16}},\ \bibinfo {pages} {327} (\bibinfo {year} {1948})}\BibitemShut {NoStop}%
\bibitem [{\citenamefont {Tossell}(1992)}]{Tossell_1992_NuclearMagneticShieldings}%
  \BibitemOpen
  \bibfield  {author} {\bibinfo {author} {\bibfnamefont {J.~A.}\ \bibnamefont {Tossell}},\ }\href {https://doi.org/10.1007/978-94-011-1652-7} {\emph {\bibinfo {title} {{Nuclear Magnetic Shieldings and Molecular Structure}}}}\ (\bibinfo  {publisher} {Springer Science+Business Media Dordrecht},\ \bibinfo {year} {1992})\BibitemShut {NoStop}%
\bibitem [{\citenamefont {Morss}\ \emph {et~al.}(2011)\citenamefont {Morss}, \citenamefont {Edelstein}, \citenamefont {Fuger},\ and\ \citenamefont {Katz}}]{Morss_2011_ChemistryActinideTransactinide}%
  \BibitemOpen
  \bibfield  {author} {\bibinfo {author} {\bibfnamefont {L.~R.}\ \bibnamefont {Morss}}, \bibinfo {author} {\bibfnamefont {N.~M.}\ \bibnamefont {Edelstein}}, \bibinfo {author} {\bibfnamefont {J.}~\bibnamefont {Fuger}},\ and\ \bibinfo {author} {\bibfnamefont {J.~J.}\ \bibnamefont {Katz}},\ }\href {https://doi.org/10.1007/978-94-007-0211-0} {\emph {\bibinfo {title} {{The Chemistry of the Actinide and Transactinide Elements}}}}\ (\bibinfo  {publisher} {Springer Netherlands},\ \bibinfo {address} {Dordrecht},\ \bibinfo {year} {2011})\BibitemShut {NoStop}%
\bibitem [{\citenamefont {Blackwell}\ \emph {et~al.}(2025)\citenamefont {Blackwell}, \citenamefont {Yamamoto}, \citenamefont {Thomas}, \citenamefont {Dioguardi}, \citenamefont {Cary}, \citenamefont {Kozimor}, \citenamefont {Bauer}, \citenamefont {Ronning},\ and\ \citenamefont {Hirata}}]{Blackwell_2025_Localagingeffects}%
  \BibitemOpen
  \bibfield  {author} {\bibinfo {author} {\bibfnamefont {S.~B.}\ \bibnamefont {Blackwell}}, \bibinfo {author} {\bibfnamefont {R.}~\bibnamefont {Yamamoto}}, \bibinfo {author} {\bibfnamefont {S.~M.}\ \bibnamefont {Thomas}}, \bibinfo {author} {\bibfnamefont {A.~P.}\ \bibnamefont {Dioguardi}}, \bibinfo {author} {\bibfnamefont {S.~K.}\ \bibnamefont {Cary}}, \bibinfo {author} {\bibfnamefont {S.~A.}\ \bibnamefont {Kozimor}}, \bibinfo {author} {\bibfnamefont {E.~D.}\ \bibnamefont {Bauer}}, \bibinfo {author} {\bibfnamefont {F.}~\bibnamefont {Ronning}},\ and\ \bibinfo {author} {\bibfnamefont {M.}~\bibnamefont {Hirata}},\ }\bibfield  {title} {\bibinfo {title} {{Local aging effects in PuB4: Growing inhomogeneity and slow dynamics of local-field fluctuations probed by ${}^{239}$Pu NMR}},\ }\href {https://doi.org/10.1103/physrevb.111.075152} {\bibfield  {journal} {\bibinfo  {journal} {Physical Review B}\ }\textbf {\bibinfo {volume} {111}},\ \bibinfo {pages} {075152} (\bibinfo {year} {2025})}\BibitemShut {NoStop}%
\end{thebibliography}%

\end{document}


\newcommand {\beq} {\begin{equation}}
	\newcommand {\eeq} {\end{equation}}
\newcommand {\bqa} {\begin{eqnarray}}
	\newcommand {\eqa} {\end{eqnarray}}
\newcommand {\ba} {\ensuremath{b^\dagger}}
\newcommand {\Ma} {\ensuremath{M^\dagger}}
\newcommand {\psia} {\ensuremath{\psi^\dagger}}
\newcommand {\psita} {\ensuremath{\tilde{\psi}^\dagger}}
\newcommand{\lp} {\ensuremath{{\lambda '}}}
\newcommand{\A} {\ensuremath{{\bf A}}}
\newcommand{\Q} {\ensuremath{{\bf Q}}}
\newcommand{\kk} {\ensuremath{{\bf k}}}
\newcommand{\qq} {\ensuremath{{\bf q}}}
\newcommand{\kp} {\ensuremath{{\bf k'}}}
\newcommand{\rr} {\ensuremath{{\bf r}}}
\newcommand{\rp} {\ensuremath{{\bf r'}}}
\newcommand {\ep} {\ensuremath{\epsilon}}
\newcommand{\nbr} {\ensuremath{\langle ij \rangle}}
\newcommand {\no} {\nonumber}
\newcommand{\up} {\ensuremath{\uparrow}}
\newcommand{\dn} {\ensuremath{\downarrow}}

\renewcommand{\thetable}{S\arabic{table}}
\renewcommand{\thefigure}{S\arabic{figure}}


\title{Supplementary Material for ``Microscopic investigation of enhanced Pauli paramagnetism in metallic Pu$_2$C$_3$''}

\author{R. Yamamoto}
\thanks{These authors contributed equally to this work.}
\affiliation{Materials Physics and Applications $-$ Quantum, Los Alamos National Laboratory, Los Alamos, New Mexico  87545, USA}

\author{M. S. Cook}
\thanks{These authors contributed equally to this work.}
\affiliation{MST-16 $-$ Nuclear Materials Science, Los Alamos National Laboratory, Los Alamos, New Mexico  87545, USA}

\author{A. R. Altenhof}
\thanks{These authors contributed equally to this work.}
\affiliation{Materials Physics and Applications $-$ Quantum, Los Alamos National Laboratory, Los Alamos, New Mexico  87545, USA}

\author{P. Sherpa}
\affiliation{Materials Physics and Applications $-$ Quantum, Los Alamos National Laboratory, Los Alamos, New Mexico 87545, USA}
\affiliation{Department of Physics and Astronomy, University of California, Davis, California 95616, USA}

\author{S. Park}
\affiliation{Materials Physics and Applications $-$ Quantum, Los Alamos National Laboratory, Los Alamos, New Mexico 87545, USA}

\author{J. D. Thompson}
\affiliation{Materials Physics and Applications $-$ Quantum, Los Alamos National Laboratory, Los Alamos, New Mexico 87545, USA}

\author{H. E. Mason}
\affiliation{Chemistry Division, Los Alamos National Laboratory, Los Alamos, New Mexico  87545, USA}

\author{D. C. Arellano}
\affiliation{MST-16 $-$ Nuclear Materials Science, Los Alamos National Laboratory, Los Alamos, New Mexico  87545, USA}

\author{D. V. Prada}
\affiliation{MST-16 $-$ Nuclear Materials Science, Los Alamos National Laboratory, Los Alamos, New Mexico  87545, USA}

\author{P. H. Tobash}
\affiliation{MST-16 $-$ Nuclear Materials Science, Los Alamos National Laboratory, Los Alamos, New Mexico  87545, USA}

\author{F. Ronning}
\affiliation{Materials Physics and Applications $-$ Quantum, Los Alamos National Laboratory, Los Alamos, New Mexico  87545, USA}

\author{E. D. Bauer}
\affiliation{Materials Physics and Applications $-$ Quantum, Los Alamos National Laboratory, Los Alamos, New Mexico  87545, USA}

\author{N. Harrison}
\affiliation{National High Magnetic Field Laboratory, Los Alamos National Laboratory, Los Alamos, New Mexico  87545, USA}

\author{W. A. Phelan}
\affiliation{MST-16 $-$ Nuclear Materials Science, Los Alamos National Laboratory, Los Alamos, New Mexico  87545, USA}

\author{A. P. Dioguardi}\thanks{\textcolor{blue}{apd@lanl.gov}} 
\affiliation{Materials Physics and Applications $-$ Quantum, Los Alamos National Laboratory, Los Alamos, New Mexico  87545, USA}

\author{M. Hirata}\thanks{mhirata@lanl.gov}
\affiliation{Materials Physics and Applications $-$ Quantum, Los Alamos National Laboratory, Los Alamos, New Mexico  87545, USA}

\date{\today}

\maketitle

\subsection{Lorentzian fitting to ${}^{13}$C NMR spectrum}
\begin{figure}[h]
	\includegraphics[width=15.6cm]{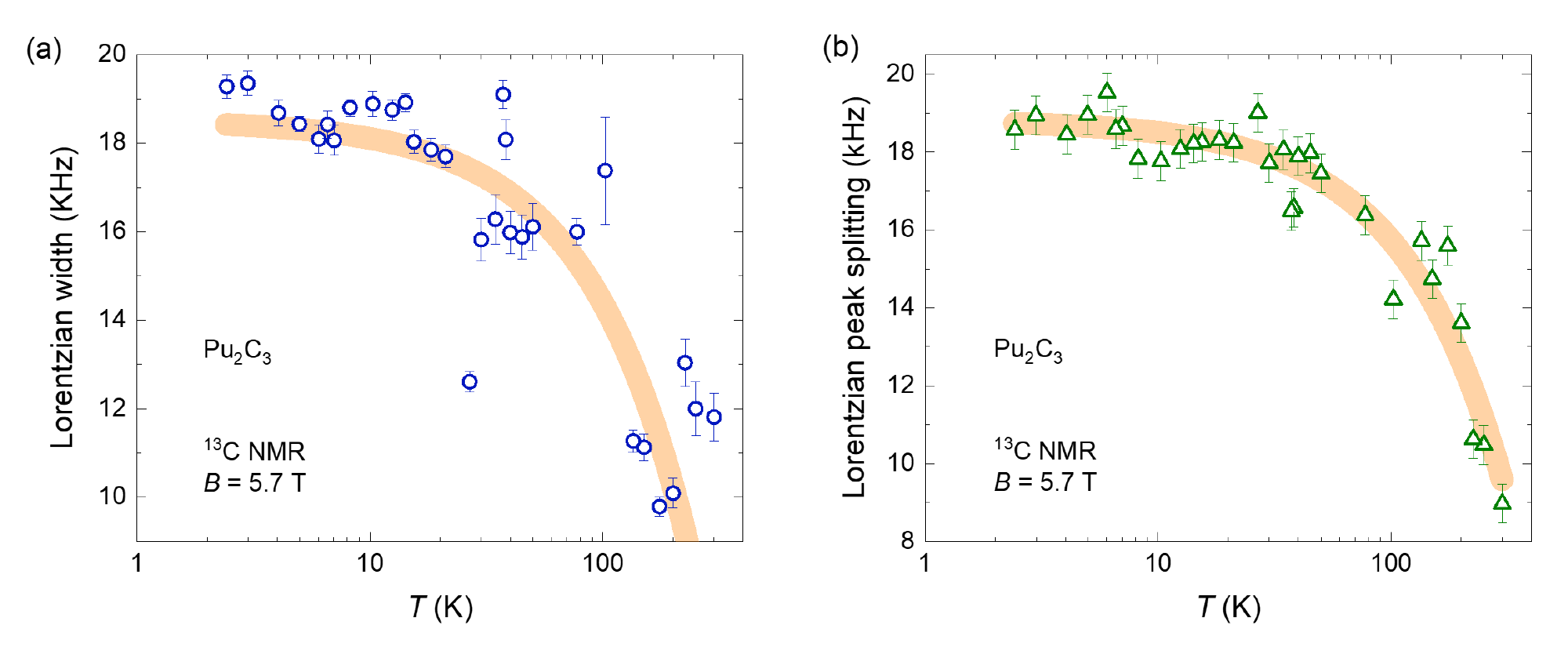}
	\caption{${}^{13}$C NMR line width and splitting of Pu$_2$C$_3$ powder, plotted as a function of temperature at 5.7~T in a semi-logarithmic scale. Spectra were fitted with two Lorentzian forms to deduce the Lorentzian width (a) and splitting (b). Solid curves (orange) are least-square fits to the data with a form $a + bT$.}
	\label{fig:width-splitting-vs-T}
\end{figure}

We extracted the line splitting and width of ${}^{13}$C NMR powder pattern of Pu$_2$C$_3$ at 5.7~T under varying temperatures by fitting the spectrum with two Lorentzian forms. Fig.~\ref{fig:width-splitting-vs-T} shows the resulting peak width [Fig.~\ref{fig:width-splitting-vs-T}(a)] and splitting [Fig.~\ref{fig:width-splitting-vs-T}(b)] plotted as a function of temperature. The temperature dependence of these quantities approximately follow that of the site-averaged isotropic shift $\overline{\delta_{\mathrm{iso}}}$ [Fig.~6(a) in the main text], suggesting that the splitting originates from two unique ${}^{13}$C sites that sit in slightly different hyperfine environments. We find that both the width and splitting linearly scale to temperature.

These quantities were also evaluated as a function of field at 35~K for lower field (from 1.0 to 5.7~T) and 150~K for higher field (from 6.0 to 12~T). To connect the two sets of data taken at different temperatures, we multiplied a constant factor of 1.225 to the latter that is deduced from Fig.~\ref{fig:width-splitting-vs-T} to account for the temperature dependence. Fig.~4 in the main text presents the resulting field dependence. The two-horn spectral feature visible at lower field becomes less clear at higher fields due to the increased broadening of ${}^{13}$C signals (Fig.~3 in the main text). The enhanced data scattering in Fig.~4(a) above 7.0~T represents this situation and is reflecting the lesser fit quality with Lorentzian.        

\subsection{${}^{13}$C NMR powder spin echo decay simulations with SIMPSON}

Spin-echo decay (SED) curves were measured with a Hahn echo pulse sequence in which clear oscillations were observed in the echo intensity as a function of the interpulse separation $\tau$, reflecting homonuclear dipole couplings between nearest-neighbor ${}^{13}$C spins. (We only expect to see ${}^{13}$C--${}^{13}$C coupling, as it has the largest coupling constant.) The spin-spin (transverse) relaxation time $T_{2}$ was determined by fitting SED curves with multiple sinusoidal terms with different frequencies and a Gaussian decay envelope function.

Both indirect and direct dipole--dipole coupling interactions manifest as secular and non-secular contributions to $T_2$~\cite{Gupta_2021_IsItTime}. The secular contribution is the result of the weak heteronuclear dipolar coupling and the weak homonuclear dipolar coupling being refocused by a Hahn-echo sequence. Therefore, the dipolar coupling manifests in $T_2$ data measured by Hahn-echo sequences with varied inter-echo spacing, often referred to as either SED oscillation measurements or $J$-resolved spectroscopy (JRES), and in Carr-Purcell-Meiboom-Gill (CPMG) pulse sequences. These measurements are often conducted in solution state NMR; however, these effects are well known to occur in the solid state even when the dipolar-coupled spins experience inhomogeneous broadening from anisotropic interactions (e.g., magnetic shielding) under static or magic-angle spinning conditions~\cite{Dementyev_2003_AnomaliesNMRsilicon,Franzoni_2005_Manifestationsabsencespin, Li_2007_GeneratingUnexpectedSpin, Li_2008_Intrinsicoriginspin, Siegel_2004_ApplicationMultiplePulse, Altenhof_2022_Reducingeffectsweak,Yannoni_1991_NMRdeterminationbond}. A phase-incremented CPMG-type sequence, known as phase incremented echo train acquisition (PIETA)~\cite{Baltisberger_2012_CommunicationPhaseincremented}, has also been used for measuring indirect dipolar coupling in solids while minimizing spectral artifacts~\cite{Srivastava_2018_Correlatinggeminal2JSi-O-Si,Srivastava_2018_Correlatingstructuraldistributions, JardonAlvarez_2021_Silicon29echo}.

Time-domain powder-averaged spin density matrix simulations were conducted in SIMPSON version 4.2.1~\cite{Bak_2011_SIMPSONgeneralsimulation, Tosner_2014_Computerintensivesimulation} to analyze and fit ${}^{13}$C-NMR powder spectra and FFTs of oscillating SED curves. Simulations use two ${}^{13}$C spins with shift tensors (primarily representing the Knight shift here) and a homonuclear dipolar coupling between them. The shift tensor, which determines the line shape here, is reported in the Haeberlen notation following the standard convention used in SIMPSON ~\cite{Haeberlen_1976_Advancesmagneticresonance}, with principal components defined as $|\delta_{zz} - \delta_{\mathrm{iso}}| \geq |\delta_{xx} -\delta_{\mathrm{iso}}| \geq |\delta_{yy} -\delta_{\mathrm{iso}}|$, which includes the isotropic shift, $\delta_{\mathrm{iso}} = (\delta_{zz} + \delta_{xx} + \delta_{yy})/3$, the reduced anisotropy, $\delta= \delta_{zz}-\delta_{\mathrm{iso}}$, and the asymmetry parameter, $\eta = (\delta_{yy} – \delta_{xx})/\delta$. The average shift of the pattern from the origin (about 1000~ppm in Fig.~3) is not taken into account in the fit, as it has no effects on the lineshape. The orientations of the two shift tensors are assumed to be coincident, with an orthogonal internuclear dipolar vector between them. Powder averaging uses the Zaremba-Conroy-Wolfsberg (ZCW) scheme with 4180 orientations. Oscillating SED curves were analyzed with the homonuculear JRES pulse sequence that is simulated with a $\pi/2$ – $\tau$ – $\pi$ – $\tau$ – acquire sequence, where $\tau$ is the inter-echo spacing that is incremented in steps of 15 $\mathrm{\mu}$s for 64 points. All processing is done in Python with custom routines.

\begin{figure}[!ht]
	\includegraphics[width=18cm]{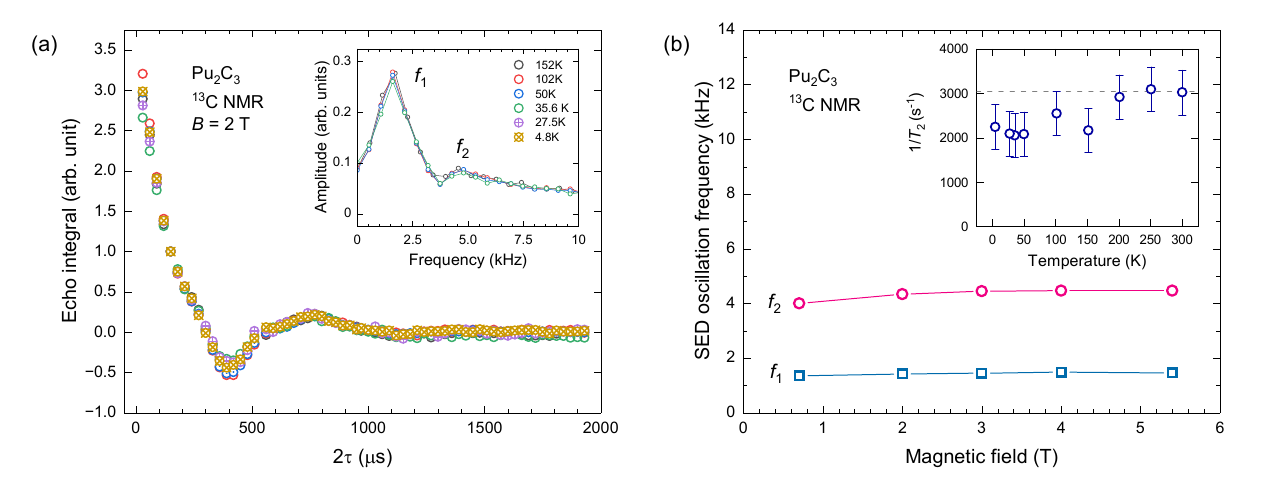}
	\caption{${}^{13}$C NMR SED oscillations in Pu$_2$C$_3$ powder. (a) Normalized echo intensity plotted as a function of interpulse spacing $2\tau$ at 2~T for varying temperatures. Inset shows the Fast Fourier Transform with two characteristic oscillation frequencies $f_{1}$ and $f_{2}$. (b) SED oscillation frequencies $f_{1}$ and $f_{2}$ plotted against magnetic field. Inset: Temperature dependence of the spin-spin relaxation rate 1/$T_{2}$ at 2~T. The horizontal dashed line stands for the calculated van-Vleck contribution coming from direct nuclear dipole--dipole interactions~\cite{Abragam_1961_PrinciplesNuclearMagnetism} considering both homonuclear (${}^{13}$C--${}^{13}$C) and heteronuclear (${}^{13}$C--${}^{239}$Pu) couplings up to 30 nearest-neighbor pairs (or for the atoms lying within a spherical radius of 11~\r{A}).}
	\label{fig:6}
\end{figure}

\begin{figure}[b]
	\includegraphics[width=13cm]{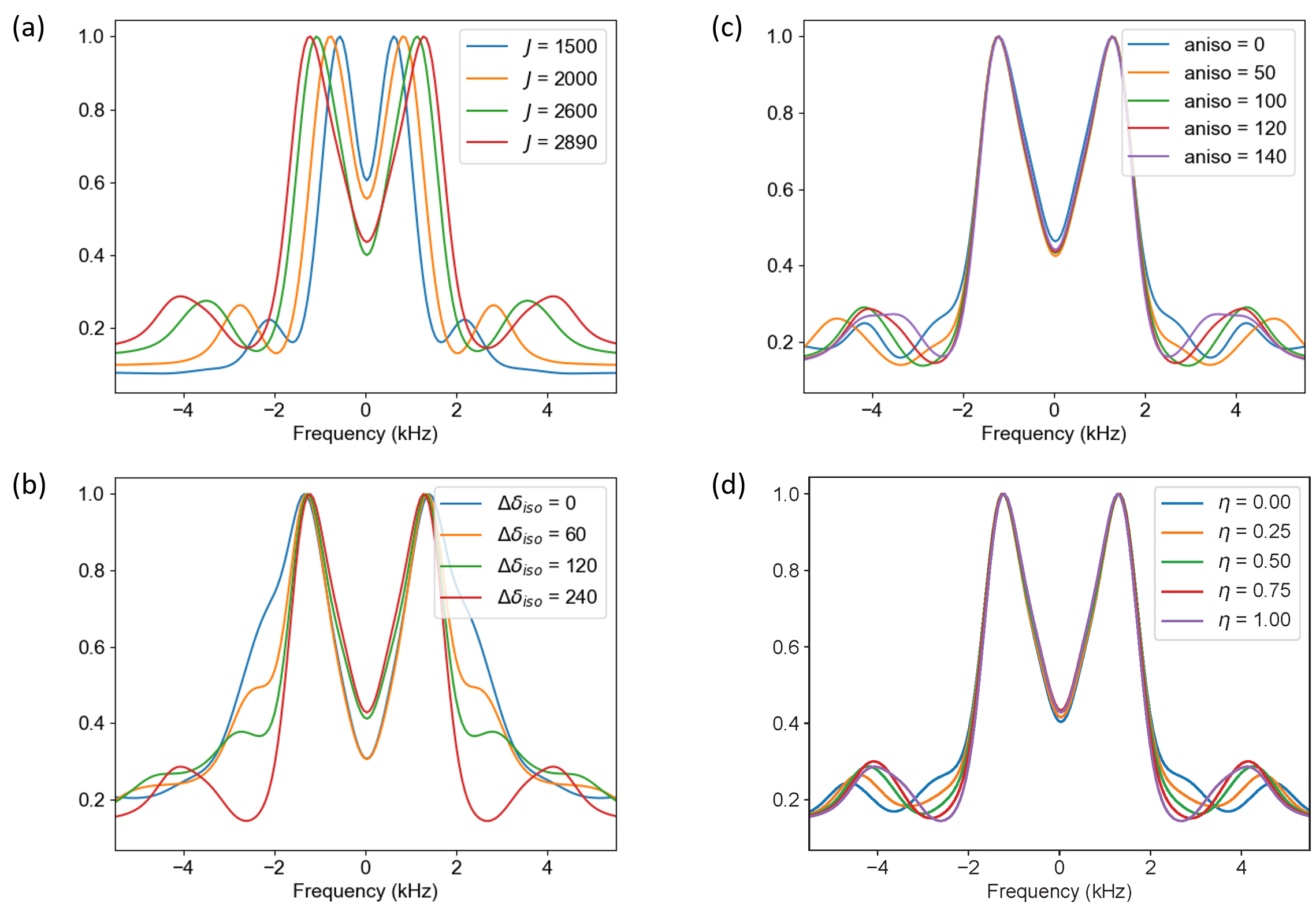}
	\caption{${}^{13}$C JRES simulations of the Fourier transform of powder SED oscillations with SIPSON. (a) With varied direct nuclear dipole--dipole coupling constant, $J$, using $\Delta\delta_{\mathrm{iso}} =$~240~ppm and $\delta =$~120~ppm. (b) With a varied difference in isotropic shifts between the two ${}^{13}$C tensors, $\Delta\delta_{\mathrm{iso}}$, using $J$ = 2.89 kHz and $\delta$ = 120 ppm. (c) With a varied reduced shift anisotropy, $\delta$, for both ${}^{13}$C shift tensors using $J$ = 2.89 kHz and $\Delta\delta_{\mathrm{iso}}$ = 240 ppm. (d) With a varied shift asymmetry, $\eta$, for both ${}^{13}$C shift tensors using $J$ = 2.89~kHz, $\Delta\delta_{\mathrm{iso}}$ = 240~ppm, and $\delta$ = 120~ppm.}
	\label{fig:JRES}
\end{figure}

\begin{figure}[h]
	\includegraphics[width=7cm]{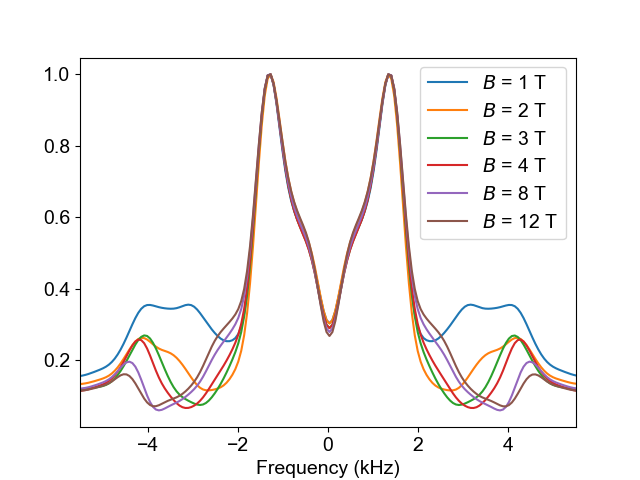}
	\caption{${}^{13}$C JRES simulations of SED oscillations with SIPSON with a varied static magnetic field, $B$, using $J$ = 2.89~kHz, $\Delta\delta_{\mathrm{iso}}$ = 240~ppm, $\delta$ = 120~ppm, and $\eta$ = 1.0. These results suggest that the shift-dependent features are most prominent at low magnetic fields, which may be onset by strong dipolar coupling, i.e., $\Delta\delta_{\mathrm{iso}} \approx J$.}
	\label{fig:field-dep}
\end{figure}

\begin{figure}[t]
	\includegraphics[width=5.6cm]{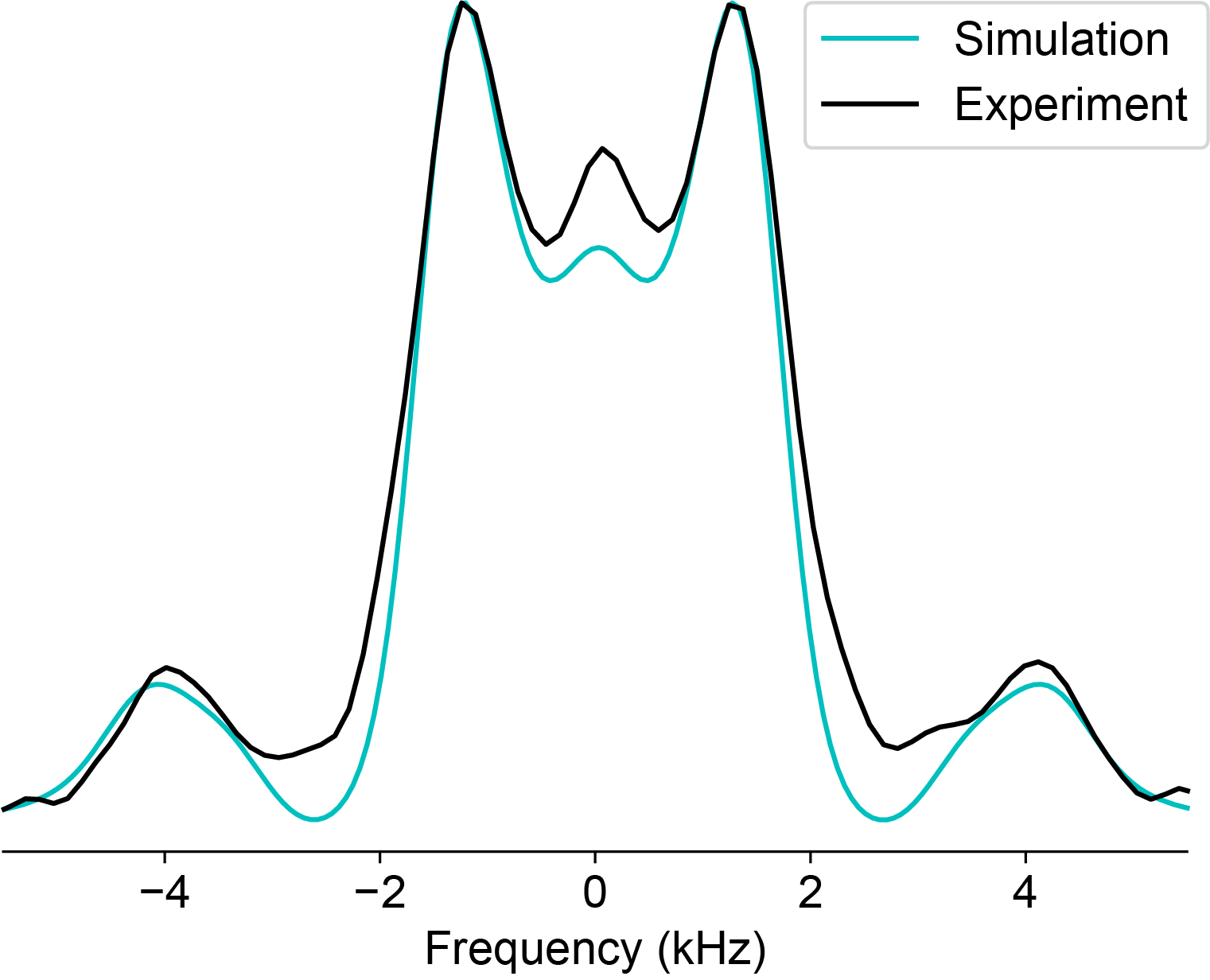}
	\caption{${}^{13}$C JRES analysis of Pu$_2$C$_3$ powder. Experimentally determined curve at 2~T based on the SED oscillation data in Fig.~5(a) in the main text (black) and simulated curve with SIMPSON (blue) are shown.}
	\label{fig:AA2}
\end{figure}

Figure~\ref{fig:6}(a) shows typical SED curves at 2~T and various $T$ plotted as a function of twice the interpulse spacing, $2\tau$, determined from integrated full echo intensity $M(2\tau)$. The SED curves have two representative frequencies ($f_{1}$ and $f_{2}$) with an overall Gaussian decay form, $M(2\tau) = M(0)[A~\mathrm{cos}(2 \pi f_{1}\cdot 2\tau) + B~\mathrm{cos}(2 \pi f_{2}\cdot 2\tau)] \cdot \mathrm{exp}[-(2\tau/T_{2})^2]$, where $T_{2}$ is the nuclear spin-spin relaxation time. The FFT of SED curves [inset of Fig.~\ref{fig:6}(a)] reveals a strong oscillating component at $f_{1} \simeq 1.8$~kHz and a secondary weak component at $f_{2} \simeq 4.5$~kHz with little $B$ and $T$ dependence [Fig.~\ref{fig:6}(b) and its inset]. 1/$T_{2}$ showed also little $T$ dependence that is accounted for by the direct nuclear dipole--dipole coupling as estimated from van Vleck's second moment method [dashed line in the inset of Fig.~\ref{fig:6}(b)]~\cite{Abragam_1961_PrinciplesNuclearMagnetism}, derived by considering both homonuclear (${}^{13}$C--${}^{13}$C) and heteronuclear (${}^{13}$C--${}^{239}$Pu) couplings up to 30 nearest-neighbor pairs. 

Simulations of various ${}^{13}$C--${}^{13}$C $J$-resolved spectroscopy (JRES) [or Fourier Transform of SED shown in a symmetrized way around the frequency origin] are conducted with SIMPSON to examine the effects of different tensor parameters on the SED for powder specimens. The direct nuclear dipole--dipole coupling constant, $J$, is shown to affect all key features in the JRES Pake-doublet-like powder pattern [Fig.~\ref{fig:JRES}(a)]. However, adjusting the other shift-tensor parameters, $\Delta\delta_{\mathrm{iso}}$, $\delta$, and $\eta$, and the magnetic field, $B$, does not affect the “horn” discontinuities of the spectral pattern. Instead, the shift-tensor and $B$ parameters affect resonances that manifest at significantly higher/lower frequencies relative to the horns. Therefore, $J$ can be fit with just the horn discontinuities [Fig.~\ref{fig:JRES}(a)]. Then, $\Delta\delta_{\mathrm{iso}}$ can be fit with the higher/lower frequency features while using a constrained $J$ [Fig.~\ref{fig:JRES}(b)]. The $\Delta\delta_{\mathrm{iso}}$ is also readily confirmed by the ${}^{13}$C NMR data acquired at various fields (Fig.~3 in the main text). The reduced anisotropy, $\delta$, can also be fit with the higher/lower frequency JRES features while using a constrained $J$ and $\Delta\delta_{\mathrm{iso}}$ [Fig.~\ref{fig:JRES}(c)]. $\delta$ is then likewise confirmed according to the pattern breadths in ${}^{13}$C NMR data (Fig.~3 in the main text). Finally, $\eta$ is fit with the JRES pattern while using the constrained $J$, $\Delta\delta_{\mathrm{iso}}$, and $\delta$ [Fig.~\ref{fig:JRES}(d)]. Fig.~\ref{fig:AA2} gives the best fit of the simulated pattern to the experimental data. 

Simulations of the JRES powder pattern were conducted at different magnetic field strengths (Fig.~\ref{fig:field-dep}). In this work, it is surprising that the JRES powder pattern shows features that depend on the shift tensors, since JRES/SED should not encode these effects. However, simulations show that these shift-based effects are most prominent at low fields (i.e., $\leq$ 4~T). This may be explained by the presence of strong homonuclear (${}^{13}$C--${}^{13}$C) direct dipole--dipole coupling at low fields. JRES/SED will no longer fully refocus all operators in the strong-coupling Hamiltonian, likely leading to “artifacts” in the resulting powder spectra (i.e., researchers may generically refer to these as artifacts when expecting a Pake-like pattern). Strong coupling effects manifest when $\Delta\delta_{\mathrm{iso}} \leq J$  or $\Delta\delta_{\mathrm{iso}} \approx J$. Weak coupling is only realized when $\Delta\delta_{\mathrm{iso}} \gg J$. Here, $\Delta\delta_{\mathrm{iso}}$ = 240~ppm, which is 2.6, 5.1, and 10.3~kHz at 1, 2, and 4~T, respectively, and $J$ = 2.89~kHz. Therefore, strong coupling is likely present and causing these additional features in the JRES powder spectra, which are captured in SIMPSON simulations.

\subsection{Analytic form of SED oscillation for a single ${}^{13}$C--${}^{13}$C dimer}
The SED oscillations [Fig.~\ref{fig:6}(a)] can be also examined based on analytic frameworks. For this, it is worth focussing on the simplest case where there is only a single, homonuclear dipole--dipole coupled spin-1/2 dimer (${}^{13}$C--${}^{13}$C dimer). Assuming a different isotropic shift for each of the ${}^{13}$C nuclei within the dimer with the difference $\Delta\delta_{\mathrm{iso}} = \Omega/(\gamma_nB)$ (where $\Omega$ is the corresponding frequency difference and $\gamma_n$ is the nuclear gyromagnetic ratio) and the direct dipole--dipole coupling constant $J$, the SED oscillation of integrated echo intensity can be expressed as~\cite{Abragam_1961_PrinciplesNuclearMagnetism}  

\begin{figure}[t]
	\includegraphics[width=13cm]{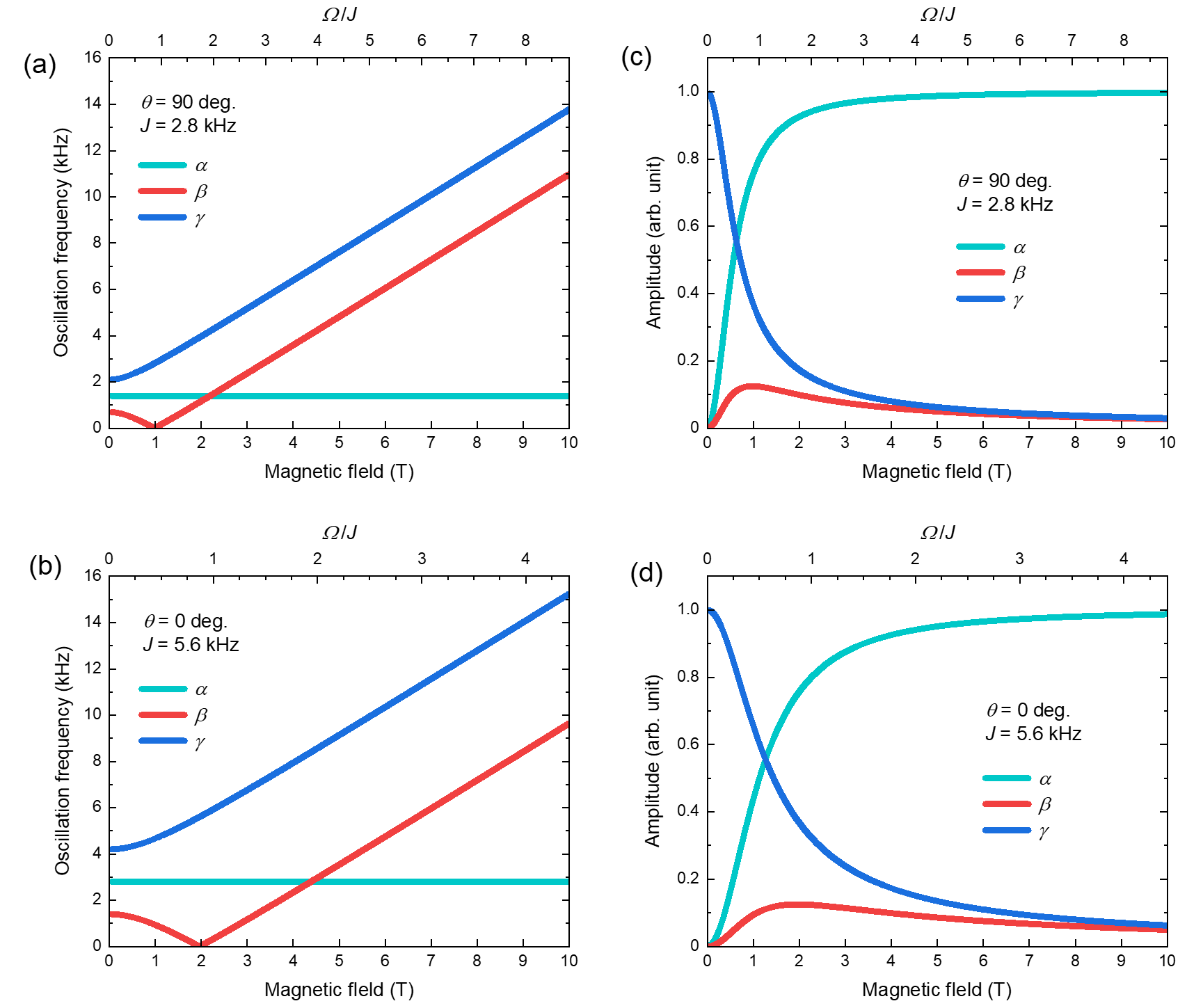}
	\caption{Simulated properties of ${}^{13}$C--${}^{13}$C SED oscillations for a single dimer plotted as a function of magnetic field for $\theta=90^{\circ}$ [(a) and (c)] and $\theta=0^{\circ}$ [(b) and (d)], with $\theta$ standing for the angle between external field and the dimer's bonding direction, the bond length $r=1.38$~\AA, and the frequency difference $\Omega$ obtained from a liner fit to the experimental splitting in Fig.~4(a) in the main text. Three components $\alpha$, $\beta$, and $\gamma$ in (a) and (b) [(c) and (d)] correspond to the frequency (normalized relative amplitude) of the first, second, and third terms of Eq.~(\ref{eq:1}), respectively.}
	\label{fig:SED_analytical}
\end{figure}

\begin{equation}
M(2\tau)=\frac{\Omega^2}{\Delta^2}\cos\left(\omega_\alpha\cdot2\tau\right)-\frac{J}{4\Delta}\left(1-\frac{J}{2\Delta}\right) \cos\left(\omega_\beta\cdot2\tau\right)+ \frac{J}{4\Delta}\left(1+\frac{J}{2\Delta}\right) \cos\left(\omega_\gamma\cdot2\tau\right).
\label{eq:1}
\end{equation}

\noindent Here, three frequency parameters are introduced: $\omega_\alpha/2\pi=J/2$, $\omega_\beta/2\pi =(J-\Delta)/2$, and $\omega_\gamma/2\pi = (J+\Delta)/2$, with an additional parameter, $\Delta=\sqrt{J^2+4\Omega^2}/2$. Note that the frequency difference $\Omega$ is proportional to $B$, and thus $\Delta$ also varies with $B$. The dipolar coupling constant is given by
\begin{equation}
J=\frac{\gamma_n^2\hbar}{r^3}(1-3\cos^2\theta),
\label{eq:2}
\end{equation}

\noindent where $r$ is the intra-dimer (${}^{13}$C--${}^{13}$C) bond length, and $\theta$ is the relative angle of $B$ to the bond direction. $J$ varies between $J(\theta =90^{\circ})=\gamma_n^2\hbar/r^3$ and $J(\theta=0^{\circ}) =-2\gamma_n^2\hbar/r^3$ as a function of $\theta$.

Fig.~\ref{fig:SED_analytical} shows the calculated oscillation frequencies ($\omega_\alpha/2\pi$, $\omega_\beta/2\pi$, and $\omega_\gamma/2\pi$) and the normalized relative intensities for each component plotted as a function of $B$ for $\theta=90^{\circ}$ [Figs.~\ref{fig:SED_analytical}(a) and \ref{fig:SED_analytical}(c)] and $\theta=0^{\circ}$ [Figs.~\ref{fig:SED_analytical}(b) and \ref{fig:SED_analytical}(d)], with $r=1.38$~\AA~and the frequency difference $\Omega$ obtained from a liner fit ($\Omega = aB$) to the experimental splitting in Fig.~4(a) in the main text. The experimental field range ($1<B<12$~T) approximately corresponds to $\Omega/J \geq 1$ for both orientations, in which the dominant oscillation comes from $\omega_\alpha$ that purely represents the direct dipole--dipole coupling $J$ and shows no field dependence. The other oscillation components, $\omega_\beta$ and $\omega_\gamma$, appear at higher frequencies due to a combined effect of the shift difference ($\Omega$) and the dipolar coupling ($J$) with a much weaker intensity and stronger field dependence. 

For polycrystalline and powder samples, the direction of dimers relative to $B$ is randomly oriented in all directions. Consequently, the size of $J$ in Eq.~(\ref{eq:1}) should vary, with a distribution of the weight following the underlying powder pattern~\cite{Pake_1948_NuclearResonanceAbsorption}. The $J$-induced $\omega_\alpha$ term, however, would remain field independent and most intense even in the powder-averaged case, which can account for the observed intense frequency component $f_1$ in the inset of Fig.~5(a) of the main text.

\subsection{Possible Aging effect seen by ${}^{13}$C NMR}

\begin{figure}[t]
	\includegraphics[width=10.0cm]{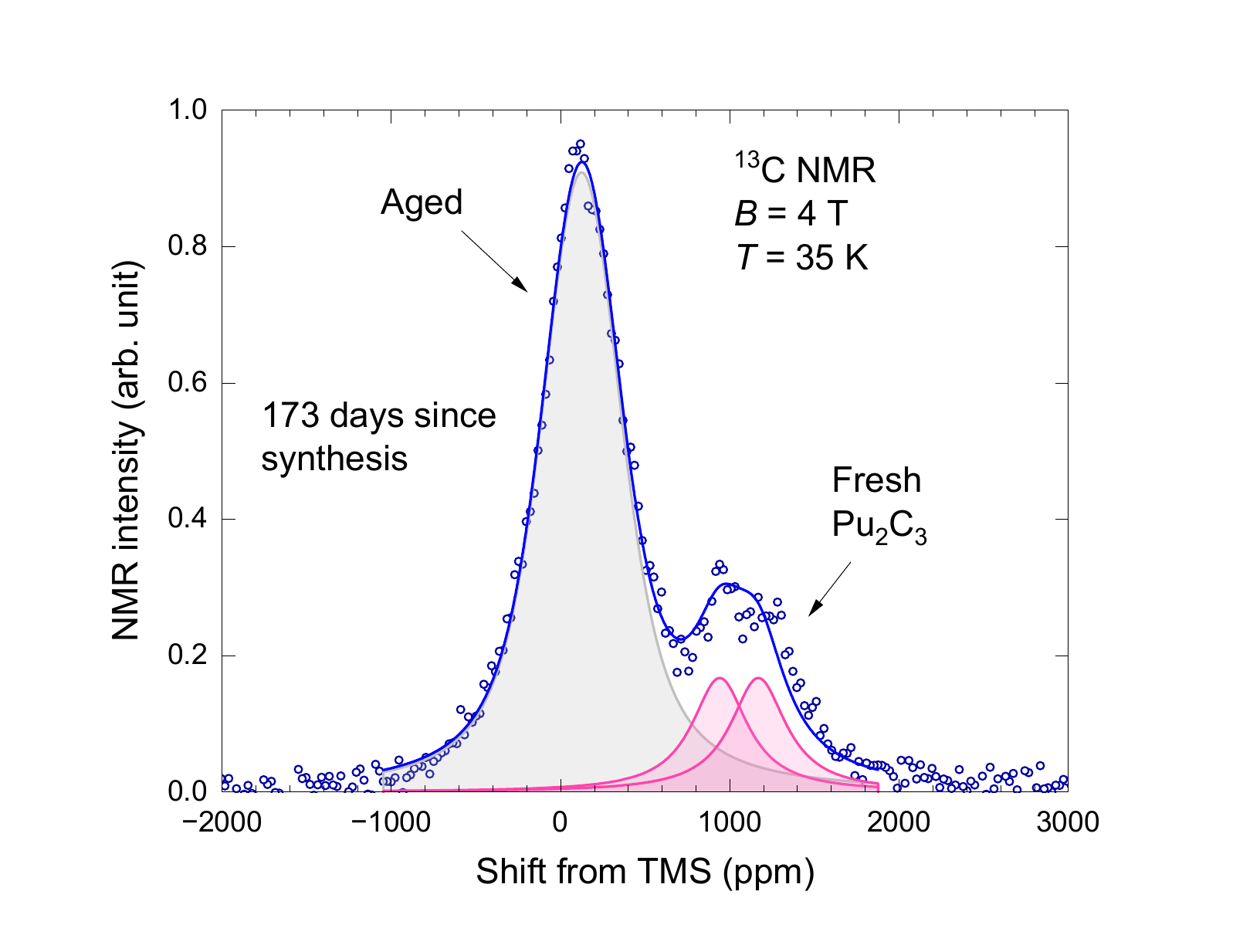}
	\caption{${}^{13}$C NMR spectrum taken a few months later in the same sample, recorded at 4~T at 35~K. On top of the signal corresponding to the fresh Pu$_{2}$C$_{3}$ (red double horn indicated as ``Fresh Pu$_{2}$C$_{3}$''), there was another much more intense peak showing up on the lower frequency side (gray peak indicated as ``Aged'').}
	\label{fig:DamagedSpectrum}
\end{figure}

In Fig.~\ref{fig:DamagedSpectrum}, we show the ${}^{13}$C spectrum measured at 4~T and 35~K in the same sample recorded 173 days after synthesis. In addition to the original signal corresponding to Pu$_{2}$C$_{3}$ (dubbed ``Fresh Pu$_{2}$C$_{3}$'' in Fig.~\ref{fig:DamagedSpectrum}, present between days 6 and 69 after synthesis), we found another more intense peak on the lower-frequency (lower-$f$) side with a center of gravity shift value of $\approx$~70~ppm (dubbed ``Aged'' in Fig.~\ref{fig:DamagedSpectrum}). A $T_1$- and $T_2$-correction to FFT intensities of each peak yields an intensity ratio of $\sim 1:8$ at 173 days later from original sample synthesis. The high-$f$ Pu$_{2}$C$_{3}$ peak retained all the original characteristics found in the fresh sample: the two-horn structure in the spectrum, the SED oscillation pattern, $\overline{\delta_{\mathrm{iso}}}$, $T_1$, and $T_2$. In contrast, the low-$f$ resonance did not share these features, suggesting that the low-$f$ line comes from ${}^{13}$C atoms in a different (likely disordered) local environment.

\begin{figure}[t]
	\includegraphics[width=15cm]{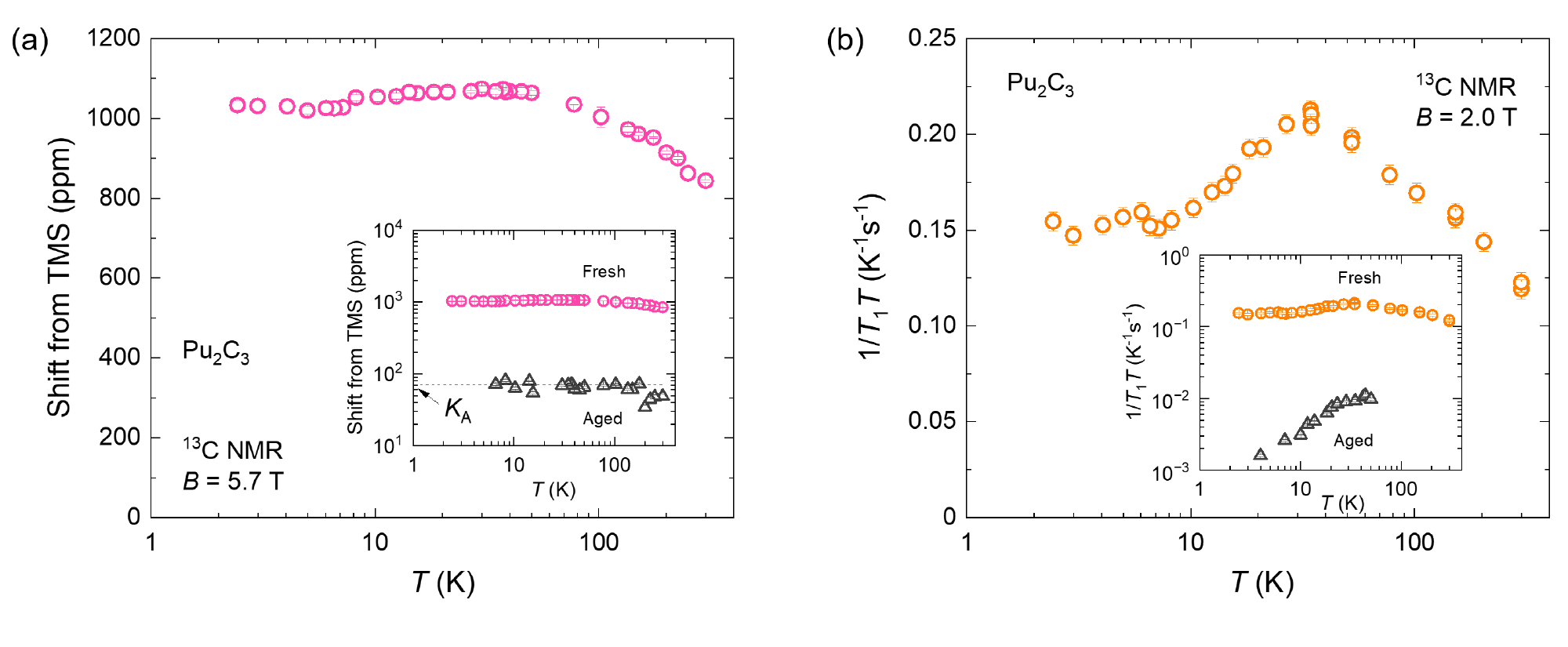}
	\caption{Temperature dependence of the average ${}^{13}$C isotropic shift $\overline{\delta_{\mathrm{iso}}}$ measured at 5.7~T (a) and spin-lattice relaxation rate divided by temperature $1/T_{1}T$ measured at 2.0~T (b) of fresh Pu$_{2}$C$_{3}$ powder (replotted from the main text). Insets show the corresponding results in the aged sample (a few months later) plotted together with the fresh sample in a logarithmic scale. The relaxation rate in the aged sample was measured at 5.7~T to reduce the overlap of two lines (see Fig.~\ref{fig:DamagedSpectrum}).}
	\label{fig:shiftT1Aged}
\end{figure}

It is interesting to note that in this low-$f$ ``Aged'' peak, we found very different temperature dependence in the shift and $1/T_{1}T$ compared to the ``Fresh Pu$_{2}$C$_{3}$'' peak, as summarized in the insets of Fig.~\ref{fig:shiftT1Aged}. The shift value shows no temperature dependence with a constant value of $K_\mathrm{A} \approx$~70~ppm [inset of Fig.~\ref{fig:shiftT1Aged}(a)]. This value falls within the typical range of shift values for ${}^{13}$C resonance in nonmagnetic insulators (between 0 and 200~ppm)~\cite{Tossell_1992_NuclearMagneticShieldings}. Activation-like behavior is observed in $1/T_{1}T$ [inset of Fig.~\ref{fig:shiftT1Aged}(b)] suggesting that the aged part of the sample is a nonmagnetic band insulator.

We note that unlike in our NMR measurements, there was only a negligibly small change in heat capacity and susceptibility after several months from the original experiment. This may be related to the different time span the samples were exposed to the air prior to the experiment. Actinide carbides are known to be sensitive to oxidation and hydrolysis. The powder samples used in NMR had larger contact surface area and might have reacted with air much quicker. As such, it may be relevant that specific heat and susceptibility samples were exposed to air less than 30 minutes before the experiment, whereas the NMR sample was exposed to air for at least one to two hours. As a result, a fraction of NMR sample may have reacted with oxygen/water in the air~\cite{Morss_2011_ChemistryActinideTransactinide}, which resulted in a generation of the low-$f$ signal in the ${}^{13}$C spectrum.

In another plutonium compound, PuB$_4$---with the same isotopic makeup of Pu---a notable NMR-signal-intensity loss of the ${}^{239}$Pu NMR signal was observed. This was associated with self-damage effects linked to radioactive $\alpha$-decay of Pu, which developed over the course of a few years, but not within a few months~\cite{Blackwell_2025_Localagingeffects}. The NMR signal of the ligand $^{11}$B was still observable in an aged sample where the ${}^{239}$Pu signal was completely lost, suggesting that ligand sites might have a lower sensitivity to self-damage effects. Due to the rapid rate at which the ${}^{13}$C spectrum changes in Pu$_2$C$_3$, it is unlikely that this effect is due to self-damage. Further investigation such as x-ray diffraction study is needed to fully understand the chemistry of this damaged part of sample.

\bibliography{Pu2C3_SI}